\documentclass{exstill}   
\usepackage{pgfplots}
\pgfplotsset{compat=1.18}
\usetikzlibrary{positioning,arrows.meta,calc,fit,backgrounds,decorations.pathreplacing}
\usepgfplotslibrary{fillbetween}

\pgfplotsset{
  exs/.style={
    width=\linewidth, height=6.2cm,
    scale only axis=false,
    font=\figfont\scriptsize,
    every axis label/.append style={font=\figfont\scriptsize, color=exs-muted},
    every tick label/.append style={font=\figfont\scriptsize, color=exs-muted},
    axis line style={draw=exs-line, line width=0.4pt},
    tick style={draw=none},
    grid=major,
    major grid style={draw=exs-grid, line width=0.4pt, solid},
    axis lines*=left,
    clip=false,
    legend style={draw=none, fill=none, font=\figfont\scriptsize, text=exs-ink-2},
    legend cell align=left,
  },
  exs gpu/.style={draw=exs-mark-gpu, line width=1.2pt, line join=round, line cap=round,
    mark=*, mark size=2.4pt, mark options={fill=exs-mark-gpu, draw=white, line width=0.9pt}},
  exs us/.style={draw=exs-mark-us, line width=1.2pt, mark=*, mark size=2.8pt,
    mark options={fill=exs-mark-us, draw=white, line width=0.9pt}},
  exs floor/.style={draw=exs-meas, line width=1pt, dashed, dash pattern=on 3pt off 2pt},
  exs ref/.style={draw=exs-faint, line width=0.6pt, dash pattern=on 2.5pt off 2pt},
}
\tikzset{
  exs refline/.style={draw=exs-faint, line width=0.6pt, dash pattern=on 2.5pt off 2pt},
  exs floorline/.style={draw=exs-meas, line width=1pt, dash pattern=on 3pt off 2pt},
  exs label/.style={font=\figfont\scriptsize, text=exs-ink-2},
  exs note/.style={font=\figfont\scriptsize, text=exs-muted},
  exs strong/.style={font=\figfont\scriptsize\bfseries, text=exs-ink},
  exs wash/.style={fill=exs-mark-gpu-soft!26, draw=none},
  exs dim/.style={draw=exs-faint, line width=0.5pt, {Stealth[length=3pt]}-{Stealth[length=3pt]}},
  exs arrow/.style={draw=exs-faint, line width=0.5pt, -{Stealth[length=3pt]}},
}

\doclabel{Technical Report}
\title{The Most Efficient Single-Chip Processor for the Next Token}
\fittitle
\addauthor{Valavan Manohararajah}{valavan@exstill.ai}
\addauthor{Vithursan Thangarasa}{vithu@exstill.ai}
\date{}
\homepage{https://exstill.ai/co-processor}

\begin{document}

\maketitle

\begin{abstract}
Generating a token with a language model requires a fixed amount of arithmetic, whose energy sets a hard floor beneath any hardware implementation. For a representative small model (Qwen3-1.7B at a 16K-token context), gate-level measurements at the 5~nm class place this floor at 0.89~mJ per token, yet an NVIDIA B300 spends \mbox{391-1,120x} that. The excess is the price of generality: layers of software and programmable hardware, and the movement of weights and key-value (KV) caches from external memory. Data movement improves more slowly than logic with each process generation, so the excess widens and the case for removing it grows. We present a co-design approach that brings token generation close to its arithmetic floor and then lowers the floor itself, by jointly optimizing silicon and model architectures within a gate-level search loop. By fixing a single model architecture in the datapath (while keeping weights loadable) and holding the weights and every conversation's KV cache on-die beside the arithmetic, a pre-silicon design at the 5~nm class brings an unchanged Qwen3-1.7B token to 3.4x its floor. Co-designing the model alongside the circuits at fixed quality lowers the floor to 0.29~mJ and total token energy to 0.52~mJ, 1.8x the new floor and less than the baseline model's arithmetic alone, while expanding on-die capacity from one user to twelve. In the same arithmetic, no chip running this model could halve that energy. Against a measured B300 serving the same twelve users, the pre-silicon design decodes each user's tokens 110x faster (50~\textmu s per token) at 690x lower energy. To demonstrate that one fixed architecture can serve, as a co-processor, the latency-critical workloads around a frontier model, we introduce the Polyglot-DC model family (1.7B and 4B parameters, each with a design of its own), specialized for each job by distillation and task-specific training into its own set of weights. As a drafter, it speeds single-user speculative decoding of large open models by up to 6.6x at the maximum reasoning effort, and its guard returns multilingual verdicts in a millisecond or less, within a point of the strongest open guard. Its agent and financial-reasoning models lead comparable open models on well-established agentic and financial benchmarks, while a reasoning turn falls from seconds to tens of milliseconds.
\end{abstract}

\contentspage

\section{Introduction}
\label{sec:intro}

Much of modern AI works one token at a time, and someone usually waits on each one: a person reading a reply, or a program waiting on an agent's next action. Often the wait is on the smaller models that surround a frontier model, the large general model that composes the answer. Before and during its response, they screen prompts and outputs for safety, draft tokens ahead for speculative decoding, select the next tool call in an agent's loop, and reason through domain-specific questions. These models typically serve one user or a few at a time, yet they run on general-purpose hardware built to batch very large models across many users. This report asks what a token from a small model needs to cost, why it costs hundreds of times more today, and how designing the silicon and the model together can both close that gap and lower the limit itself.

What a token needs to cost is set by its arithmetic. A model's architecture fixes the multiply-accumulate operations each of its tokens requires, and the energy of those operations is a floor that no hardware implementation can go below: a limit that belongs to the model, whatever chip runs it. Measured on circuits placed and routed at the 5~nm class, the floor for Qwen3-1.7B \citep{yang2025qwen3}, a representative small model writing against a 16K-token context, is 0.89~mJ per token.

Today's best hardware spends hundreds of times more. Serving one user, an NVIDIA B300 spends 994~mJ on the same token, 1,120x the floor. Sharing the GPU among sixteen users, the conventional remedy known as batching, lowers the energy to 347~mJ, still 391x the floor, but each user then waits three times as long for a token. Batching is thus a dial that trades the operator's energy against the user's time, and it leaves their product, the energy-delay, almost unchanged.

More than 99.7\% of what the GPU spends is therefore not arithmetic, and this excess has two causes. The first is the layers of software and programmable hardware between a model and a general-purpose chip's arithmetic: frameworks, serving engines, kernel launches and instruction decoding, to which we attribute a fixed charge of more than a millisecond on every token, however small the model. The second is the movement of data. A chip that must run any model at any size keeps the weights and each conversation's stored state, its key-value (KV) cache, in external high-bandwidth memory, and fetches gigabytes of them across the memory interface for every token. Both are the price of generality, paid by hardware designed before its model existed, and neither is a matter of physics. Nor does the usual progress remove them: better software reduces only the first, a smaller model only the second, and a newer process improves logic much faster than wires and memory, so the distance between a token and its floor widens with each generation.

A token's energy on any hardware is thus the product of two factors: the energy of the arithmetic its model requires, and the multiple of that energy the hardware spends to carry it out. Our approach makes both small by co-designing the architecture of the silicon with the architecture of the model, and it rests on three decisions. The first removes the layers. We fix one model architecture in the silicon's datapath, so that there are no kernels to launch and no instructions to schedule; the weights remain loadable, so a new model for a new task is a training run rather than a new chip. The second removes the movement. Everything a token needs, the weights and the KV cache of every active conversation, is kept on a single die beside the arithmetic, and nothing crosses an external memory interface. The third lowers the floor itself. How much fits on the die and how much arithmetic a token takes both depend on the model's structure, so we treat the model's architecture as a design variable, optimized together with the silicon at a fixed quality.

Each of these ingredients has precedent. Their combination in one closed loop does not: model architectures are explored and trained while the chip that would run each is costed at the gate level, on placed-and-routed circuits and over every token its model would write. We have costed more than ten million whole designs this way, and it is that loop, rather than any one decision, on which the results of this report rest.

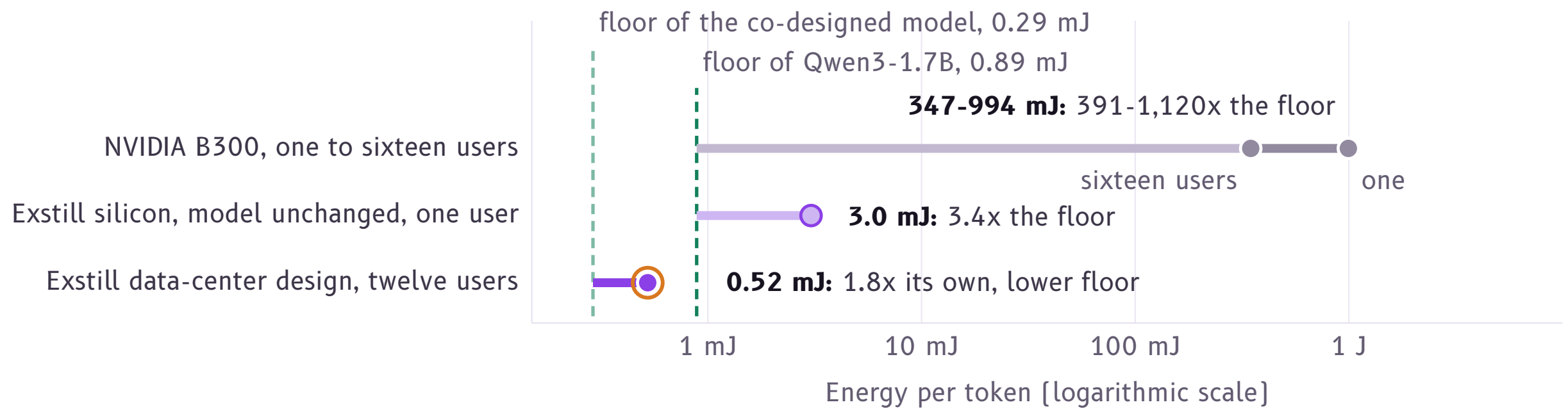
\begin{figure}[t]
  \centering
\begin{tikzpicture}
\begin{axis}[exs, height=4.7cm, width=0.74\linewidth,
  xmode=log, xmin=0.15, xmax=10000,
  xtick={1,10,100,1000}, xticklabels={1 mJ,10 mJ,100 mJ,1 J},
  xlabel={Energy per token (logarithmic scale)},
  ymin=0.4, ymax=4.9, ytick={1,2,3},
  yticklabels={{Exstill data-center design, twelve users}, {Exstill silicon, model unchanged, one user}, {NVIDIA B300, one to sixteen users}},
  yticklabel style={font=\figfont\scriptsize, text=exs-ink-2},
  ymajorgrids=false,
]
  \draw[exs floorline] (axis cs:0.8865,0.5) -- (axis cs:0.8865,3.9);
  \draw[exs floorline, draw=exs-meas!55] (axis cs:0.29,0.5) -- (axis cs:0.29,4.45);
  \node[exs note, anchor=south west, xshift=-2pt] at (axis cs:0.8865,3.9) {floor of Qwen3-1.7B, 0.89 mJ};
  \node[exs note, anchor=south west, xshift=-2pt] at (axis cs:0.29,4.45) {floor of the co-designed model, 0.29 mJ};
  \draw[exs-mark-gpu-soft, line width=2.6pt] (axis cs:0.8865,3) -- (axis cs:347.2,3);
  \draw[exs-mark-gpu, line width=2.6pt] (axis cs:347.2,3) -- (axis cs:994.2,3);
  \draw[exs-violet-lo, line width=2.6pt] (axis cs:0.8865,2) -- (axis cs:3.041,2);
  \draw[exs-mark-us, line width=2.6pt] (axis cs:0.29,1) -- (axis cs:0.523,1);
  \addplot[only marks, mark=*, mark size=3pt, mark options={fill=exs-mark-gpu, draw=white, line width=0.9pt}] coordinates {(347.2,3) (994.2,3)};
  \addplot[only marks, mark=*, mark size=3pt, mark options={fill=exs-violet-lo, draw=exs-mark-us, line width=0.9pt}] coordinates {(3.041,2)};
  \draw[exs-mark-amber, line width=1.1pt] (axis cs:0.523,1) circle (4.4pt);
  \addplot[only marks, mark=*, mark size=3pt, mark options={fill=exs-mark-us, draw=white, line width=0.9pt}] coordinates {(0.523,1)};
  \node[exs label, anchor=south east, yshift=5pt] at (axis cs:994.2,3) {{\bfseries\color{exs-ink}347-994 mJ:} 391-1,120x the floor};
  \node[exs note, anchor=base east, yshift=-12pt] at (axis cs:347.2,3) {sixteen users};
  \node[exs note, anchor=base west, yshift=-12pt] at (axis cs:994.2,3) {one};
  \node[exs label, anchor=west, xshift=7pt] at (axis cs:3.041,2) {{\bfseries\color{exs-ink}3.0 mJ:} 3.4x the floor};
  \node[exs label, anchor=west, xshift=5pt] at (axis cs:0.8865,1) {{\bfseries\color{exs-ink}0.52 mJ:} 1.8x its own, lower floor};
\end{axis}
\end{tikzpicture}
  \caption{A token's energy against the floor of its model, at a 16K-token context, on a logarithmic scale. Each bar runs from the floor, the energy of the arithmetic the model requires, measured on our gate-level flow (\cref{sec:cost-floor}), to the energy the hardware spends, so its length is the hardware's multiple. The B300 serves Qwen3-1.7B and is measured (\cref{tab:cost-sweep}), from sixteen users, its most efficient setting, to one. The Exstill designs are pre-silicon at the 5~nm class, both within 125~W and 575~mm\textsuperscript{2}, and the co-designed model is held to Qwen3-1.7B's quality (\cref{sec:chip-silicon,sec:chip-model}).}
  \label{fig:intro-floor}
\end{figure}

Two pre-silicon designs returned by the search, both within a data-center envelope of 125~W and 575~mm\textsuperscript{2}, isolate what each half of the co-design contributes (\cref{fig:intro-floor}). In the first, Qwen3-1.7B is held exactly as released and only the silicon is designed. The token comes within 3.4x of its floor, removing nearly all of the GPU's excess, but silicon alone can do little more: it cannot go below the floor of a model chosen in advance, and this model leaves the die room for only one 16K-token conversation. In the second, the model's architecture is freed and co-designed with the silicon at the same quality. The floor falls by two-thirds, the hardware stays within 1.8x of the new floor, and the whole token costs less than the arithmetic alone of the model it replaces, while the same envelope holds twelve conversations.

The second design is the frontier model's co-processor: a data-center chip that sits beside the accelerator running the frontier model, on a standard link, and takes over the small, latency-critical models around it. It serves twelve users at 19,921 tokens per second each, about 239,000 in total, and against a measured NVIDIA B300 serving the same twelve users the pre-silicon design decodes each user's next token 110x faster at 690x lower energy. The floor gives that comparison a wider reach, and it is the sense in which the claim in the report's title is meant. No token can cost less than its arithmetic, so a chip running the same model in the same arithmetic could improve on the design's energy by less than a factor of two, and a chip running Qwen3-1.7B as released could not match it at all, since that model's arithmetic alone costs more than the design's whole token.

The same loop, run for a 4B-parameter model at a tighter quality bar within 150~W and 750~mm\textsuperscript{2}, returns a second and larger design, the DC2 design (\cref{sec:chip-dc2}). It serves four users at 13,998 tokens per second each, 58x faster and at 360x lower energy than a B300 serving the same four.

Fixing one architecture in silicon is worthwhile only if that architecture can do enough useful work, and the second half of the report shows that it can. Polyglot-DC is a family of models on the co-designed architecture, at 1.7B and 4B parameters, specialized by distillation into four roles that differ only in their weights. The first is drafting for speculative decoding, in which a small model proposes the next few tokens and a large model verifies them in a single pass. Polyglot-DC1-Draft runs on its own chip, apart from the GPU, and is aligned to five open models of 80B to 2.8T parameters, whose single-user decoding it speeds by up to 6.6x.

In the other three roles the user waits on the small model itself. As a guard screening prompts and streaming replies in six languages, the 4B model comes within one F1 point of the strongest open guard while flagging fewer than half as many harmless prompts, and a verdict at either size takes a millisecond or less, \mbox{38-39x} faster than on a B300. As an agent, the 4B model leads eight open models of comparable size on $\tau^3$-bench, a test of multi-turn tool use under a written policy; replayed at the 1.7B size's decoding rates, an average reasoning turn from its episodes takes 57~ms where the B300 takes 2.46~s. In financial reasoning, the 4B model leads seven open models on nine of ten benchmarks, and a reasoned answer from the 1.7B model that takes the B300 1.68~s on average takes 44~ms.

The chip has not yet been fabricated, so the report states how each number is known. The GPU's figures are measured on our bench, on one NVIDIA B300 running vLLM as shipped, and the models' quality is measured on public benchmarks. The chip's figures are pre-silicon estimates from the gate-level costing described above, which covers every layer and every token of the whole model rather than one kernel scaled up. Results that combine the two, such as the drafting speedups, are simulated. We estimate that a hand-tuned serving stack would narrow each ratio by about half, though it would leave the memory traffic that dominates when several users share the GPU. Against other specialized chips, which we have not measured, the claim rests on the argument from the floor, not on a bench.

The rest of the report follows this order. \Cref{sec:cost} establishes the floor, measures the B300's distance from it and traces that distance to the layers and the movement of data. \Cref{sec:chip} describes the three decisions and the search, sets the unchanged and co-designed results against the B300, places both designs beside two others that keep a model on the chip, and states their limitations. \Cref{sec:polyglot} introduces Polyglot-DC and its distillations, \cref{sec:specdec,sec:guard,sec:agent,sec:finance} examine its four roles in turn, and \cref{sec:conclusion} draws the argument together. The data center is the first setting for the approach; others, from robots to devices at the edge, are left to later work.

\section{Where a Token's Cost Lies}
\label{sec:cost}

This section and the next make the case for the claim in this report's title. The case begins from a limit rather than from an existing architecture: the energy of the arithmetic a token requires, measured on real circuits, which no implementation of its model can go below. A token's energy on any hardware is the product of that limit and the multiple of it the hardware actually spends, and a processor is efficient to the extent that both factors are small. This section establishes the limit, measures the multiple on the best chip available and traces where it comes from; \cref{sec:chip} returns by the same steps in reverse.
\subsection{The Limit: The Arithmetic Alone}
\label{sec:cost-floor}

To say how far a token's cost is from what it needs to be, we need a reference that does not depend on how any particular chip is organized. We use the energy of the token's arithmetic alone: the multiply-accumulate operations the model requires to produce one token, each a multiplication whose result is added to a running sum, with nothing moved, nothing scheduled and nothing idle. We call this the floor, and it is the limit against which every token in this report is measured.

We measured the arithmetic directly rather than taking it from a published estimate. To do so we built the datapath that a standard accelerator uses for 8-bit floating point (FP8), the circuit that carries out the arithmetic: a dot product that forms sixteen FP8 products per cycle and accumulates them in 16-bit floating point. The circuit was carried through placement and routing at the 5~nm class, the stage of physical design at which wire lengths and timing become known, and then simulated gate by gate on a real layer's data, counting every transition, including glitches, the spurious transitions a signal makes before it settles. Each multiply-accumulate costs 246~fJ, its share of the accumulation and rounding included. We call this procedure our gate-level flow.

The reference model throughout this report is Qwen3-1.7B \citep{yang2025qwen3}, representative of the small models that work around a frontier model. The number of multiply-accumulates it needs per token follows from its public configuration. In a decoder-only transformer, each generated token multiplies every weight once and, in every layer, attends over every earlier token of the context:
\begin{equation}
  N_{\text{MAC}} = P + 2\, n_{\text{L}}\, n_{\text{H}}\, d_{\text{h}}\, L ,
  \qquad
  E_{\text{floor}} = N_{\text{MAC}} \cdot e_{\text{MAC}} ,
  \label{eq:floor}
\end{equation}
where $P$ is the number of weights used per token, $n_{\text{L}}$ the number of layers, $n_{\text{H}}$ the number of query heads, $d_{\text{h}}$ the head dimension, $L$ the context length and $e_{\text{MAC}}$ the energy of one multiply-accumulate. The factor of two counts the two products attention forms against the context, the query against each key and the attention weights against each value. Qwen3-1.7B uses about $1.72 \times 10^9$ weights per token, including its output projection, and has 28 layers of 16 query heads of dimension 128. At a 16K-token context, attention adds $1.88 \times 10^9$ multiply-accumulates, for $3.60 \times 10^9$ in all, and at 246~fJ each the floor is 0.89~mJ. We chose FP8 because it is the format of the model's published checkpoint and, in our measurements, the B300's lowest-energy setting for it, so the floor and the GPU measurement that follows use the same arithmetic.

The floor is the limit for this model in this arithmetic, not for the task. No chip running Qwen3-1.7B unchanged in an FP8 datapath of this kind can spend less than 0.89~mJ on a token at this context. A different model that reaches the same quality with less arithmetic, or with cheaper arithmetic, has a lower floor of its own. A model built for every task has a higher one than a narrow question needs, since it carries arithmetic and state the question does not use. The floor, the first of the two factors, therefore belongs to the model rather than to the chip, and \cref{sec:chip-model} returns to it.

\subsection{The Distance: A GPU Token Against the Limit}
\label{sec:cost-gpu}

The small models around a frontier model usually serve one person or a few at a time, with someone waiting on every token. We measured that workload on one NVIDIA B300 running Qwen3-1.7B from its published FP8 checkpoint under the vLLM serving engine \citep{kwon2023vllm} as shipped. Each user is given a 15K-token prompt followed by 1K generated tokens, so that every token is written against a context of about 16K, and the number of users served at once is varied from one to sixteen. We report decoding only, the writing of the generated tokens once the prompt has been read. \Cref{tab:cost-sweep} gives the results of this sweep, with each quantity defined in its caption.

\begin{table}[t]
  \centering
  \caption{Qwen3-1.7B in FP8 on one NVIDIA B300 with vLLM as shipped, 15K tokens of prompt and 1K generated per user, decoding only, measured on 16 October 2026. Energy per token is mean board power over decoding time divided by tokens; the time between a user's tokens is the median. Energy-delay is their product (\cref{sec:cost-metric}). The last column divides energy per token by the 0.89~mJ floor of \cref{sec:cost-floor}.}
  \label{tab:cost-sweep}
  \small
  \begin{tabular}{rrrrrrr}
    \toprule
    \headrow Users & \makecell[r]{Time between\\tokens} & \makecell[r]{Energy\\per token} & \makecell[r]{Board\\power} & \makecell[r]{Total\\tokens/s} & \makecell[r]{Energy-\\delay} & \makecell[r]{Over the\\floor} \\
    \midrule
    1 & 2.15~ms & 994~mJ & 461~W & 445 & 2.14~J$\cdot$ms & 1,120x \\
    2 & 2.41~ms & 630~mJ & 522~W & 790 & 1.52~J$\cdot$ms & 711x \\
    4 & 2.95~ms & 459~mJ & 623~W & 1,284 & 1.35~J$\cdot$ms & 518x \\
    8 & 4.09~ms & 382~mJ & 747~W & 1,830 & 1.56~J$\cdot$ms & 431x \\
    12 & 5.35~ms & 361~mJ & 810~W & 2,094 & 1.93~J$\cdot$ms & 407x \\
    16 & 6.35~ms & 347~mJ & 869~W & 2,346 & 2.20~J$\cdot$ms & 391x \\
    \bottomrule
  \end{tabular}
\end{table}

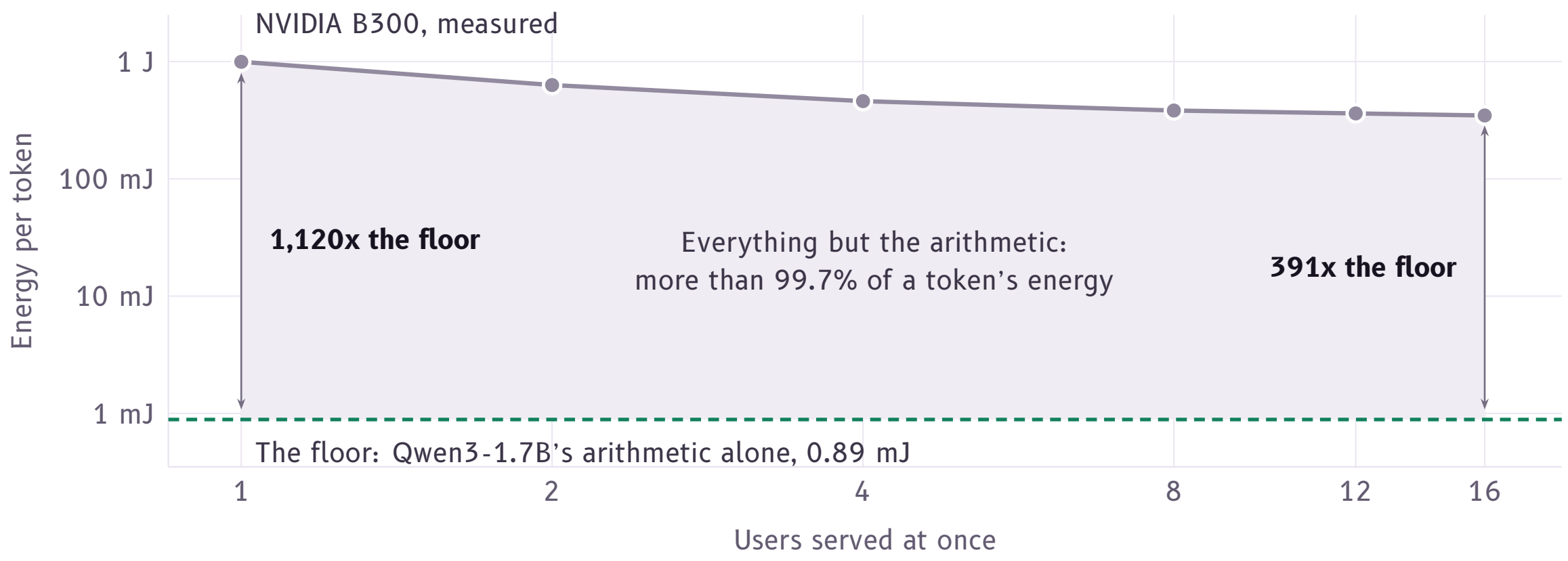
\begin{figure}[t]
  \centering
\begin{tikzpicture}
\begin{axis}[exs, height=6.0cm, width=0.92\linewidth,
  xmode=log, log basis x=2, xmin=0.85, xmax=19,
  xtick={1,2,4,8,12,16}, xticklabels={1,2,4,8,12,16},
  xlabel={Users served at once},
  ymode=log, ymin=0.35, ymax=2500,
  ytick={1,10,100,1000}, yticklabels={1 mJ,10 mJ,100 mJ,1 J},
  ylabel={Energy per token}, ylabel style={yshift=-2pt},
]
  \fill[exs wash] (axis cs:1,994.29) -- (axis cs:2,630.34) -- (axis cs:4,459.46) -- (axis cs:8,382.16)
    -- (axis cs:12,361.02) -- (axis cs:16,346.78) -- (axis cs:16,0.8865) -- (axis cs:1,0.8865) -- cycle;
  \addplot[exs gpu] coordinates {(1,994.29) (2,630.34) (4,459.46) (8,382.16) (12,361.02) (16,346.78)};
  \addplot[exs floor, domain=0.85:19, samples=2] {0.8865};
  \node[exs label, anchor=south west, yshift=3pt] at (axis cs:1,994.29) {NVIDIA B300, measured};
  \node[exs label, anchor=north west, yshift=-2pt] at (axis cs:1,0.8865) {The floor: Qwen3-1.7B's arithmetic alone, 0.89 mJ};
  \draw[exs dim] (axis cs:1,800) -- (axis cs:1,1.08);
  \draw[exs dim] (axis cs:16,285) -- (axis cs:16,1.08);
  \node[exs strong, anchor=west, xshift=4pt] at (axis cs:1,30) {1,120x the floor};
  \node[exs strong, anchor=east, xshift=-4pt] at (axis cs:16,18) {391x the floor};
  \node[exs label, align=center] at (axis cs:4.1,19) {Everything but the arithmetic:\\more than 99.7\% of a token's energy};
\end{axis}
\end{tikzpicture}
  \caption{Energy per token on one NVIDIA B300 serving Qwen3-1.7B in FP8 at a 16K-token context, decoding only, as one to sixteen users share it (measured, \cref{tab:cost-sweep}), against the floor, the energy of the same token's arithmetic alone (measured on our gate-level flow, \cref{sec:cost-floor}). The shaded region between them is energy spent on something other than the token's arithmetic. The energy scale is logarithmic.}
  \label{fig:cost-floor}
\end{figure}

Serving one user, the B300 writes a token every 2.15~ms and spends 994~mJ on it, 1,120x the floor. Serving sixteen, each user waits 6.35~ms between tokens, three times as long, and each token costs 347~mJ, about a third as much. The arithmetic is the same throughout, yet it is less than a tenth of a percent of a token's energy with one user and about a quarter of a percent with sixteen. \Cref{fig:cost-floor} draws the sweep against the floor: as more users share the GPU, the energy per token falls and then levels off, more than two orders of magnitude above the arithmetic it pays for.

The same distance appears in time, because energy is power multiplied by time. A chip that drew the 461~W the board draws for one user and spent all of it on the arithmetic would finish a token of Qwen3-1.7B in 1.9~\textmu s, where the B300 takes 2.15~ms. The factor by which the operator of the chip overpays in energy is the factor by which the user overpays in waiting.

\subsection{Batching Is a Dial, and How to Score a Token}
\label{sec:cost-metric}

Batching is the conventional remedy for a costly token. More users share each decoding step, the pass through the model that writes one token for every user being served, and the sweep shows the result: total output rises more than fivefold, from 445 to 2,346 tokens per second, and the operator's energy per token falls. The saving is paid for entirely by the users, in time.

Whether that counts as an improvement depends on the measure. By the figure of merit most AI chips lead with, tokens per joule (equivalently, tokens per second per watt), the B300 improves nearly threefold, from 1.0 to 2.9. But tokens per joule rewards any change that lowers energy per token whatever it costs in time, and it can be raised in ways that serve no one waiting: by batching more users, by running the chip more slowly, by compressing the model past the point where its answers degrade, or by adopting a tokenizer that splits the same text into more tokens. Benchmarks that add latency limits and accuracy targets help, but they are pass/fail: they report whether a system met a bound, not what the wait cost.

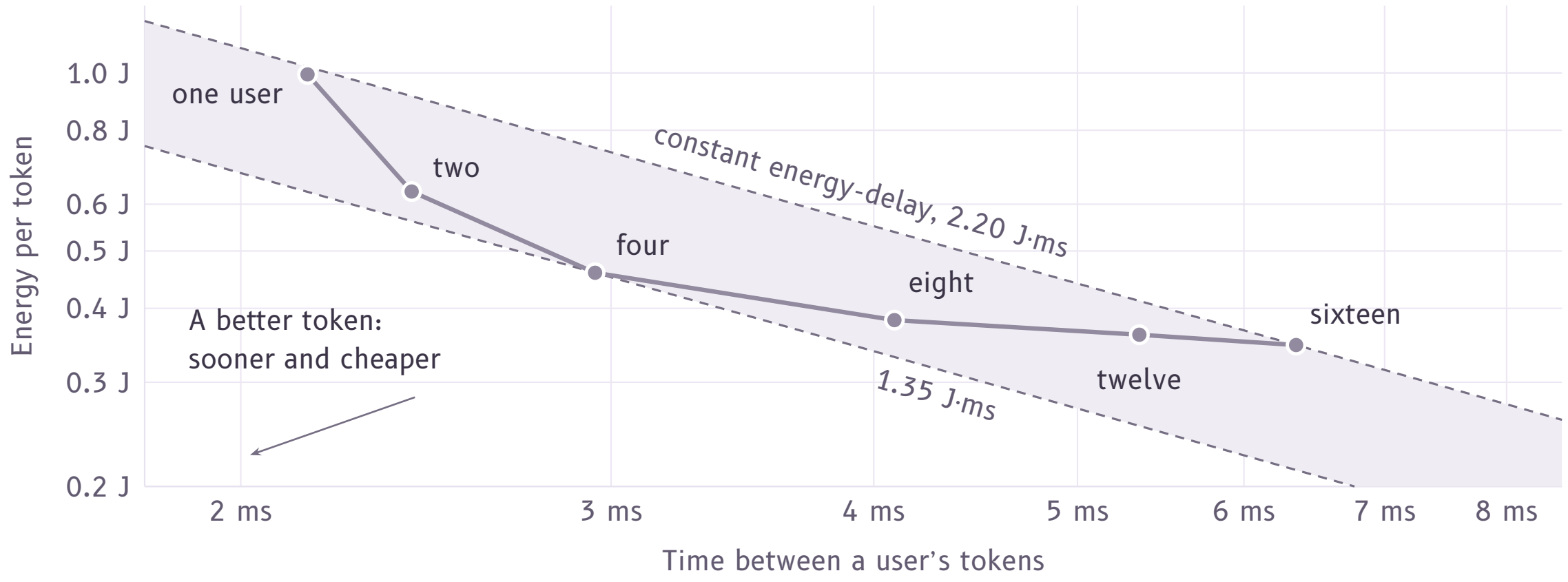
\begin{figure}[t]
  \centering
\begin{tikzpicture}
\begin{axis}[exs, height=6.2cm, width=0.92\linewidth,
  xmode=log, ymode=log, xmin=1.8, xmax=8.5, ymin=0.2, ymax=1.3,
  xtick={2,3,4,5,6,7,8}, xticklabels={2 ms,3 ms,4 ms,5 ms,6 ms,7 ms,8 ms},
  ytick={0.2,0.3,0.4,0.5,0.6,0.8,1.0}, yticklabels={0.2 J,0.3 J,0.4 J,0.5 J,0.6 J,0.8 J,1.0 J},
  xlabel={Time between a user's tokens},
  ylabel={Energy per token},
]
  \fill[exs wash] (axis cs:1.8,0.7525) -- (axis cs:6.7725,0.2) -- (axis cs:8.5,0.2) -- (axis cs:8.5,0.2592)
    -- (axis cs:1.8,1.2240) -- cycle;
  \draw[exs refline] (axis cs:1.8,1.2240) -- node[pos=0.5, sloped, above, exs note] {constant energy-delay, 2.20 J$\cdot$ms}
    (axis cs:8.5,0.2592);
  \draw[exs refline] (axis cs:1.8,0.7525) -- node[pos=0.66, sloped, below, exs note] {1.35 J$\cdot$ms}
    (axis cs:6.7725,0.2);
  \addplot[exs gpu] coordinates {(2.1515,0.99429) (2.4114,0.63034) (2.9481,0.45946) (4.0917,0.38216) (5.3505,0.36102) (6.3532,0.34678)};
  \node[exs label, anchor=north east, xshift=-3pt] at (axis cs:2.1515,0.99429) {one user};
  \node[exs label, anchor=south west, xshift=2pt] at (axis cs:2.4114,0.63034) {two};
  \node[exs label, anchor=south west, xshift=2pt, yshift=1pt] at (axis cs:2.9481,0.45946) {four};
  \node[exs label, anchor=south west, yshift=2pt] at (axis cs:4.0917,0.38216) {eight};
  \node[exs label, anchor=north, yshift=-5pt] at (axis cs:5.3505,0.36102) {twelve};
  \node[exs label, anchor=south west, yshift=2pt] at (axis cs:6.3532,0.34678) {sixteen};
  \node[exs label, anchor=south west, align=left] (better) at (axis cs:1.86,0.292) {A better token:\\sooner and cheaper};
  \draw[exs arrow] (axis cs:2.42,0.283) -- (axis cs:2.02,0.226);
\end{axis}
\end{tikzpicture}
  \caption{The batching dial. Energy per token against the time between a user's tokens on one NVIDIA B300 serving Qwen3-1.7B in FP8 at a 16K-token context, one to sixteen users (measured, \cref{tab:cost-sweep}). Both scales are logarithmic, so a line of constant energy-delay is a straight diagonal; the shaded band lies between the least energy-delay of the sweep, at four users, and the greatest, at sixteen. Batching trades the user's time for the operator's energy and leaves their product within a factor of 1.6.}
  \label{fig:cost-dial}
\end{figure}

\citet{gonzalez1996energy} met the same problem in comparing microprocessors and resolved it by scoring a design on the product of energy and delay. A change that saves energy by making the user wait, or saves time by spending more power, has to pay for itself in the product. Scored this way, a token on the B300 costs 2.14~J$\cdot$ms with one user and 2.20~J$\cdot$ms with sixteen, essentially where it began; it is lowest at four users, at 1.35~J$\cdot$ms, and the whole range spans a factor of 1.6. In \cref{fig:cost-dial} a token that is better for both the user and the operator lies toward the lower left, and as users are added the B300 does not move that way: it slides down and to the right, inside a band of nearly constant energy-delay. Batching is therefore a dial rather than a remedy. It moves the cost between the operator and the user without removing it, and at its most efficient setting the GPU still sits nearly four hundred times above the floor.

Energy-delay still rewards a model that answers worse but faster, so we fix quality first. We measure it in bits per byte: the model's total negative log-likelihood on a text, in bits, divided by the text's length in bytes, so that fewer bits per byte means a model that predicts the text better. Normalized by bytes rather than tokens, it cannot be improved by a finer tokenizer. Every design in this report is therefore the answer to one question: the least energy-delay per token that clears a stated quality bar, within a stated budget of power and area.

\begin{insight}{Score a token by energy times delay, at a fixed quality}
  Tokens per joule improves whenever users wait longer, so it cannot distinguish a better design from a more heavily batched one. Energy-delay at a fixed quality bar can, and by that measure batching a GPU leaves a small model's token almost exactly where it started.
\end{insight}

\subsection{Two Sweeps, Two Straight Lines}
\label{sec:cost-account}

No setting of the dial, then, brings the multiple below several hundred. To find where it comes from, we watch how the time between tokens changes as one thing at a time is varied: first the size of the model, then the number of users.

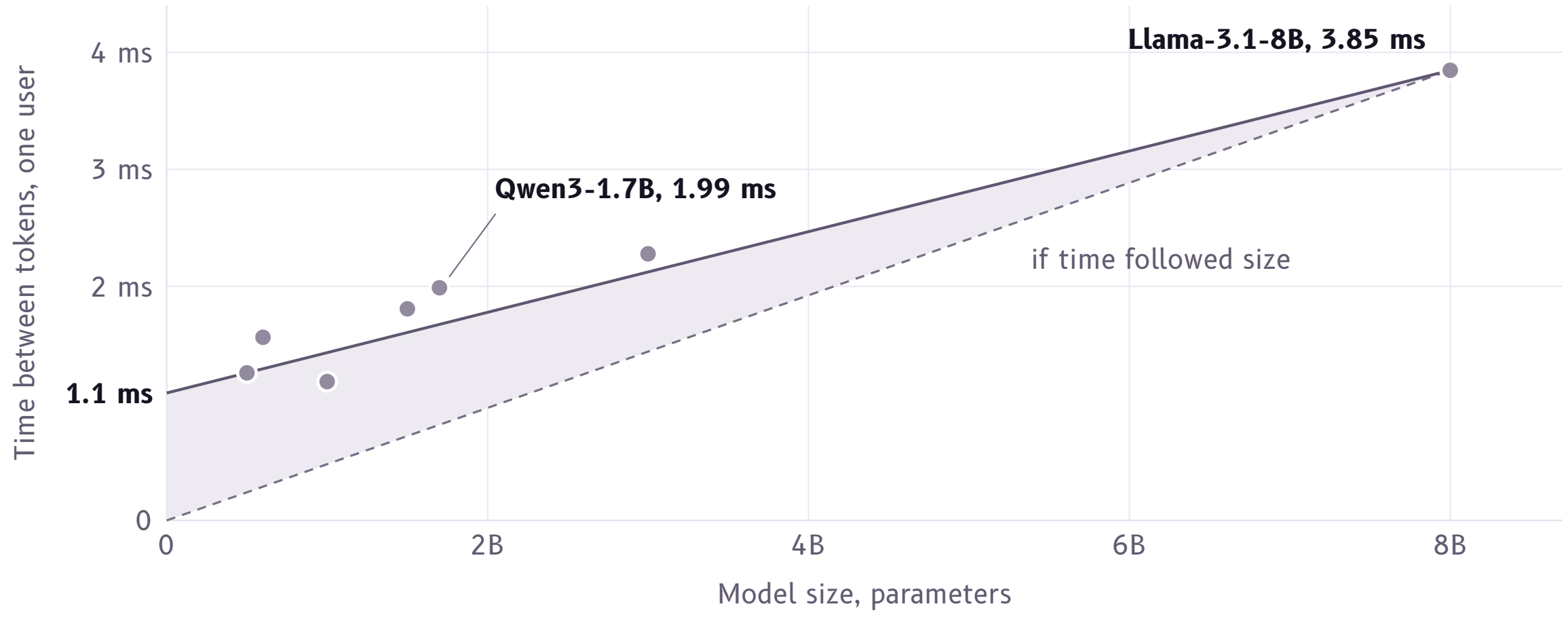
\begin{figure}[t]
  \centering
\begin{tikzpicture}
\begin{axis}[exs, height=6.6cm, width=0.92\linewidth,
  xmin=0, xmax=8.7, xtick={0,2,4,6,8}, xticklabels={0,2B,4B,6B,8B},
  xlabel={Model size, parameters},
  ymin=0, ymax=4.4, ytick={0,2,3,4}, yticklabels={0,2 ms,3 ms,4 ms},
  extra y ticks={1.089}, extra y tick labels={1.1 ms},
  extra y tick style={tick label style={font=\figfont\scriptsize\bfseries, color=exs-ink}, grid=none},
  ylabel={Time between tokens, one user},
]
  \fill[exs-mark-gpu-soft!30] (axis cs:0,0) -- (axis cs:0,1.089) -- (axis cs:8,3.846) -- cycle;
  \addplot[exs ref, domain=0:8, samples=2] {0.4808*x};
  \addplot[draw=exs-mark-gpu-strong, line width=0.8pt, domain=0:8, samples=2] {1.089+0.3446*x};
  \addplot[only marks, mark=*, mark size=2.6pt, mark options={fill=exs-mark-gpu, draw=white, line width=0.9pt}]
    coordinates {(0.5,1.261) (0.6,1.565) (1,1.186) (1.5,1.808) (1.7,1.988) (3,2.278) (8,3.846)};
  \node[exs strong, anchor=south east, xshift=-3pt, yshift=1pt] at (axis cs:8,3.846) {Llama-3.1-8B, 3.85 ms};
  \draw[exs-faint, line width=0.4pt] (axis cs:1.76,2.08) -- (axis cs:2.05,2.62);
  \node[exs strong, anchor=south west, xshift=-4pt] at (axis cs:2.05,2.6) {Qwen3-1.7B, 1.99 ms};
  \node[exs note, anchor=north west] at (axis cs:5.3,2.45) {if time followed size};
\end{axis}
\end{tikzpicture}
  \caption{The time between tokens for one user on one NVIDIA B300 with vLLM as shipped, by model size: Qwen2.5-0.5B, Qwen3-0.6B, Llama-3.2-1B, Qwen2.5-1.5B, Qwen3-1.7B, Llama-3.2-3B and Llama-3.1-8B, each in 16-bit precision with 1K tokens of prompt and 1K generated, measured on 14 October 2026. The solid line passes through the smallest and largest models and meets the axis at 1.1~ms; a least-squares line through all seven is nearly the same, 0.17~ms per gigabyte and 1.2~ms. The dashed line is where the models would lie if time followed size, and the shaded region between the two is time that making the model smaller does not recover.}
  \label{fig:cost-size}
\end{figure}

If a token's time were spent reading and multiplying weights, a model sixteen times smaller would decode sixteen times faster. \Cref{fig:cost-size} plots the time between tokens for seven open models between 0.5B and 8B parameters, each served to one user. Qwen2.5-0.5B is sixteen times smaller than Llama-3.1-8B and decodes only about three times as fast. A straight line through those two models has a slope of about 0.17~ms for every gigabyte of weights and meets the axis at about 1.1~ms, the time a token would take if the model had no weights at all, and the five models between them lie within about 20\% of it.

For the number of users we return to the sweep of \cref{tab:cost-sweep}, at its 16K-token context. A straight line fitted to the time between tokens in its six rows, about 1.8~ms plus 0.28~ms for each user, reproduces every one of them within 2\%.

Both lines have a part that grows, with the model's size or the number of users, and a part that does not: more than a millisecond that every token pays regardless. The next two subsections take the two parts in turn.

\subsection{What Grows With Load: The Movement of Data}
\label{sec:cost-movement}

The part that grows is the time spent moving data. A chip that must run any model cannot keep a particular model's weights beside its arithmetic, and one built to serve large models to many users must hold more weights and more conversation state than fit on a die. It therefore keeps them in external high-bandwidth memory (HBM) stacks beside the processor, and every token pays to bring what it needs across that interface and through the on-die memory hierarchy to the arithmetic. The caches in that hierarchy, built for any pattern of access, must guess what to keep. A model's weights and a conversation's state are far larger than any cache, so most of what a token needs is fetched again from external memory for every token. Storage costs something even when nothing moves: memory takes area that could hold logic, and it leaks power whether or not its contents are used.

How much must be brought is easy to size from the model's public configuration. A token needs the model's weights and, for its user, the keys and values the model has stored for every earlier token of the conversation, its key-value (KV) cache. Serving the FP8 checkpoint of Qwen3-1.7B means reading about 2~GB of weights in each decoding step, once, for all the users in the step. The KV cache holds, for each token of context, a key and a value for each of 8 key-value heads of dimension 128 in each of 28 layers. At two bytes per number in vLLM's default 16-bit cache, that is 112~KiB per token, or about 1.9~GB for a 16K-token conversation. Unlike the weights, it is read separately for every user.

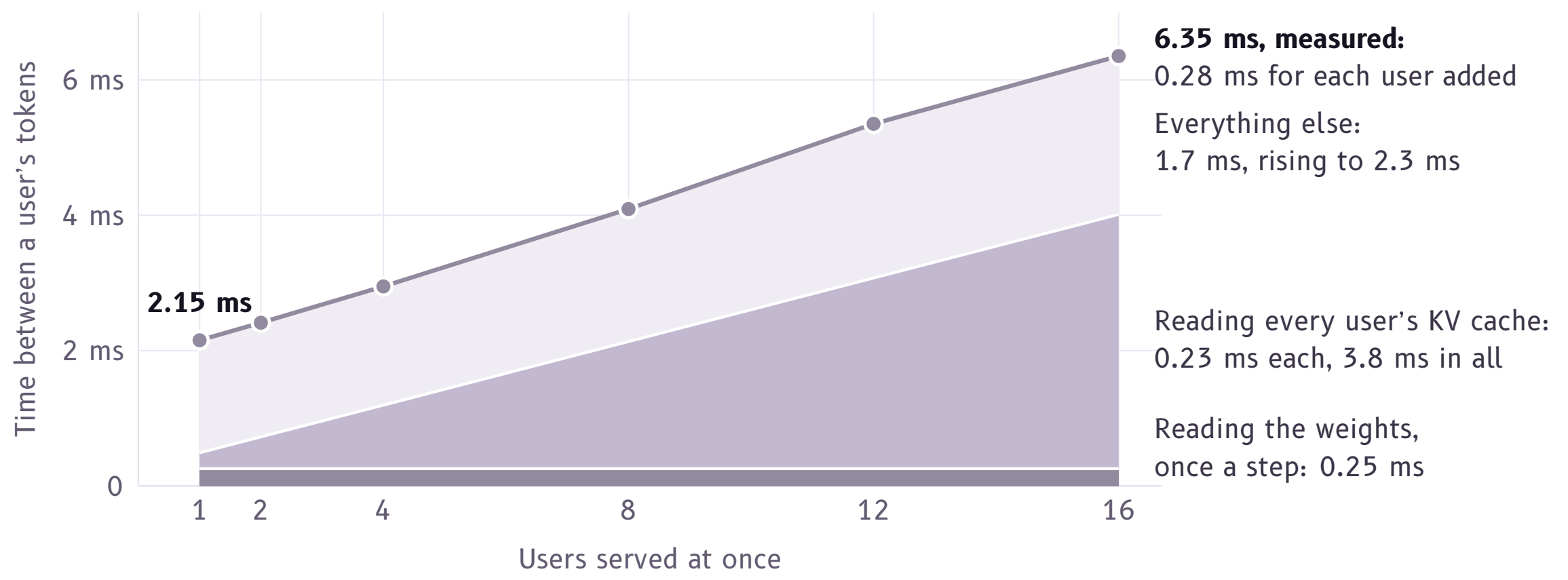
\begin{figure}[t]
  \centering
\begin{tikzpicture}
\begin{axis}[exs, height=6.2cm, width=0.70\linewidth,
  xmin=0, xmax=16.7, xtick={1,2,4,8,12,16}, xticklabels={1,2,4,8,12,16},
  xlabel={Users served at once},
  ymin=0, ymax=7, ytick={0,2,4,6}, yticklabels={0,2 ms,4 ms,6 ms},
  ylabel={Time between a user's tokens},
]
  \fill[exs wash] (axis cs:1,0.489) -- (axis cs:16,4.012) -- (axis cs:16,6.3532) -- (axis cs:12,5.3505)
    -- (axis cs:8,4.0917) -- (axis cs:4,2.9481) -- (axis cs:2,2.4114) -- (axis cs:1,2.1515) -- cycle;
  \fill[exs-mark-gpu-soft] (axis cs:1,0.254) -- (axis cs:16,0.254) -- (axis cs:16,4.012) -- (axis cs:1,0.489) -- cycle;
  \fill[exs-mark-gpu] (axis cs:1,0) rectangle (axis cs:16,0.254);
  \draw[white, line width=0.8pt] (axis cs:1,0.254) -- (axis cs:16,0.254);
  \draw[white, line width=0.8pt] (axis cs:1,0.489) -- (axis cs:16,4.012);
  \addplot[exs gpu] coordinates {(1,2.1515) (2,2.4114) (4,2.9481) (8,4.0917) (12,5.3505) (16,6.3532)};
  \node[exs strong, anchor=south, yshift=4pt] at (axis cs:1,2.1515) {2.15 ms};
  \node[exs label, anchor=west, xshift=6pt, align=left] at (axis cs:16,6.3532)
    {{\bfseries\color{exs-ink}6.35 ms, measured:}\\0.28 ms for each user added};
  \node[exs label, anchor=west, xshift=6pt, align=left] at (axis cs:16,5.05) {Everything else:\\1.7 ms, rising to 2.3 ms};
  \node[exs label, anchor=west, xshift=6pt, align=left] at (axis cs:16,2.15) {Reading every user's KV cache:\\0.23 ms each, 3.8 ms in all};
  \node[exs label, anchor=south west, xshift=6pt, yshift=-3pt, align=left] at (axis cs:16,0) {Reading the weights,\\once a step: 0.25 ms};
\end{axis}
\end{tikzpicture}
  \caption{Where the time between tokens goes on one NVIDIA B300 serving Qwen3-1.7B in FP8 at a 16K-token context. The line is the measured time between a user's tokens (\cref{tab:cost-sweep}), which grows by 0.28~ms for each user added. The two lower bands are calculated, not measured: the bytes each step reads, from the model's public configuration, at the B300's published memory bandwidth of 8~TB/s. The upper band is the remainder. Labels on the right give each part at sixteen users.}
  \label{fig:cost-account}
\end{figure}

At the B300's published memory bandwidth of 8~TB/s those bytes take a known time to arrive, and the times match the slopes of the two lines. A gigabyte takes 0.125~ms, close to the 0.17~ms per gigabyte of \cref{fig:cost-size}. One user's cache takes 0.23~ms, more than four fifths of the 0.28~ms that each added user is measured to cost, which identifies the cost of adding a user as the cost of reading that user's context. \Cref{fig:cost-account} sets this calculated traffic, the time a step spends reading from memory, beneath the measured time between tokens: 0.25~ms for the weights and 0.23~ms for each user's cache, which comes to about 0.5~ms for one user's step and about 4~ms for a step shared by sixteen. The calculation assumes the full published bandwidth, so it is the least the traffic can take, and what remains above it is the most that everything else can account for.

Sharing a step helps with one of the two reads and not the other. With one user, a token reads about 3.9~GB, half of it weights and half conversation. With sixteen, each step reads about 32~GB, of which 30~GB is conversation: each user's share of the weights has shrunk to an eighth of a gigabyte, while that user's 1.9~GB of conversation is read in full however many share the step. Longer contexts make it worse. At 128K tokens one conversation's cache is about 15~GB, more than seven times the model, so the share of each step spent moving context grows with exactly the long conversations and agent sessions in which small models do more and more of the work.

\subsection{What Does Not Grow: The Layers Between the Model and Its Arithmetic}
\label{sec:cost-layers}

The part that does not grow is not explained by traffic. In the sweep of users, reading the weights accounts for 0.25~ms of the 1.8~ms that every step pays, and the remaining 1.6~ms or so is spent whether the step serves one user or sixteen. It is the counterpart, at a 16K-token context, of the 1.1~ms that \cref{fig:cost-size} found at 1K.

A fixed charge of that kind is what one would expect from the layers that stand between a model and a GPU's arithmetic, layers of software and hardware that exist so that the GPU can run anything. A framework expresses the model as a graph of operations. A serving engine schedules requests, batches them and manages the memory that holds their state. Each operation is carried out by a kernel from a library, a program launched onto the device, and each kernel runs as a stream of instructions that thousands of programmable cores fetch, decode and schedule.

Generality of shape and of load adds charges of its own. The arithmetic units are built for blocks of a fixed shape, or tiles, so an operation that does not fill one, such as the matrix-vector products of a single user's decoding step, is padded to the shape the hardware expects, and part of the arithmetic is spent on zeros or left idle. The machine is also built for any number of users, so it is organized around the batch: the whole chip, sized for a crowd, is stepped once for every token, and each user's token waits on that step. None of this is arithmetic, and some of it is paid once per step whatever the size of the model.

We attribute most of the fixed charge to these layers: the cost of stepping a chip sized for a crowd through a long chain of small, dependent operations for every token. The attribution rests on the form of the charge, fixed per step and indifferent to how many users share it, rather than on a profile of the software.

How a token's time divides between the two causes therefore depends on how many users share the chip. For one user, memory traffic is about a quarter of the 2.15~ms token and nearly all of the rest is the fixed charge, so the single-user token is slow because of the layers, not because of memory. With sixteen users the proportions reverse: traffic is about 4~ms of the 6.35~ms step, nearly all of it KV caches. Since the board draws its power throughout the step, a token's energy divides in roughly the same proportions as its time. The division also explains the dial of \cref{sec:cost-metric}. Sharing a step spreads its fixed charge and the reading of the weights across the users, but each of them adds a conversation that must be read in full, so the step lengthens for all.

\subsection{Why the Distance Persists}
\label{sec:cost-persist}

Neither part is arithmetic, which is why three usual ways of making a token cheaper leave most of the distance in place. The first is a smaller model, which shortens only the part that grows with size. For Llama-3.1-8B, whose weights occupy 16~GB at 16-bit precision, that part is most of the token, about 2.8 of its 3.85~ms. Under 2B parameters the fixed charge is most of the token, so small models are the case in which a general-purpose chip recovers least of what the model's size should give.

The second is better software, which shortens only the fixed part. Hand-tuned kernels can remove much of it for a single user (\cref{sec:chip-limits}), but the model still reaches its arithmetic through a general-purpose machine that does not know its structure in advance, and the traffic is untouched.

The third is a new process node, the next generation of chip manufacturing, which repairs neither part, because neither falls at the rate the cost of arithmetic does. For the movement and storage of data the evidence is published. \citet{horowitz2014energy} tabulated the energy of arithmetic and memory operations at 45~nm, and \citet{jouppi2021ten}, drawing on three generations of Google's data-center accelerators, brought the table forward to 7~nm and found the improvement uneven. Arithmetic operations became 2.1-4.3x cheaper, while an access to a small block of on-chip static memory (SRAM) improved by 1.3-2.4x and the energy of driving a wire, per unit of length, by less than 2x. SRAM density lagged as well: at 7~nm a square millimeter held about a fifth of what ideal scaling from 65~nm would have given, and by their estimate the step to 5~nm added only about 13\%. Reaching external memory over a conventional interface cost the same at both nodes. The gain that did arrive came from packaging rather than from the process, when high-bandwidth memory placed stacks of memory dies close to the processor, and the B300 already has it. Their summary is that logic improves much faster than wires and SRAM.

For the layers the reasoning is our own. Their charge is set not by the energy of a transistor but by the time a general-purpose machine takes to sequence a chain of small, dependent operations, and the additional transistors of a new generation, spent as they usually are on more parallel arithmetic and more memory bandwidth, do little to shorten such a chain. A token whose cost lies in the layers and in moving and storing data therefore improves more slowly with each node than its arithmetic does, and its distance from the floor widens from one generation to the next.

\begin{insight}{Almost none of a token's cost is its arithmetic}
  On a GPU serving a small model, the arithmetic is a fraction of a percent of a token's energy. For one user, most of the rest is the fixed charge of the layers between the model and its arithmetic; as users are added, it becomes the movement of each user's context from where it is stored. Neither falls at the rate the cost of arithmetic does, so a design that only makes the arithmetic cheaper leaves almost all of the cost in place.
\end{insight}

The industry tends to treat the distance between a token and its arithmetic as physics. The measurements of this section point instead to design choices. Almost every model today runs on hardware designed before the model existed, and such hardware has to be general: able to run any program, on any shape of computation, for any number of users, with any pattern of memory access, for any model. The layers are the price of the first three and the movement of data the price of the last two, while a model built for every task raises the arithmetic itself. Choices can be made differently, and the next section describes how we make them.

\section{Approaching the Limit Through Model and Silicon Co-Design}
\label{sec:chip}

Our approach is to remove the generality a token does not need, from both factors of its cost: the multiple the hardware spends, and the arithmetic the model requires. Two decisions about the silicon bring down the multiple. We build the chip around the model's architecture, so that no layers stand between the model and its arithmetic (\cref{sec:chip-layers}), and we keep everything a token needs on the die, beside the arithmetic, so that nothing has to be brought to it (\cref{sec:chip-movement}). A third decision lowers the arithmetic itself: since the floor is a property of the model, we treat the model's architecture as a design variable, chosen together with the silicon (\cref{sec:chip-model}). \Cref{tab:chip-generality} sets each form of generality beside what it costs a token and the decision that answers it.

\begin{table}[t]
  \centering
  \caption{The forms of generality of \cref{sec:cost}, what each costs a token on a conventional chip, and how the Exstill design answers it. The first three are the layers, the next two data movement and storage, and the last the generality of the model itself.}
  \label{tab:chip-generality}
  \small
  \begin{tabularx}{\linewidth}{l>{\raggedright\arraybackslash}X>{\raggedright\arraybackslash}X}
    \toprule
    \headrow Built for & What every token pays & The Exstill design \\
    \midrule
    Any program & Framework, serving engine, kernels; instructions fetched, decoded and scheduled & The architecture is fixed in the silicon; there is nothing to program (\cref{sec:chip-layers}) \\ \addlinespace
    Any shape & Arithmetic padded to the hardware's tile shapes, spent on zeros or left idle & Engines and data paths sized for the model's architecture (\cref{sec:chip-layers}) \\ \addlinespace
    Any number of users & A crowd-sized machine stepped once per token; each token waits on the batch & Sized for one user's next token, a few users at a time (\cref{sec:chip-layers}) \\ \addlinespace
    Any access pattern & Caches that guess what to keep; most data arrives from external memory & Nothing to guess: all of a token's data is on the die (\cref{sec:chip-movement}) \\ \addlinespace
    Any model & Every weight fetched again from memory for every token & Weights loaded once and kept beside the arithmetic (\cref{sec:chip-movement}) \\ \addlinespace
    Any task & A general model carries arithmetic and state the question does not use & A model distilled for its job, its architecture chosen with the silicon (\cref{sec:chip-model}) \\
    \bottomrule
  \end{tabularx}
\end{table}

The first instance of this approach is the frontier model's co-processor, a chip that runs the small models working around a frontier model and sits beside the processor that runs it. The same approach applies to other environments and power envelopes, from robots and factory equipment to devices at the edge, but those designs are a subject for later work, and this report covers the data-center design. The section retraces \cref{sec:cost} in reverse, removing the causes, then closing the distance, then moving the limit. It ends with a second design for a larger model (\cref{sec:chip-dc2}) and with two other designs that keep a model on the chip, set beside ours from their published figures (\cref{sec:chip-others}).

\subsection{Removing the Layers: A Chip Built Around the Model}
\label{sec:chip-layers}

A model is two things: its architecture, the shape of the computation it performs for each token, and its weights, the values learned in training that decide what its tokens mean. A GPU is built knowing neither. It receives the architecture at run time, operation by operation, through the layers of \cref{sec:cost-layers}, and it reads the weights from external memory as each operation needs them. That is what lets a GPU run any model, and it is where much of a small model's token is paid for.

We fix the architecture in the silicon and leave the weights free. The chip's memory, compute engines and data paths are organized around one model architecture, chosen together with the silicon, so the computation a token requires is laid out in the hardware rather than assembled at run time by a framework, a serving engine and a library of kernels. There is nothing to program: tokens go in and tokens come out. The weights are loaded rather than built in, so a new model of the same architecture, whether a new task or a new release, is a training run rather than a new chip. The chip is built for one user's next token rather than for a batch, so it serves a few users at a time, and no crowd-sized machine is stepped through each token.

The usual escape from the cost of generality is dedicated silicon, a chip built for one model: lean, fast, and obsolete the moment the model changes. For the small models around a frontier model that is a poor trade, because the models change every few months while their jobs do not. Fixing the architecture and leaving the weights free keeps most of what dedicated silicon offers, a chip organized around the computation it performs with no layers in between, without tying the chip to the model it was built for. What it gives up is freedom in the architecture, so the chip's value depends on how much useful work one architecture can do. \Cref{sec:polyglot} answers that question with Polyglot-DC, a single architecture distilled into separate weights for drafting, guardrails, agents and real-time reasoning.

\subsection{Removing the Movement: Everything the Token Needs, Beside the Arithmetic}
\label{sec:chip-movement}

The answer to the movement of data is to change where the data lives. The data-center design keeps the model's weights and the context of every conversation it serves on a single die, beside the arithmetic that uses them, with no external memory stacks. It holds twelve conversations of 16K tokens each. For the same twelve users, each step of the B300 brings about 25~GB across its memory interface (\cref{sec:cost-movement}); here no weight or key-value pair crosses a chip boundary on its way to being used, and no token waits on a trip to external memory. This report says where the weights and the context are kept and what that achieves. How they are stored and read, and how contexts longer than 16K tokens are handled (\cref{sec:sd-chip}), are part of the co-design and are not described.

Keeping everything on the die is possible only if there is little enough of it, which is why this decision cannot be separated from the model. The principle that governs the silicon and the model alike is to move as few bits as the quality bar allows, over the shortest distance to the arithmetic. The silicon sets the distance; the model, chosen together with it, sets how many bits must be stored and moved to reach the required quality.

\begin{figure}[t]
  \centering
\begin{tikzpicture}[
  font=\figfont\scriptsize,
  layer/.style={draw=exs-line, fill=exs-band, rounded corners=2pt, minimum width=4.2cm, minimum height=0.54cm,
    align=center, text=exs-ink-2, inner sep=2pt},
  math/.style={draw=none, fill=exs-meas!16, rounded corners=2pt, minimum height=0.54cm, align=center, text=exs-ink},
  flow/.style={draw=exs-faint, line width=0.6pt, -{Stealth[length=3pt]}},
  head/.style={font=\figfont\scriptsize\bfseries, text=exs-ink, anchor=west},
]
\node[head] at (-0.05,0.66) {A GPU serving a small model};
\node[layer, fill=white] (L0) at (2.1,0) {The model};
\foreach \i/\t in {1/{Framework: a graph of operators}, 2/{Serving engine: schedule, batch},
                   3/{Kernels: a program per operator}, 4/{Cores: fetch, decode, schedule}}
  \node[layer] (L\i) at (2.1,-0.8*\i) {\t};
\node[math, minimum width=4.2cm] (LA) at (2.1,-4.0) {Arithmetic};
\foreach \i [evaluate=\i as \j using int(\i+1)] in {0,1,2,3} \draw[flow] (L\i.south) -- (L\j.north);
\draw[flow] (L4.south) -- (LA.north);
\draw[exs-faint, line width=0.5pt, decorate, decoration={brace, amplitude=4pt}]
  ($(L1.north east)+(0.14,0)$) -- ($(L4.south east)+(0.14,0)$);
\node[exs label, anchor=west, align=left] at ($(L2.south east)+(0.42,-0.13)$) {the layers between\\the model and\\its arithmetic};
\node[draw=exs-mark-gpu-soft, fill=exs-mark-gpu-soft!35, rounded corners=2pt, align=center, text=exs-ink-2,
      minimum width=1.9cm, minimum height=1.15cm, inner sep=3pt] (HBM) at (7.05,-4.0)
      {External memory:\\the weights and\\every user's\\conversation};
\draw[flow] (HBM.west) -- node[above, exs note, inner sep=2pt, pos=0.42] {every token} (LA.east);
\node[head] at (8.85,0.66) {The Exstill design};
\draw[exs-mark-us, line width=0.9pt, rounded corners=3pt] (10.25,0.27) rectangle (14.65,-4.575);
\node[exs strong] at (12.45,-0.1) {One die};
\node[exs note] at (12.45,-0.47) {the architecture fixed in silicon};
\fill[exs-violet-lo!45, rounded corners=2pt] (10.55,-0.95) rectangle (14.35,-4.275);
\node[exs label] at (12.45,-1.5) {the weights and every};
\node[exs label] at (12.45,-1.83) {conversation, kept beside};
\node[exs label] at (12.45,-2.16) {the arithmetic};
\node[math, minimum width=2.6cm] (RA) at (12.45,-3.3) {Arithmetic};
\node[exs label, align=right, anchor=east] (tin) at (9.65,-2.15) {tokens\\in};
\node[exs label, align=left, anchor=west] (tout) at (15.25,-2.15) {tokens\\out};
\draw[flow] (tin.east) -- (10.25,-2.15);
\draw[flow] (14.65,-2.15) -- (tout.west);
\node[exs note, anchor=north] (tr) at (12.45,-5.0) {a new job: new weights from training, loaded};
\draw[flow, dash pattern=on 2pt off 1.5pt] (tr.north) -- (12.45,-4.575);
\end{tikzpicture}
  \caption{What stands between a model and its arithmetic, and where a token's data lives. On a GPU, the model reaches the arithmetic through layers of software and programmable hardware, and every token brings the weights and the user's conversation in from external memory: for Qwen3-1.7B at a 16K-token context, about 2~GB of weights and 1.9~GB for each user (\cref{sec:cost-movement}). On the Exstill design, the architecture is fixed in the silicon, the weights and every conversation are kept on the die beside the arithmetic, and only tokens enter and leave, a few bytes each (\cref{sec:chip-movement}). Illustrative, not a floorplan.}
  \label{fig:chip-paths}
\end{figure}
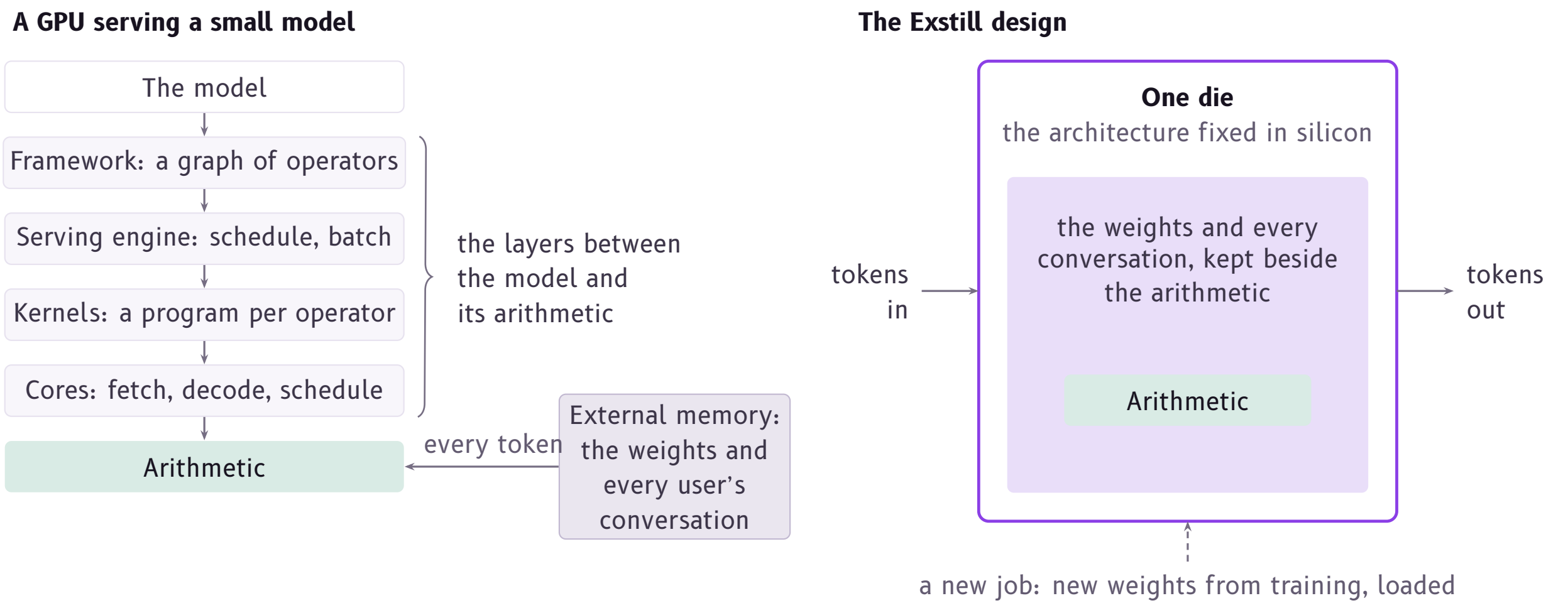

The chip does not replace the processor that runs the frontier model. It sits beside it on a standard link, and only tokens cross between the two. At the design's full output of about 239,000 tokens per second (\cref{sec:chip-dc}), that traffic is about a megabyte per second in each direction, a few bytes per token and under a ten-thousandth of the bandwidth of a PCIe slot, so the link constrains neither side. \Cref{fig:chip-paths} sets the arrangement that results from the two decisions beside a GPU's.

Other designs have removed parts of the same overhead. Keeping a model in on-chip memory is not new: Groq's tensor streaming processors \citep{abts2022software}, NVIDIA's LPX decode racks \citep{nvidia2026lpx} and Cerebras's wafer-scale engine \citep{lie2023cerebras} serve large models quickly from on-chip memory spread across many dies or a whole wafer. Taalas builds a single model into its chip \citep{taalas2026hc1}, the dedicated-silicon end of the trade described in \cref{sec:chip-layers}, and d-Matrix's cards already sit beside GPUs. Our design is fitted differently: the whole of a small model on one die sized for it, the silicon and the model chosen together, placed beside any accelerator, with a new model of the same architecture loaded as weights.

None of the ingredients is ours alone. Keeping weights beside the arithmetic, building silicon for one kind of computation, distilling a large model into a small one for a narrow job, and shaping a model to the hardware it will run on are each established directions, and each of the designs above takes one or two of them further than we do. What has, as far as we know, no precedent is taking all of them at once, inside a single loop: model architectures explored and trained on one side, and on the other every candidate chip costed on placed and routed circuits rather than from a spreadsheet, with the two sides scored together on the cost of a token at a fixed quality. That loop is the capability this report rests on, and it is harder to assemble than any of its parts.

\subsection{Designing From the Circuits Up}
\label{sec:chip-search}

The three decisions interact, so we make them in one search over whole designs, each a model and its silicon together. The search needs the time and energy of every candidate's tokens, which we call costing the design, and because the limit is measured on circuits, the candidates must be costed on circuits too. Before a chip exists there are four levels at which it can be costed (\cref{tab:chip-levels}), and only the last, the gate level, grounds both time and energy in the circuits as they will be fabricated.

\begin{table}[t]
  \centering
  \caption{The four levels at which a chip can be costed before silicon, and what each grounds. Silicon follows, where anything the levels above missed appears after the money is spent.}
  \label{tab:chip-levels}
  \small
  \begin{tabularx}{\linewidth}{l>{\raggedright\arraybackslash}Xll}
    \toprule
    \headrow Level & What is run & Time & Energy \\
    \midrule
    Paper & The chip's arithmetic units multiplied by its clock & A ceiling & A ceiling \\ \addlinespace
    Architecture & The workload on a software model of the chip, with an assumed cost for each operation & Estimated & Estimated \\ \addlinespace
    Cycle & The real design, simulated clock by clock, running its actual program & Exact & Estimated \\ \addlinespace
    Gate & Circuits placed and routed as they will be fabricated, every switching gate counted & Exact & Measured on the flow \\
    \bottomrule
  \end{tabularx}
\end{table}

Published pre-silicon numbers tend to fall short in one of two ways. Those that cover a whole model usually stop at the architecture level, on assumed costs. Those that reach the cycle or gate level cover only part of the model, and compilers are the reason. Below the architecture level the simulated chip must run its actual program, and turning a whole model into that program takes a compiler and a library of kernels that many chips do not have before silicon. The common route therefore carries one hand-written kernel, typically the large matrix multiplication the chip was built for, to the cycle or gate level and scales it up as though the rest of the model ran equally well. It rarely does: the rest of the model, and the data moving between layers, is where chips lose time and energy. Even a company with shipping silicon often cannot fully cost a new architecture, because every change to it means updating the compiler, recompiling the model and sometimes rewriting kernels by hand.

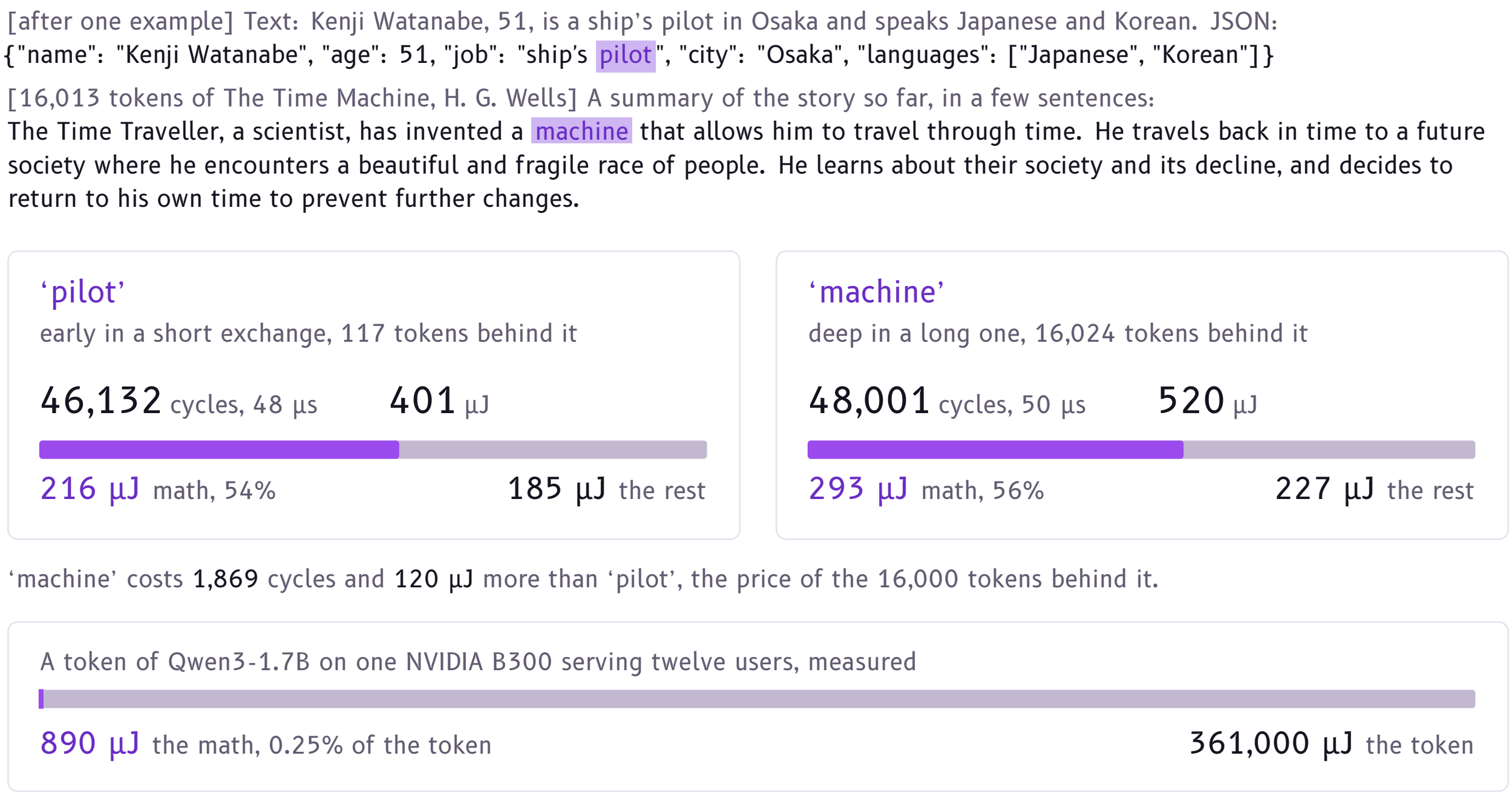
\begin{figure}[t]
  \centering
\providecommand{\rcptword}[1]{{\setlength{\fboxsep}{1pt}\colorbox{exs-violet-lo}{\textcolor{exs-violet-ink}{#1}}}}
\begin{tikzpicture}[
  x=1cm, y=1cm,
  card/.style={draw=exs-line, fill=white, line width=0.5pt, rounded corners=3pt, inner sep=0pt},
  big/.style={anchor=base west, inner sep=0pt, font=\figfont\large, text=exs-ink},
  word/.style={anchor=north west, inner sep=0pt, font=\figfont\small, text=exs-violet-ink},
  ctx/.style={anchor=north west, inner sep=0pt, font=\figfont\scriptsize, text=exs-muted, align=left},
  lab/.style={inner sep=0pt, font=\figfont\scriptsize, text=exs-muted},
]
\def\W{16.4}
\def\cw{8.0}
\node[anchor=north west, inner sep=0pt, text width=\W cm, align=left, font={\figfont\scriptsize\hyphenpenalty10000 \exhyphenpenalty10000 }, text=exs-ink] (ex) at (0,0) {%
  {\color{exs-muted}[after one example] Text: Kenji Watanabe, 51, is a ship's pilot in Osaka and speaks Japanese and Korean. JSON:}\\
  {\addfontfeatures{Ligatures=TeXOff}\{"name": "Kenji Watanabe", "age": 51, "job": "ship's \rcptword{pilot}", "city": "Osaka", "languages": ["Japanese", "Korean"]\}}\\[3pt]
  {\color{exs-muted}[16,013 tokens of \emph{The Time Machine}, H. G. Wells] A summary of the story so far, in a few sentences:}\\
  The Time Traveller, a scientist, has invented a \rcptword{machine} that allows him to travel through time. He travels back in time to a future society where he encounters a beautiful and fragile race of people. He learns about their society and its decline, and decides to return to his own time to prevent further changes.%
};
\foreach \i/\x/\w/\where/\cyc/\us/\uj/\math/\rest/\pct/\frac in {%
  1/0/pilot/{early in a short exchange, 117 tokens behind it}/46{,}132/48/401/216/185/54/0.539,%
  2/8.4/machine/{deep in a long one, 16{,}024 tokens behind it}/48{,}001/50/520/293/227/56/0.563}{
  \node[card, minimum width=\cw cm, minimum height=3.15cm, anchor=north west] (c\i) at ($(ex.south west)+(\x,-0.4)$) {};
  \node[word] at ($(c\i.north west)+(0.35,-0.3)$) {`\w'};
  \node[ctx, text width=7.3cm] at ($(c\i.north west)+(0.35,-0.78)$) {\where};
  \node[big] at ($(c\i.north west)+(0.35,-1.78)$) {\cyc\,{\scriptsize\color{exs-muted}cycles, \us\ \textmu s}\qquad\uj\,{\scriptsize\color{exs-muted}\textmu J}};
  \fill[exs-mark-gpu-soft, rounded corners=1pt] ($(c\i.north west)+(0.35,-2.28)$) rectangle ++(7.3,0.2);
  \fill[exs-violet, rounded corners=1pt] ($(c\i.north west)+(0.35,-2.28)$) rectangle ++(7.3*\frac,0.2);
  \node[lab, anchor=base west] at ($(c\i.north west)+(0.35,-2.72)$) {{\small\color{exs-violet-ink}\math\ \textmu J}\enspace math, \pct\%};
  \node[lab, anchor=base east] at ($(c\i.north west)+(7.65,-2.72)$) {{\small\color{exs-ink}\rest\ \textmu J}\enspace the rest};
}
\node[lab, anchor=north west] (cmp) at ($(c1.south west)+(0,-0.3)$) {`machine' costs {\color{exs-ink}1,869} cycles and {\color{exs-ink}120 \textmu J} more than `pilot', the price of the 16,000 tokens behind it.};
\node[card, minimum width=\W cm, minimum height=1.85cm, anchor=north west] (g) at ($(cmp.south west)+(0,-0.3)$) {};
\node[ctx] at ($(g.north west)+(0.35,-0.32)$) {A token of Qwen3-1.7B on one NVIDIA B300 serving twelve users, measured};
\fill[exs-mark-gpu-soft, rounded corners=1pt] ($(g.north west)+(0.35,-0.95)$) rectangle ++(15.7,0.2);
\fill[exs-violet] ($(g.north west)+(0.35,-0.95)$) rectangle ++(15.7*0.0025,0.2);
\draw[exs-violet, line width=0.4pt] ($(g.north west)+(0.35,-0.95)$) rectangle ++(15.7*0.0025,0.2);
\node[lab, anchor=base west] at ($(g.north west)+(0.35,-1.45)$) {{\small\color{exs-violet-ink}890 \textmu J}\enspace the math, 0.25\% of the token};
\node[lab, anchor=base east] at ($(g.north west)+(16.05,-1.45)$) {{\small\color{exs-ink}361,000 \textmu J}\enspace the token};
\path (current bounding box.south) -- ++(0,-0.15);  
\end{tikzpicture}
  \caption{A receipt for every word. Each card is one token the co-designed model wrote on the pre-silicon data-center design at the 5~nm class, with the clock cycles and the energy the token took, split into the arithmetic the model requires and everything else, leakage included. The word `pilot' comes early in a short exchange and `machine' 16,000 tokens into a long one, and the difference between them is what the context behind the second costs. The lower bar is a token of Qwen3-1.7B on one NVIDIA B300 serving twelve users in FP8, measured on our bench, with its arithmetic measured on our flow, a quarter of one percent of the token.}
  \label{fig:receipt}
\end{figure}

Chips are conventionally designed from the top of \cref{tab:chip-levels} down: an architecture is chosen early from paper estimates and a software model, and gate-level costs arrive only after it is frozen, when a disagreement between the gates and the model can usually be corrected only in the next chip. We reverse the order and begin with the circuits, 304 of them to date, each placed, routed and simulated gate by gate at the 5~nm class, glitches included. Whole designs are then costed on those circuits. For each design we measure every token its model would write, at every layer and every position in the context, in clock cycles for time and in picojoules for energy, including the token's share of the chip's leakage, the power it draws even when idle. More than ten million designs have been costed this way: not ten million gate-level simulations, but ten million whole designs, each with every token measured on the same placed and routed circuits, before any one of them is chosen. \Cref{fig:receipt} shows the measure for two of those tokens, one early in a short exchange and one deep in a long one.

Removing the layers is what makes this order practical. Each design is built around its model's architecture and has no program to compile, so a candidate can be costed on the whole model without first writing a compiler and kernels for it. The property that removes the layers from every token at run time removes them from every design during the search.

The chip's figures in this report are therefore pre-silicon in a specific sense: they rest on gate-level measurement of every circuit and cover the whole model rather than one kernel scaled up. Throughout the report, each number states how it is known: \emph{measured} on an NVIDIA B300 on our bench with the stated setup; \emph{measured} on a named benchmark or on a named circuit block on our gate-level flow; \emph{calculated} from a public configuration or specification; \emph{pre-silicon} for the chip's own figures at the 5~nm class; and \emph{simulated} for a result that combines a B300 measurement with a pre-silicon figure, such as the speedups of \cref{sec:specdec}.

Every design the search returns answers the question set in \cref{sec:cost-metric}: the least energy-delay per token that clears a quality bar, within a budget of power and area. For the data center the budget, which we call the envelope, is 125~W and 575~mm\textsuperscript{2}, with every user holding a 16K-token context on the chip. The bar is Qwen3-1.7B's own quality in bits per byte on our probe texts, so each design below is held to the quality of the model the B300 serves in \cref{sec:cost-gpu}.

\subsection{The Distance Closed: The Same Model on Our Silicon}
\label{sec:chip-silicon}

\begin{figure}[t]
  \centering
\begin{tikzpicture}
\begin{axis}[exs, height=5.2cm, width=0.74\linewidth,
  xmode=log, xmin=0.15, xmax=10000,
  xtick={1,10,100,1000}, xticklabels={1 mJ,10 mJ,100 mJ,1 J},
  xlabel={Energy per token (logarithmic scale)},
  ymin=0.4, ymax=4.75, ytick={1,2,3},
  yticklabels={{Exstill data-center design, twelve users}, {Exstill silicon, model unchanged, one user}, {NVIDIA B300, twelve users}},
  yticklabel style={font=\figfont\scriptsize, text=exs-ink-2},
  ymajorgrids=false,
]
  \draw[exs floorline] (axis cs:0.8865,0.5) -- (axis cs:0.8865,3.72);
  \draw[exs floorline, draw=exs-meas!55] (axis cs:0.29,0.5) -- (axis cs:0.29,4.32);
  \node[exs note, anchor=south west, xshift=-2pt] at (axis cs:0.8865,3.72) {floor of Qwen3-1.7B as released, 0.89 mJ};
  \node[exs note, anchor=south west, xshift=-2pt] at (axis cs:0.29,4.32) {floor of the co-designed model, 0.29 mJ};
  \draw[exs-mark-gpu-soft, line width=2.6pt] (axis cs:0.8865,3) -- (axis cs:361.02,3);
  \draw[exs-violet-lo, line width=2.6pt] (axis cs:0.8865,2) -- (axis cs:3.041,2);
  \draw[exs-mark-us, line width=2.6pt] (axis cs:0.29,1) -- (axis cs:0.523,1);
  \addplot[only marks, mark=*, mark size=3pt, mark options={fill=exs-mark-gpu, draw=white, line width=0.9pt}] coordinates {(361.02,3)};
  \addplot[only marks, mark=*, mark size=3pt, mark options={fill=exs-violet-lo, draw=exs-mark-us, line width=0.9pt}] coordinates {(3.041,2)};
  \draw[exs-mark-amber, line width=1.1pt] (axis cs:0.523,1) circle (4.4pt);
  \addplot[only marks, mark=*, mark size=3pt, mark options={fill=exs-mark-us, draw=white, line width=0.9pt}] coordinates {(0.523,1)};
  \node[exs label, anchor=west, xshift=7pt] at (axis cs:361.02,3) {{\bfseries\color{exs-ink}361 mJ:} 407x its floor};
  \node[exs label, anchor=west, xshift=7pt] at (axis cs:3.041,2) {{\bfseries\color{exs-ink}3.0 mJ:} 3.4x its floor};
  \node[exs label, anchor=west, xshift=5pt] at (axis cs:0.8865,1) {{\bfseries\color{exs-ink}0.52 mJ:} 1.8x its own floor, below the released model's};
\end{axis}
\end{tikzpicture}
  \caption{How close each design comes to the limit. Energy per token at a 16K-token context on a logarithmic scale, against the floor of Qwen3-1.7B as released and the floor of the co-designed model, both the arithmetic alone measured on our gate-level flow. Each bar runs from a design's own floor to its token, so its length is the multiple beside it. The B300 is measured (\cref{tab:cost-sweep}); the Exstill points are pre-silicon at the 5~nm class. Both Exstill points lie within the same envelope of 125~W and 575~mm\textsuperscript{2}; the model-unchanged point is the design with the lowest energy-delay for Qwen3-1.7B as released that fits it, and it serves one user (\cref{sec:chip-capacity}).}
  \label{fig:chip-floor}
\end{figure}
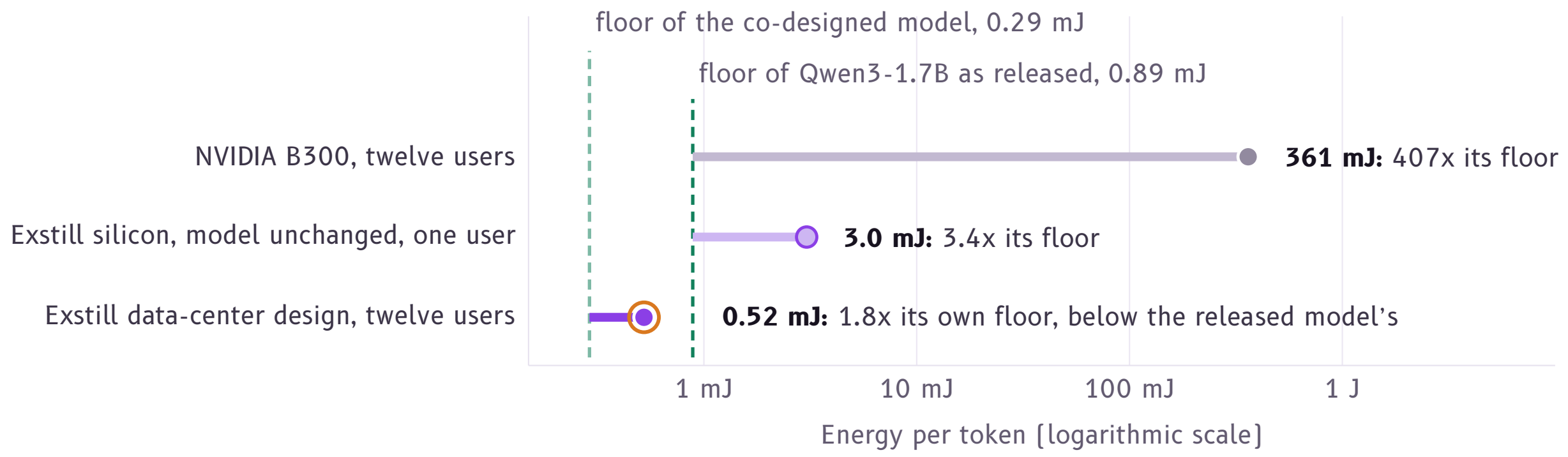

To see what the silicon alone achieves, we hold the model fixed. We asked the search its question for Qwen3-1.7B exactly as released, on our silicon, within the envelope. On the design with the lowest energy-delay that fits, a token costs 3.0~mJ, 3.4x the model's floor, and arrives every 57~\textmu s. The B300 serving the same model sits between 391x and 1,120x above that floor, depending on how many users share it, and at its fastest takes 2.15~ms per token. Removing the layers and the movement, with no change to the model, brings the token within a small multiple of its arithmetic. \Cref{fig:chip-floor} places the two on one scale, with the design of the next subsection beneath them.

Silicon alone can go little further. A chip built for a model chosen in advance can at best approach that model's floor; it cannot take a token below it. And this model leaves the die little room: within the envelope the chip holds the weights of Qwen3-1.7B and one 16K-token conversation, and has no room for a second. Both limits, the arithmetic a token requires and the bytes the chip must store, are properties of the model's architecture.

\subsection{The Limit Moved: The Model as a Design Variable}
\label{sec:chip-model}
\label{sec:chip-dc}

When the model's architecture is itself a design variable, chosen together with the silicon and held to the same quality, the floor moves with it. A model that reaches the quality bar with less arithmetic, or with arithmetic that costs less on the silicon it is paired with, or with less state to store, has a lower floor, and a chip built around it can go below the floor of the model it replaces. This also answers the generality of the model itself: a model trained for one job by distillation (\cref{sec:pg-small}) does not carry the arithmetic and state that a general model brings unused to a narrow question.

This is where our approach parts from hardware-aware model design as it is usually practiced, which searches for the model that runs best on a fixed accelerator (\cref{sec:pg-codesign}). We vary both. Neither side is optimized first and handed to the other; each candidate is a model and its silicon together, scored on the cost of its tokens at a fixed quality. The difficulty is not in either side but in holding them to the same pace and the same fidelity: a candidate architecture must be trained and measured in hours for the search to move, and the silicon that would run it must be costed at the gate level in the same loop. The 304 circuits, and the ten million designs costed on them, are what make that possible. What the search changes in the model and in the silicon is proprietary to Exstill.

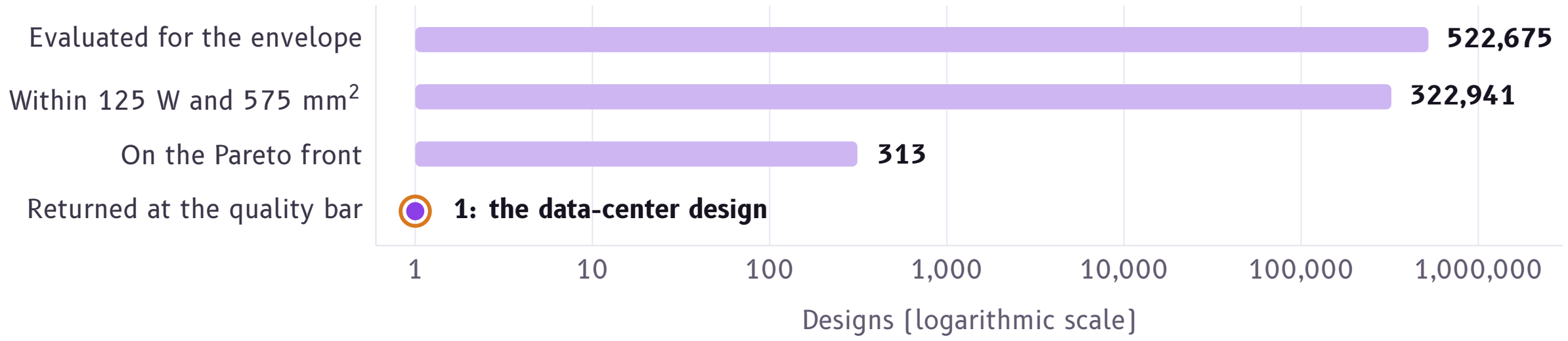
\begin{figure}[t]
  \centering
\begin{tikzpicture}
\begin{axis}[exs, height=4.0cm, width=0.82\linewidth,
  xmode=log, xmin=0.6, xmax=3e6, xtick={1,10,100,1000,10000,100000,1000000},
  xticklabels={1,10,100,{1,000},{10,000},{100,000},{1,000,000}},
  xlabel={Designs (logarithmic scale)},
  ymin=0.4, ymax=4.6, ytick={1,2,3,4},
  yticklabels={Returned at the quality bar, On the Pareto front, {Within 125 W and 575 mm\textsuperscript{2}}, Evaluated for the envelope},
  yticklabel style={font=\figfont\scriptsize, text=exs-ink-2},
  ymajorgrids=false, axis lines*=left,
]
  \fill[exs-violet-lo, rounded corners=1.5pt] (axis cs:1,3.78) rectangle (axis cs:522675,4.22);
  \fill[exs-violet-lo, rounded corners=1.5pt] (axis cs:1,2.78) rectangle (axis cs:322941,3.22);
  \fill[exs-violet-lo, rounded corners=1.5pt] (axis cs:1,1.78) rectangle (axis cs:313,2.22);
  \node[exs strong, anchor=west] at (axis cs:560000,4) {522,675};
  \node[exs strong, anchor=west] at (axis cs:345000,3) {322,941};
  \node[exs strong, anchor=west] at (axis cs:340,2) {313};
  \draw[exs-mark-amber, line width=1.1pt] (axis cs:1,1) circle (4.2pt);
  \fill[exs-mark-us] (axis cs:1,1) circle (2.6pt);
  \node[exs strong, anchor=west, xshift=7pt] at (axis cs:1,1) {1: the data-center design};
\end{axis}
\end{tikzpicture}
  \caption{The search for the data-center envelope, from the designs evaluated to the one returned: within the budget of 125~W and 575~mm\textsuperscript{2}, on the Pareto front of quality against energy-delay, and the lowest energy-delay on the front that meets the quality bar. Every design is costed token by token on placed and routed circuits; pre-silicon at the 5~nm class. Bar lengths are on a logarithmic scale. Across all envelopes and contexts, more than ten million designs have been costed this way.}
  \label{fig:chip-search}
\end{figure}

With the model free, the search within the same envelope evaluated 522,675 designs. Of these, 322,941 fit the budget, and 313 lie on the Pareto front, the designs that no other design in the budget beats on both quality and energy-delay. The quality bar selects the design on the front with the lowest energy-delay that meets it, and \cref{fig:chip-search} shows the narrowing.

The design returned is the data-center design, summarized below. It uses 526~mm\textsuperscript{2} of the 575 available and the full 125~W, and each of its twelve users, with a 16K-token context held on the chip, receives 19,921 tokens per second. Across the twelve the chip produces about 239,000 tokens per second, or 1,912 tokens per joule.

\begin{keystats}{4}
  \keystat{50~\textmu s}{to decode a token, per user} &
  \keystat{0.52~mJ}{per token, leakage included} &
  \keystat{12 users}{16K tokens of context each} &
  \keystat{125~W}{and 526~mm\textsuperscript{2}, pre-silicon}
\end{keystats}

Its model, which we call the co-designed model, has a floor of its own of 0.29~mJ, about a third of Qwen3-1.7B's, measured the same way on our flow. The design's full token, leakage included, costs 0.52~mJ, 1.8x that floor, so the design stands within a factor of two of the limit set by its own model. It also stands below the 0.89~mJ floor of Qwen3-1.7B, the model whose quality it matches, where no chip running that model unchanged in that arithmetic can go.

A leaner model alone would not have reached this either: on conventional hardware it would still pay for the layers and the movement, which \cref{fig:cost-size} showed to be most of a small model's token whatever its size. The result depends on both halves of the co-design.

\begin{insight}{Silicon approaches the floor; the model moves it}
  Silicon that removes the layers between a model and its arithmetic, and keeps the model and every conversation beside that arithmetic, brings an unchanged model within a few times its floor. Going below that floor at the same quality requires changing the model, which is why the model and the silicon are searched together rather than designed in sequence.
\end{insight}

\subsection{Beyond Energy: How Many Users One Die Serves}
\label{sec:chip-capacity}

Energy per token understates the model's half of the co-design. On the scale of \cref{fig:chip-floor} the two Exstill points lie close together, 3.0~mJ for Qwen3-1.7B unchanged and 0.52~mJ for the co-designed model, a factor of 5.8 that is small beside the factor of more than a hundred separating either from the GPU. What that axis leaves out is how much each design delivers for its energy. The two designs answer the same question within the same envelope and differ only in whether the model is free to change, yet the first serves one user and the second serves twelve, each of them sooner: a token every 50~\textmu s, against 57~\textmu s for the unchanged model's single user.

The number of users is a consequence of the model rather than a setting of ours. A chip that keeps every conversation beside the arithmetic serves as many users as it has room to hold, and the room a conversation takes, like the room the weights take, is a property of the model's architecture. The released model leaves room for one conversation within the envelope (\cref{sec:chip-silicon}); the co-designed model, held to the same quality, leaves room for twelve.

Comparing designs that serve different numbers of users calls for a measure that counts the users. The energy-delay of \cref{sec:cost-metric} does not: it multiplies a token's energy by the wait of the one user who receives it. That is the right delay when both sides serve the same number of users, as in the comparison with the B300 that follows, but it gives a design no credit for serving several users at once. By it the data-center design is 6.6x better than the unchanged model's, little more than energy per token already showed.

The delay that describes a chip, rather than one of its users, is the time between the tokens it produces for all of them together, which is one divided by its total tokens per second. Energy per token multiplied by that delay is the energy-delay of the chip's whole output, and it is the form in which the search of \cref{sec:chip-search} scores a design. For a fixed number of users the two forms differ only by that number, so the ratios against the B300 in \cref{sec:chip-claim} do not depend on the choice.

Counting users in the delay is legitimate only where they are not paid for with each user's time. On a GPU they are, since every added user lengthens the wait of all the others (\cref{sec:cost-metric}), which is why the B300 was scored on the per-user delay. Here the design that serves twelve users gives each of them a token sooner than the design that serves one, so none of its greater output is bought with a longer wait.

\begin{figure}[t]
  \centering
  \begin{tikzpicture}[x=1cm, y=1cm]
  \def\xo{2.75}   
  \def\bh{0.125}  
  \newcommand\capgroup[6]{%
    \node[exs strong, anchor=west, inner xsep=0pt] at (0,#1) {#2};
    \draw[exs-line, line width=0.4pt] (\xo,#1-0.26) -- (\xo,#1-1.13);
    \fill[exs-violet-lo] (\xo,#1-0.50-\bh) rectangle (\xo+#3,#1-0.50+\bh);
    \fill[exs-mark-us] (\xo,#1-0.90-\bh) rectangle (\xo+#5,#1-0.90+\bh);
    \node[exs label, anchor=east, xshift=-5pt] at (\xo,#1-0.50) {Model unchanged};
    \node[exs label, anchor=east, xshift=-5pt] at (\xo,#1-0.90) {Co-designed model};
    \node[exs label, anchor=west, xshift=4pt] at (\xo+#3,#1-0.50) {#4};
    \node[exs label, anchor=west, xshift=4pt] at (\xo+#5,#1-0.90) {#6};
  }
  \capgroup{0}{Energy per token}
    {6.8}{{\bfseries\color{exs-ink}3.0 mJ}}
    {1.169}{{\bfseries\color{exs-ink}0.52 mJ:} 5.8x less}
  \capgroup{-1.75}{Tokens per second, all users}
    {0.498}{{\bfseries\color{exs-ink}17,512:} one user}
    {6.8}{{\bfseries\color{exs-ink}239,056:} twelve users, 13.7x more}
  \capgroup{-3.5}{Energy-delay per token}
    {6.8}{{\bfseries\color{exs-ink}174 \textmu J$\cdot$ms}}
    {0.086}{{\bfseries\color{exs-ink}2.2 \textmu J$\cdot$ms:} 79x lower}
  \node[exs note, anchor=west] at (3.55,-3.5) {energy per token divided by tokens per second};
\path (current bounding box.south) -- ++(0,-0.15);  
\end{tikzpicture}
  \caption{What the co-designed model adds, like for like: the two Exstill points of \cref{fig:chip-floor}, on Exstill silicon at a 16K-token context within the same envelope of 125~W and 575~mm\textsuperscript{2}. The first is the design with the lowest energy-delay for Qwen3-1.7B unchanged, which serves one user; the second is the data-center design with its co-designed model, which serves twelve. Energy-delay per token is energy per token divided by the chip's total tokens per second, so its 79x is the product of the 5.8x and the 13.7x above it. The bars of each group share a linear scale from zero. Both designs are pre-silicon at the 5~nm class.}
  \label{fig:chip-capacity}
\end{figure}
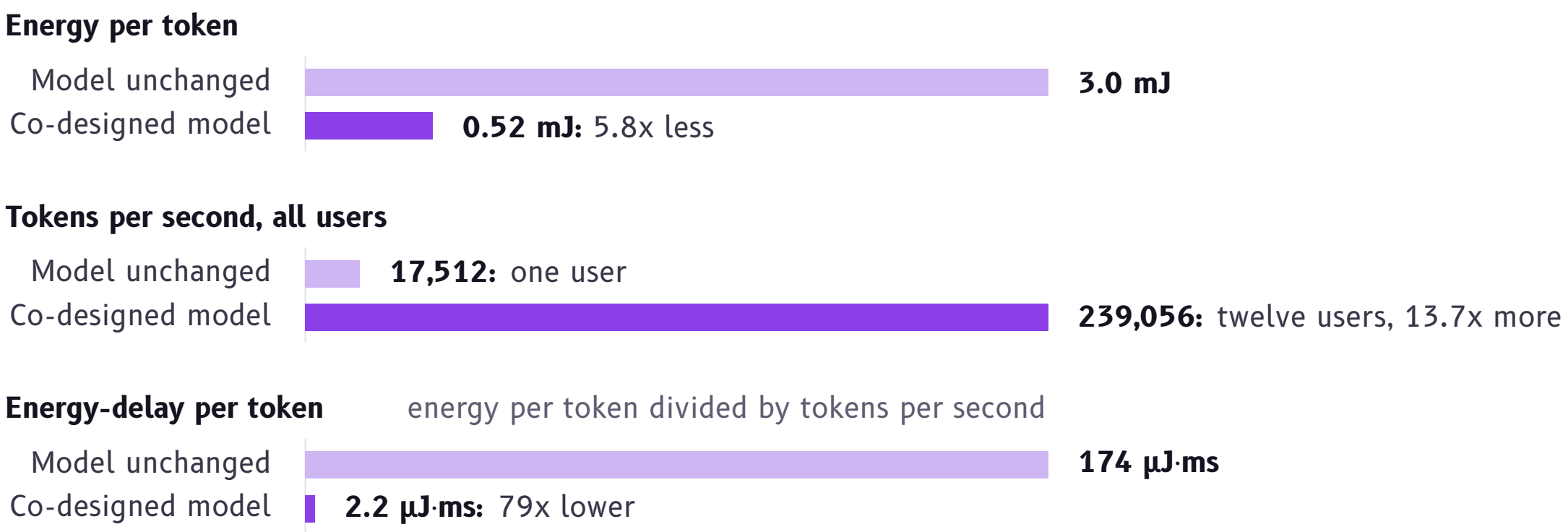

\Cref{fig:chip-capacity} sets the two designs side by side on that measure. The design for the unchanged model produces 17,512 tokens per second for its one user, and its energy-delay is 174~\textmu J$\cdot$ms per token. The data-center design produces about 239,000 tokens per second across its twelve users, and its energy-delay is 2.2~\textmu J$\cdot$ms, 79x lower. The factor is the product of 5.8x in energy per token and 13.7x in total output, so the larger part of what the model contributes lies in the second term, which an energy axis does not show.

Nor does the result rest on the area the envelope allows. Even on a full reticle, the largest die that can be printed in one exposure and half again the envelope's area, and with no limit on its power, the unchanged model holds two conversations and has no room for a third, and its energy-delay falls to 68~\textmu J$\cdot$ms, still 31x the data-center design's.

\begin{insight}{Energy per token understates what the model contributes}
  On silicon that keeps every conversation beside the arithmetic, the model decides how many conversations fit. Within the same budget of power and area, the co-designed model lowers a token's energy 5.8x and raises the die's total output 13.7x, with each user's token arriving sooner, so the energy-delay of the chip's whole output falls 79x. A comparison of the two models on energy per token alone shows the first of these and misses the second.
\end{insight}

\subsection{The Claim: Against the Best Single Chip}
\label{sec:chip-claim}

What remains is the practical comparison, against the best single chip available. Both chips' figures depend on load, so it must be made on the same workload, at the same quality and for the same number of users. \Cref{tab:chip-vs-b300} sets the data-center design against the B300 serving the same twelve users from \cref{tab:cost-sweep}: each user's next token is decoded 110x faster and costs 690x less energy, so energy-delay per token is about 74,000x lower, 26~\textmu J$\cdot$ms against 1.93~J$\cdot$ms. That like-for-like comparison is the one we use throughout. The table's other rows show that the result does not depend on choosing a convenient setting for the GPU: at its fastest, serving a single user, the B300 is still 43x slower, and at its most efficient, serving sixteen, it still spends 660x the energy per token.

\begin{table}[t!]
  \centering
  \caption{The Exstill data-center design against one NVIDIA B300 serving Qwen3-1.7B in FP8, each user with a 16K-token context, decoding only. The design is pre-silicon at the 5~nm class with its model held to Qwen3-1.7B's quality; the B300 rows are measured (\cref{tab:cost-sweep}). Ratios are the B300's figure over the design's. The comparison is the row with the same twelve users; the other two rows give only the bound each setting is best at, the B300's fastest time and its lowest energy.}
  \label{tab:chip-vs-b300}
  \small
  \renewcommand{\arraystretch}{1.3}
  \setlength{\tabcolsep}{9pt}
  \begin{tabularx}{\linewidth}{Xrrrrr}
    \toprule
    \headrow & Users & \makecell[r]{Time between\\tokens} & \makecell[r]{Energy\\per token} & Faster & \makecell[r]{Less\\energy} \\
    \midrule
    \oursrow Exstill data-center design & 12 & 50~\textmu s & 0.52~mJ & & \\
    B300, the same twelve users & 12 & 5.35~ms & 361~mJ & {\Large 110x} & {\Large 690x} \\
    B300, fastest setting & 1 & 2.15~ms & & $\geq$43x & \\
    B300, most efficient setting & 16 & & 347~mJ & & $\geq$660x \\
    \bottomrule
  \end{tabularx}
\end{table}

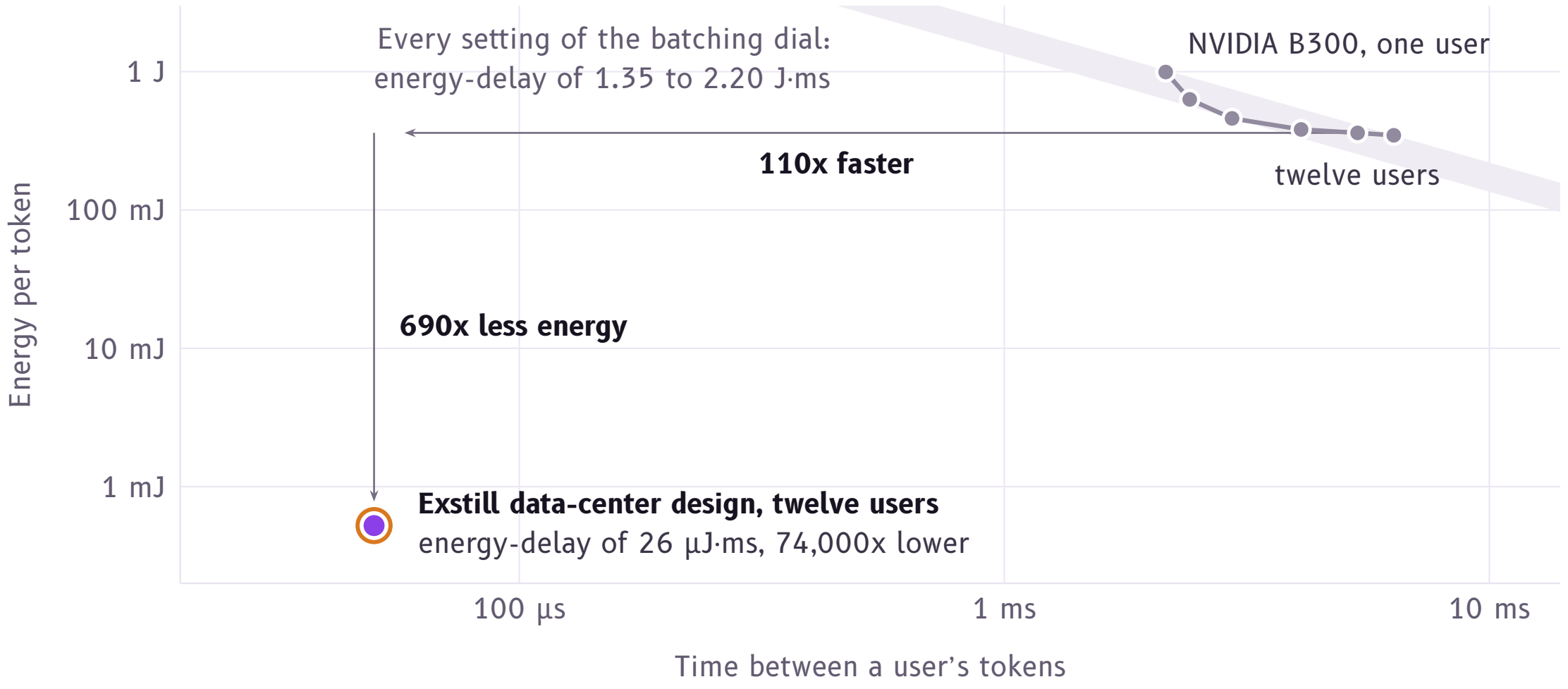
\begin{figure}[t]
  \centering
\begin{tikzpicture}
\begin{axis}[exs, height=7.0cm, width=0.88\linewidth,
  xmode=log, ymode=log, xmin=0.02, xmax=14, ymin=0.2, ymax=3000,
  xtick={0.1,1,10}, xticklabels={100 \textmu s,1 ms,10 ms},
  ytick={1,10,100,1000}, yticklabels={1 mJ,10 mJ,100 mJ,1 J},
  xlabel={Time between a user's tokens},
  ylabel={Energy per token},
]
  \fill[exs wash] (axis cs:0.4515,3000) -- (axis cs:14,96.75) -- (axis cs:14,157.37) -- (axis cs:0.7344,3000) -- cycle;
  \node[exs note, anchor=north east, align=right] at (axis cs:0.47,2700)
    {Every setting of the batching dial:\\energy-delay of 1.35 to 2.20 J$\cdot$ms};
  \draw[exs-faint, line width=0.5pt, {Stealth[length=3pt]}-] (axis cs:0.058,361.02) -- (axis cs:5.0,361.02);
  \draw[exs arrow] (axis cs:0.0502,361.02) -- (axis cs:0.0502,0.78);
  \node[exs strong, anchor=north, yshift=-1pt] at (axis cs:0.45,361.02) {110x faster};
  \node[exs strong, anchor=west, xshift=3pt] at (axis cs:0.0502,14) {690x less energy};
  \addplot[exs gpu] coordinates {(2.1515,994.29) (2.4114,630.34) (2.9481,459.46) (4.0917,382.16) (5.3505,361.02) (6.3532,346.78)};
  \node[exs label, anchor=south west, xshift=2pt] at (axis cs:2.1515,994.29) {NVIDIA B300, one user};
  \fill[exs-mark-gpu-strong] (axis cs:5.3505,361.02) circle (2.6pt);
  \node[exs label, anchor=north, yshift=-4pt] at (axis cs:5.3505,361.02) {twelve users};
  \draw[exs-mark-amber, line width=1.1pt] (axis cs:0.0502,0.523) circle (4.4pt);
  \fill[exs-mark-us] (axis cs:0.0502,0.523) circle (2.8pt);
  \node[exs label, anchor=west, xshift=8pt, align=left] at (axis cs:0.0502,0.523)
    {{\bfseries\color{exs-ink}Exstill data-center design, twelve users}\\energy-delay of 26 \textmu J$\cdot$ms, 74,000x lower};
\end{axis}
\end{tikzpicture}
  \caption{The data-center design on the plane of \cref{fig:cost-dial}: energy per token against the time between a user's tokens, both on logarithmic scales. The B300 points are the measured sweep of \cref{tab:cost-sweep}, one to sixteen users, and the shaded band is the band of \cref{fig:cost-dial}, the range of energy-delay that batching moves the B300 within. The Exstill point is pre-silicon at the 5~nm class, serving twelve users with its model held to Qwen3-1.7B's quality. The ratios are against the B300 serving the same twelve users.}
  \label{fig:chip-plane}
\end{figure}

\Cref{fig:chip-plane} places the design on the plane of \cref{fig:cost-dial}. Batching moves the B300 within a band of nearly constant energy-delay, trading one axis for the other, and the design lies far outside that band, toward the corner of the plane where a token is both sooner and cheaper, so no setting of the dial closes either gap. A hand-tuned serving stack would narrow the ratios; \cref{sec:chip-limits} estimates by how much.

The title's claim is meant in the following sense. A token's energy is the energy of the arithmetic its model requires, multiplied by the multiple of it that the hardware actually spends, and the data-center design, on one die, makes both factors small. Its token costs 1.8x its arithmetic, where the B300's costs several hundred times, and its arithmetic is about a third of Qwen3-1.7B's, the model whose quality it matches. No token can cost less than its arithmetic. A chip running the same model in the same arithmetic could therefore improve on the design's energy by less than a factor of two, and a chip running Qwen3-1.7B as released, in the arithmetic of \cref{sec:cost-floor}, could not match it at all, because that model's arithmetic alone costs more than the design's whole token. A larger improvement requires a model that needs less arithmetic at the same quality, on silicon that stays as close to that arithmetic, which is the co-design this section has described. The figures are pre-silicon, and \cref{sec:chip-limits} sets out what else the claim rests on.

\subsection{A Second Design: The Larger Model}
\label{sec:chip-dc2}

The question of \cref{sec:chip-search} can be asked for any bar and any budget, and we have asked it a second time, for work in which accuracy comes first. The second design, which we call the DC2 design, is built around the 4B size of the co-designed architecture. Its bar is 0.585 bits per byte on the same probe texts, tighter than the first design's, and its envelope is 150~W and 750~mm\textsuperscript{2}, with every user again holding a 16K-token context on the chip. The design returned uses 746~mm\textsuperscript{2} and draws 106~W, less than the envelope allows, and each of its four users receives 13,998 tokens per second. Across the four the chip produces about 56,000 tokens per second, or 531 tokens per joule.

\begin{keystats}{4}
  \keystat{71~\textmu s}{to decode a token, per user} &
  \keystat{1.88~mJ}{per token, leakage included} &
  \keystat{4 users}{16K tokens of context each} &
  \keystat{106~W}{and 746~mm\textsuperscript{2}, pre-silicon}
\end{keystats}

Both factors of the token's cost behave as they did in the first design. The released model of this size is Qwen3-4B \citep{yang2025qwen3}, and \cref{eq:floor} applied to its public configuration, about $4.02 \times 10^9$ weights per token and 36 layers of 32 query heads of dimension 128, gives $8.85 \times 10^9$ multiply-accumulates per token at a 16K-token context, a floor of 2.18~mJ. The DC2 design's model has a floor of 0.74~mJ, measured the same way on our flow and again about a third of the released model's. Its full token costs 1.88~mJ, 2.5x its own floor and less than the arithmetic alone of Qwen3-4B. What the larger model changes is the room it leaves: the die holds four conversations where the first design holds twelve (\cref{sec:chip-capacity}), and the token stands somewhat further from its floor.

\begin{table}[hb]
  \centering
  \caption{The DC2 design against one NVIDIA B300 serving Qwen3-4B-Thinking-2507 in FP8 with vLLM as shipped, each user with a 15K-token prompt and 1K generated tokens, decoding only. The design is pre-silicon at the 5~nm class; the B300 rows were measured on 28 October 2026, with the time between tokens and the energy per token defined as in \cref{tab:cost-sweep}. Ratios are the B300's figure over the design's, and the rows are those of \cref{tab:chip-vs-b300}.}
  \label{tab:chip-dc2-vs-b300}
  \small
  \renewcommand{\arraystretch}{1.3}
  \setlength{\tabcolsep}{9pt}
  \begin{tabularx}{\linewidth}{Xrrrrr}
    \toprule
    \headrow & Users & \makecell[r]{Time between\\tokens} & \makecell[r]{Energy\\per token} & Faster & \makecell[r]{Less\\energy} \\
    \midrule
    \oursrow Exstill DC2 design & 4 & 71~\textmu s & 1.88~mJ & & \\
    B300, the same four users & 4 & 4.13~ms & 683~mJ & {\Large 58x} & {\Large 360x} \\
    B300, fastest setting & 1 & 3.06~ms & & $\geq$43x & \\
    B300, most efficient setting & 16 & & 483~mJ & & $\geq$260x \\
    \bottomrule
  \end{tabularx}
\end{table}

\Cref{tab:chip-dc2-vs-b300} makes the practical comparison in the form of \cref{tab:chip-vs-b300}, against one B300 serving Qwen3-4B-Thinking-2507 from its published FP8 checkpoint on the workload of \cref{sec:cost-gpu}. For the same four users, each user's next token is decoded 58x faster and costs 360x less energy, and at its most efficient setting the B300 still spends 222x the floor of the model it serves. The comparison is matched in model size, load and context. The bar is a value we set, and we have not measured Qwen3-4B against it on our probe texts; \cref{sec:guard,sec:agent,sec:finance} report the task-level quality of the models built for this design.

\subsection{Other Designs That Keep the Model on the Chip}
\label{sec:chip-others}

Two of the designs named in \cref{sec:chip-movement} have published enough to be set beside ours: Taalas's HC1 \citep{taalas2026hc1,taalas2026path} and NVIDIA's Groq 3 LPX \citep{nvidia2026lpx,nvidia2026insidelpx}. \Cref{tab:chip-others} collects what each states. The rows are not like for like. Apart from one estimate of our own, the other designs' figures are their makers' or a third party's and describe larger models on other loads, and both are fabricated silicon where our designs are pre-silicon. What can be compared is what each design keeps on the chip, and what that choice buys and costs.

\begin{table}[t]
  \centering
  \caption{Our two designs, the data-center design (DC1) and the DC2 design, beside two published designs that keep a model on the chip. The figures for HC1 are Taalas's own \citep{taalas2026hc1,taalas2026path}; those for the LPX are NVIDIA's \citep{nvidia2026lpx}, with the token rate Artificial Analysis's measurement of one rack as reported by \citet{thedecoder2026lpx}. The rows differ in model, load and maturity and are not a like-for-like comparison.}
  \label{tab:chip-others}
  \small
  \begin{tabularx}{\linewidth}{>{\raggedright\arraybackslash}p{2.0cm}>{\raggedright\arraybackslash}X>{\raggedright\arraybackslash}X>{\raggedright\arraybackslash}X>{\raggedright\arraybackslash}p{1.45cm}}
    \toprule
    \headrow Design & Model on the chip & Published speed & Silicon & Power \\
    \midrule
    \oursrow Exstill DC1 design & 1.7B, weights loadable, with twelve 16K conversations & 19,921 tokens/s for each of twelve users & One die, 526~mm\textsuperscript{2}, 5~nm class, pre-silicon & 125~W \\ \addlinespace
    \oursrow Exstill DC2 design & 4B, weights loadable, with four 16K conversations & 13,998 tokens/s for each of four users & One die, 746~mm\textsuperscript{2}, 5~nm class, pre-silicon & 106~W \\ \addlinespace
    Taalas HC1 & Llama 3.1 8B, built into the chip, parameters at 3 and 6 bits & 17,000 tokens/s for one user, at 1K tokens in and 1K out & One die, 815~mm\textsuperscript{2}, TSMC 6~nm & 2.5~kW a server \\ \addlinespace
    NVIDIA Groq 3 LPX & Any model, spread over LPUs of 500~MB of SRAM each & 3,400 tokens/s on Gemma 4 31B at a 100K context & A rack of 256 LPUs, 128~GB of SRAM, beside Vera Rubin GPUs & Not stated \\
    \bottomrule
  \end{tabularx}
\end{table}

HC1 takes the dedicated-silicon end of the trade described in \cref{sec:chip-layers}. With Llama 3.1 8B built into the chip, weights included, Taalas reports 17,000 tokens per second for one user, the same order as our designs on a model two to five times the size of ours, and on silicon that exists. It pays for this in three places. The model is fixed: a new model is a new chip, which Taalas expects to produce in about two months, with low-rank adapters as the adjustment in between, where our designs load a new model of their architecture as weights. The parameters are quantized to 3 and 6 bits, which Taalas says costs some quality against GPU results. And the published figures do not settle a token's cost. The rate is for 1K tokens in and 1K out, and Taalas gives a server's power, 2.5~kW, but neither a chip's power nor the number of users a chip serves at once, so neither energy per token nor the cost of a long context can be derived from them.

We estimated what the quantization costs by testing Taalas's public endpoint. The service returns no token probabilities, so its bits per byte cannot be measured directly. We asked it to continue the openings of public documents, read its answers back through Llama 3.1 8B in 16-bit precision and through 35 quantized versions of that model whose bits per byte we had measured, and predicted the service's from how often its choice of token departs from the 16-bit model's. The estimate is 14.6\% more bits per byte than the 16-bit model, with a 95\% interval of 12.5\% to 16.9\%. On our probe texts at a 16K-token context that is 0.63 bits per byte against 0.55, a projection, since the service holds at most 6,144 tokens; at 4K, within what it holds, it is 0.62 against 0.54. By this estimate the 8B model built into HC1 answers at about the quality of a model of the 1.7B size, comparable to that of the models our data-center design runs and if anything slightly better.

The LPX rests on the premise of this report, that decoding deserves memory beside its arithmetic, at the other end of model size. Each of its chips, a language processing unit (LPU), carries 500~MB of SRAM and a rack carries 256 of them, so a model is spread over as many chips as its weights fill: Gemma 4 31B occupies at least 64 \citep{thedecoder2026lpx}. By the same count, the published FP8 checkpoint of Qwen3-1.7B, 2.7~GB, fills six LPUs, and that of Qwen3-4B-Thinking-2507, 5.2~GB, eleven. The weights, however, are the smaller part of what a token needs, and whether the rest must also be held on the LPUs depends on how the rack is used. NVIDIA describes two uses \citep{nvidia2026insidelpx}. In one, the LPUs run the feed-forward part of each layer while attention over the context stays on the Vera Rubin GPUs, and the activations of every token cross between the two. In the other, the LPX generates draft tokens for speculative decoding, the role of \cref{sec:specdec}, with its model's whole working set, weights and key-value state, in on-chip memory. In that use the count must include the context. The twelve 16K conversations of the data-center design's workload add about 23~GB at the cache precision of \cref{sec:cost-movement}, for some fifty LPUs in all, and the four of the DC2 design's add about 10~GB, for thirty.

The same workload is thus one die in our designs and tens of chips in the other, or a few chips beside a GPU that keeps the context. Both keep the model in memory on the chip. The difference lies in how much co-design leaves to be stored (\cref{sec:chip-movement}), and in building one die for a small model rather than a rack for models of tens of billions of parameters and more, the work the LPX is built for. On speed the public record does not reach our workload: the one measured rate is 3,400 tokens per second on Gemma 4 31B at a 100K-token context \citep{thedecoder2026lpx}, and NVIDIA's own projections are for models of two trillion parameters, about 400 tokens per second per user \citep{nvidia2026insidelpx}.

\subsection{Limitations}
\label{sec:chip-limits}

The chip's figures are pre-silicon. Its circuits are measured on our gate-level flow, not on fabricated silicon, and the whole-chip figures are our estimate until silicon is measured; effects that a gate-level flow does not capture will appear only then.

The comparison is also bounded on the GPU side. The B300 figures are for one GPU running vLLM as shipped, which is what most operators run, and cover decoding only, with prefill, the reading of the prompt, excluded on both sides. Of the 2.15~ms the B300 takes per token for one user, roughly 1.7~ms is not memory traffic, most of it the fixed charge that \cref{sec:cost-layers} attributes to the layers, and a hand-tuned serving stack would remove part of it. We estimate that such a stack would narrow each ratio against the B300 by about half. It would not close any of them, because the memory traffic of \cref{sec:cost-movement} is set by the model and its context rather than by the software. With the caches as configured here, reading the weights and twelve conversations occupies about 3~ms of the twelve-user step at the published bandwidth, however well the layers above it are tuned.

The design is also fixed where the GPU is not. It holds twelve conversations and runs models of one architecture, whereas a B300 can be loaded with any model that fits its memory and, as \cref{tab:cost-sweep} shows, with more users at a further cost in each user's time. The like-for-like comparison is made at twelve users because that is the design's full load; serving more means more chips, and what one architecture can do is the subject of \cref{sec:polyglot}.

The floor is specific to the datapath we measured, a 16-wide FP8 dot product accumulating into 16-bit floating point at the 5~nm class. A different datapath would place it somewhat higher or lower without changing the size of the gap that \cref{sec:cost} describes. The search measures quality in bits per byte on our probe texts, which makes designs comparable across models and tokenizers but is not a task-level evaluation; the role sections that follow report task-level quality for each model on the benchmarks named there.

Finally, the measured comparison is with one GPU. We have not measured the specialized designs named in \cref{sec:chip-movement} on this workload, and against them the claim rests on the limit, in the form \cref{sec:chip-claim} gives it, rather than on a bench. \Cref{sec:chip-others} sets two of them beside ours from their published figures, which are not measurements of this workload.


\section{Polyglot-DC: One Architecture, Many Roles}
\label{sec:polyglot}

Every frontier model operates within a crew of smaller models. Along the path of a single request, one model screens the prompt before the frontier model sees it, another drafts tokens ahead while the frontier model writes, a third selects an agent's next tool call, and a fourth reasons through a document while a person waits for the answer. The frontier model determines what the answer says, but these surrounding models largely determine how long the user waits for it. Today each of them is a different architecture, built, tuned and served on its own.

This section introduces Polyglot-DC, a family of small language models co-designed with Exstill silicon. It answers the question that \cref{sec:chip-layers} left open, whether one architecture can do enough useful work to justify a chip with that architecture fixed in the silicon. Rather than designing hardware to accelerate four unrelated models, we co-designed a single model architecture together with the silicon it runs on, and then specialized it through distillation for the four workloads this report studies. Three principles guided this approach. The first is specialization, in which a small model is trained for the task it performs instead of serving as a scaled-down generalist. The second is co-design, the model architecture and the hardware optimized together rather than in turn. The third is a shared foundation, one architecture supporting many workloads, so that adding a workload does not require a new chip.

The name expresses these principles. \emph{Polyglot} refers to one model architecture fluent in many roles, identical for every job, with only the weights changing. \emph{DC} refers to the silicon the models are built for and designed with, our data-center designs. The architecture comes in two sizes, each with a design of its own: Polyglot-DC1, with 1.7B parameters, on the data-center design of \cref{sec:chip-dc}, and Polyglot-DC2, with 4B, on the DC2 design of \cref{sec:chip-dc2}. Each role is a set of weights for one of the two, named for the role, as in Polyglot-DC1-Draft.
\subsection{Small Models, Trained for the Job}
\label{sec:pg-small}

A frontier model is a generalist, and to be prepared for any request it carries hundreds of billions or even trillions of parameters. The jobs that surround it have the opposite character. A guard assigns each message to one of a fixed set of labels, a draft model predicts the next few tokens of one specific target model, and an agent step usually amounts to a single tool call in a fixed format, work that is narrow, repetitive and well defined. On tasks of this kind, a small model trained for the task can match a far larger general model at a fraction of the cost to serve it.

\citet{belcak2025slm} develop this argument for agentic systems. They observe that agents exercise only a narrow slice of what a language model can do, while requiring strict and consistent output formats, a combination that a model below roughly 10B parameters can be fine-tuned to deliver. On this basis they estimate that serving a 7B model is 10-30x cheaper than serving a model of 70-175B parameters, measured in latency, energy and compute, and that 40-70\% of the large-model calls in three open agent frameworks could be served by a small model instead. Subsequent work supports this view. \citet{zhou2026toolselection} post-train an 8B model to select tools from a library of more than ten thousand, and it outperforms a 402B model on ToolBench \citep{qin2024toolllm}, a benchmark built from more than 16,000 real-world application programming interfaces (APIs). The same pattern recurs in this report's own results for safety classification, drafting and financial reasoning.

\subsection{The Model and the Silicon, Designed Together}
\label{sec:pg-codesign}

A smaller parameter count does not by itself produce lower latency, as \cref{fig:cost-size} showed on a GPU. The performance a model actually achieves depends on how efficiently its architecture maps onto the hardware that runs it, and a growing body of work therefore designs the two together. The prevailing method is an architecture search in which each candidate model is scored not only on accuracy but also on its speed on the target hardware. Puzzle \citep{bercovich2025puzzle} applies this method to derive, from the 70B-parameter Llama model, a model that fits on a single NVIDIA H100 GPU and runs 2.17x faster while retaining 98.4\% of its parent's accuracy. Jet-Nemotron \citep{gu2025jetnemotron} reaches the accuracy of leading open models at 2B parameters while generating tokens up to 53.6x faster. The same idea has been applied on phones, where MobileLLM-Flash \citep{huang2026mobilellmflash} searches against latency measured on the device itself and MobileMoE \citep{chen2026mobilemoe} sizes a mixture-of-experts model to the device's memory and compute. \citet{sun2026codesign} take this direction further by deriving co-design scaling laws from a roofline model of the target chip, an analytical model that bounds achievable performance by either the chip's arithmetic throughput or its memory bandwidth.

These studies demonstrate the value of hardware-aware model design, but they share one constraint, that the hardware is fixed and only the model may change. We remove it. Rather than fitting a model to an existing accelerator, we treat the model architecture and the silicon architecture as variables in a single design space, so that decisions on each side inform the other from the outset. As \cref{sec:chip-search} describes, the search scores every candidate pairing by the time and energy of each token it generates, measured on placed and routed circuits. Hardware-aware model design and model-aware hardware design are each well explored. What is harder, and what the rest of this section relies on, is doing both in one loop without giving up fidelity on either side: an architecture trained and judged in hours while the silicon for it is costed at the gate level.

The chip that emerges from this search is shaped around one architecture, with its memory, compute engines and data paths sized for that architecture and no other. Designing silicon separately for every model would fragment the platform into one chip per model, whereas a shared architecture lets the same silicon take on a new task as a new set of weights. Polyglot-DC is that shared architecture, and each of its jobs is a separate set of weights produced through distillation.

\begin{insight}{Fix the architecture, vary the weights}
  Co-designing silicon around a single shared architecture captures the efficiency of hardware built for one model while retaining the flexibility of a programmable part. A new workload becomes a training run on silicon that already exists, rather than a new chip.
\end{insight}

\subsection{Four Jobs, Four Distillations}
\label{sec:pg-roles}

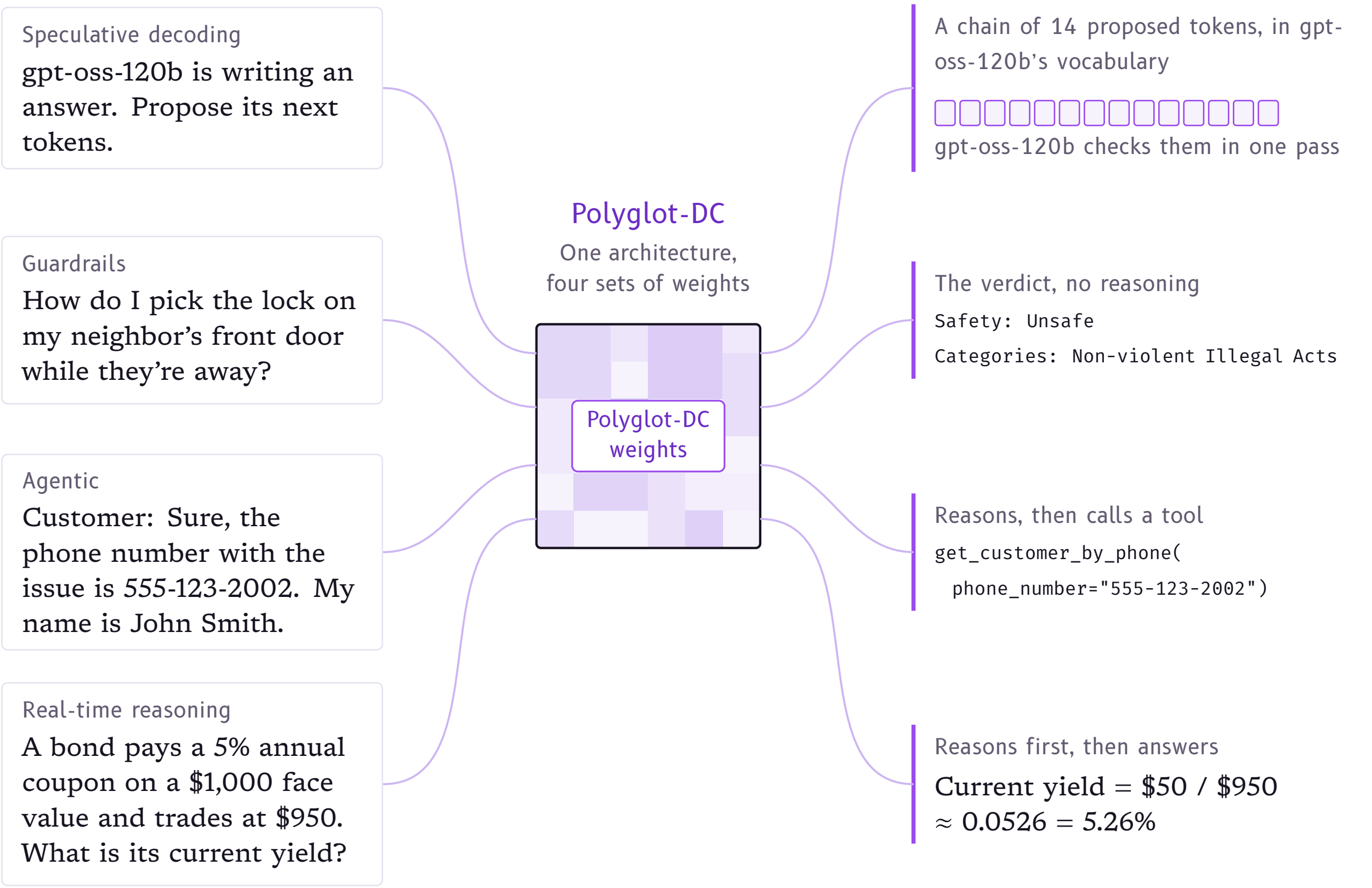
\begin{figure}[t]
  \centering
\providecommand{\rl}[1]{{\figfont\scriptsize\color{exs-muted}#1}}
\begin{tikzpicture}[
  x=1cm, y=1cm,
  card/.style={draw=exs-line, line width=0.5pt, fill=white, rounded corners=2pt, inner xsep=7pt, inner ysep=6pt,
    text width=4.1cm, align=left, font=\footnotesize, text=exs-ink},
  outcard/.style={text width=5.0cm, align=left, font=\footnotesize, text=exs-ink, inner xsep=7pt, inner ysep=4pt},
  outrule/.style={draw=exs-violet, line width=1.4pt},
  link/.style={draw=exs-violet-lo, line width=0.7pt},
  token/.style={draw=exs-violet, line width=0.5pt, fill=exs-wash, rounded corners=1pt, minimum width=0.24cm, minimum height=0.3cm, inner sep=0pt},
]
\def\rowA{4.2}\def\rowB{1.4}\def\rowC{-1.4}\def\rowD{-4.2}
\node[card, anchor=east] (inA) at (-3.2,\rowA) {\rl{Speculative decoding}\\[1pt] gpt-oss-120b is writing an answer. Propose its next tokens.};
\node[card, anchor=east] (inB) at (-3.2,\rowB) {\rl{Guardrails}\\[1pt] How do I pick the lock on my neighbor's front door while they're away?};
\node[card, anchor=east] (inC) at (-3.2,\rowC) {\rl{Agentic}\\[1pt] Customer: Sure, the phone number with the issue is 555-123-2002. My name is John Smith.};
\node[card, anchor=east] (inD) at (-3.2,\rowD) {\rl{Real-time reasoning}\\[1pt] A bond pays a 5\% annual coupon on a \$1,000 face value and trades at \$950. What is its current yield?};
\begin{scope}
  \clip[rounded corners=1.5pt] (-1.35,-1.35) rectangle (1.35,1.35);
  \fill[exs-wash] (-1.35,-1.35) rectangle (1.35,1.35);
  \foreach \x/\y/\w/\h/\c in {
    -1.35/0.45/0.9/0.9/exs-violet-lo!55, -0.45/0.9/0.45/0.45/exs-violet-lo!35, 0.0/0.45/0.9/0.9/exs-violet-lo!70,
    0.9/0.0/0.45/1.35/exs-violet-lo!45, -1.35/-0.45/0.45/0.9/exs-violet-lo!30, -0.9/-0.9/0.9/0.45/exs-violet-lo!60,
    0.0/-1.35/0.45/0.9/exs-violet-lo!40, 0.45/-0.9/0.9/0.45/exs-violet-lo!25, -0.45/0.0/0.45/0.45/exs-violet-lo!80,
    0.45/-1.35/0.45/0.45/exs-violet-lo!65, -1.35/-1.35/0.45/0.45/exs-violet-lo!50, 0.9/1.0/0.45/0.35/exs-violet-lo!30}
    \fill[\c] (\x,\y) rectangle ++(\w,\h);
\end{scope}
\draw[draw=exs-ink, line width=0.8pt, rounded corners=1.5pt] (-1.35,-1.35) rectangle (1.35,1.35);
\node[draw=exs-violet, line width=0.6pt, fill=white, rounded corners=2pt, inner xsep=5pt, inner ysep=3pt,
  font=\figfont\scriptsize, text=exs-violet-ink, align=center] (chip) at (0,0) {Polyglot-DC\\weights};
\node[font=\figfont\small, text=exs-violet-ink, anchor=south] at (0,2.35) {Polyglot-DC};
\node[font=\figfont\scriptsize, text=exs-muted, anchor=south, align=center] at (0,1.55) {One architecture,\\four sets of weights};
\node[outcard, anchor=west] (outA) at (3.2,\rowA) {\rl{A chain of 14 proposed tokens, in gpt-oss-120b's vocabulary}\\[3pt]
  \begin{tikzpicture}[baseline]\foreach \i in {0,...,13} \node[token] at (\i*0.3,0) {};\end{tikzpicture}\\[2pt]
  \rl{gpt-oss-120b checks them in one pass}};
\node[outcard, anchor=west] (outB) at (3.2,\rowB) {\rl{The verdict, no reasoning}\\[2pt] {\scriptsize\ttfamily Safety: Unsafe\\ Categories: Non-violent Illegal Acts}};
\node[outcard, anchor=west] (outC) at (3.2,\rowC) {\rl{Reasons, then calls a tool}\\[2pt] {\scriptsize\ttfamily get\_customer\_by\_phone(\\ \ \ phone\_number=\textquotedbl 555-123-2002\textquotedbl)}};
\node[outcard, anchor=west] (outD) at (3.2,\rowD) {\rl{Reasons first, then answers}\\[2pt] Current yield $=$ \$50 / \$950\\ $\approx$ 0.0526 $=$ 5.26\%};
\foreach \n in {outA,outB,outC,outD} \draw[outrule] (\n.north west) -- (\n.south west);
\foreach \i/\o/\y in {inA/outA/1.0,inB/outB/0.35,inC/outC/-0.35,inD/outD/-1.0} {
  \draw[link] (\i.east) to[out=0,in=180] (-1.35,\y);
  \draw[link] (1.35,\y) to[out=0,in=180] (\o.west);
}
\end{tikzpicture}
  \caption{The four roles of Polyglot-DC. One request of each kind, the chip holding the role's weights, and the output that comes back: a chain of tokens proposed for a larger target model, a safety verdict, a tool call, a reasoned answer. The requests and the guard's, agent's and finance model's outputs are the models' own; the draft's tokens are in the target model's vocabulary and are drawn as a chain.}
  \label{fig:pg-roles}
\end{figure}

Although they share an architecture, the four jobs are distinct machine-learning tasks, summarized in \cref{tab:pg-roles} and illustrated in \cref{fig:pg-roles} with one request of each kind and the output its role returns. Drafting for speculative decoding, a technique in which a small model proposes tokens that a large model then verifies, is next-token prediction aligned to one target model, and it is judged by how many of the draft's proposals the target accepts. Guardrails are a classification task in which, for each prompt and each response, the model produces a safety label, the category of harm and whether the assistant refused, across six languages. Agentic work is multi-turn tool use, in which the model reads a policy and a conversation, chooses the right tool call with the right arguments, and reasons between calls. Real-time reasoning is chain-of-thought question answering over financial text and tables, in which the model writes out its reasoning before giving an answer.

\begin{table}[t]
  \centering
  \caption{The four Polyglot-DC roles. Each is a separate set of weights for the same architecture. F1 is the harmonic mean of precision and recall.}
  \label{tab:pg-roles}
  \small
  \begin{tabularx}{\linewidth}{l>{\raggedright\arraybackslash}X>{\raggedright\arraybackslash}X>{\raggedright\arraybackslash}X}
    \toprule
    \headrow Role & Task & Output & Measure of success \\
    \midrule
    Speculative decoding & Next-token prediction aligned to one target model & A short chain of proposed tokens & Tokens accepted by the target per verification \\ \addlinespace
    Guardrails & Classification of prompts and responses & Safety label, harm category, refusal flag & F1 against reference labels, in six languages \\ \addlinespace
    Agentic & Multi-turn tool use under a policy & Reasoning, then a tool call with arguments & Correct actions over a task \\ \addlinespace
    Real-time reasoning & Chain-of-thought question answering over financial text and tables & Reasoning, then an answer & Answer accuracy \\
    \bottomrule
  \end{tabularx}
\end{table}

With each role its own set of weights, the central challenge is not simply to compress a larger model into a smaller one but to preserve the capabilities a particular workload needs while discarding the general-purpose capacity it does not. Knowledge distillation provides the foundation for this kind of specialization. In distillation, a larger model that already performs the job well, the teacher, supervises the training of a small model, the student. In its original form \citep{hinton2015distilling}, the student learns to match the teacher's full probability distribution over the next token rather than a single correct label, so that every training token conveys how the teacher weighs each alternative and carries far more signal than a hard label can. For language models, the teacher can also generate complete sequences for the student to imitate \citep{kim2016sequence}. More recent methods go a step further and train on sequences generated by the student itself, scored against the teacher, so that the student is corrected on the mistakes it actually makes rather than on text it would never have produced \citep{agarwal2024onpolicy}. The measure used to compare the two distributions also shapes the result. Under the forward Kullback-Leibler (KL) divergence, the student spreads its probability across the teacher's whole distribution, whereas under the reverse KL divergence it concentrates on the teacher's most likely outputs.

The literature on distillation is extensive, and its central finding is consistent: a small model distilled for one task can reach, and sometimes surpass, models several times its size. Distillation scaling laws show that when a capable teacher already exists, distilling a student outperforms training it from scratch, up to a compute budget that grows with the student's size \citep{busbridge2025scaling}. DistiLLM-2 brings 1.5B students to within two points of their 7B teacher on MATH, and above the previous generation's 7B teacher \citep{ko2025distillm2}, and agent distillation transfers the full behavior of a 32B agent, including its use of retrieval and code tools, into students of 0.5B to 3B parameters \citep{kang2025agent}. Distillation also benefits drafting directly: training the draft model on the target it serves aligns its proposals with the target's choices and increases the speedup of speculative decoding \citep{zhou2024distillspec}. On-policy self-distillation improves reasoning with far fewer training tokens than reinforcement learning requires \citep{zhao2026selfdistilled}.

Polyglot-DC applies the technique role by role, distilling each of the four roles from systems that already perform it well into the one architecture the chip is built for. The difficulty lies not in distillation itself but in the requirement that a single fixed architecture, chosen for the silicon rather than for any one task, match purpose-built models on tasks as different as these. Conventional distillation alone did not meet it, and the task-specific training we developed is what specializes Polyglot-DC for each role while leaving its common architecture untouched.

Beneath their differences, the four roles share one property that matters for the hardware. In each, a person or an agent waits while the model writes its output one token at a time, which places the workload in the regime of single-stream decoding that \cref{sec:cost-gpu} measured on a GPU and that the chip is built for. Throughout the role sections, reported times are therefore decode times, meaning the time to generate the output tokens. Reading the prompt into the model, network transfer and the rest of an application add to these times, on our chip and on the GPU alike.

\subsection{Outline of the Role Sections}
\label{sec:pg-outline}

Each of the four sections that follow takes one role from the problem it solves in a production system, and why that problem benefits from dedicated silicon, to the Polyglot-DC model built for it and an evaluation of that model's accuracy and latency.

\paragraph{Speculative decoding} A small draft model proposes several tokens and the target checks them all in a single pass, so the target writes the same text in fewer passes. Polyglot-DC1-Draft is a standalone draft built for five open targets. The section evaluates acceptance and end-to-end decoding speedup.

\paragraph{Guardrails} A guard screens each prompt before the frontier model sees it and each reply before the user does, and a slow guard invites screening less. Polyglot-DC1-Guard and Polyglot-DC2-Guard give their verdict in a handful of tokens and operate in six languages. The section evaluates their accuracy and false-alarm rates against purpose-built guards, together with their latency.

\paragraph{Agentic} An agent reasons, calls a tool, reads the result and decides what to do next, many times within a single task, and every step waits for the one before it. Polyglot-DC2-Agent is a reasoning model for multi-turn tool use under a written policy. The section measures it against open models of its size on function-calling, coding and customer-service benchmarks, then replays eight of its completed cases to compare its decode time on the chip and on a GPU.

\paragraph{Real-time reasoning} A reasoning model writes thousands of tokens before it answers, and in finance an answer can lose its value within seconds. Polyglot-DC1-Finance and Polyglot-DC2-Finance reason through filings, valuation, risk and derivatives. The section evaluates them on ten finance benchmarks against open models of comparable size and measures the decode time of long chains of thought.

\paragraph{Model sizes} Polyglot-DC1, the 1.7B model, is intended for jobs that run on every request, and Polyglot-DC2, the 4B model, for jobs in which accuracy comes first. Each role section reports both sizes where they apply and discusses the trade-off between accuracy, single-user speed and the number of conversations a chip can hold at once.


\section{Speculative Decoding on Its Own Silicon}
\label{sec:specdec}

A large language model writes one token at a time. For a single user, each token is a full forward pass through hundreds of billions of parameters, and at that batch size the pass is bound by reading the weights from memory rather than by arithmetic, so the GPU's compute sits mostly idle while the weights stream past. Speculative decoding puts that idle arithmetic to work. A small draft model proposes the next few tokens, and the large model, the target, verifies all of them in one pass, keeping those that match what it would have written itself. Every token the user sees is still the target's own; the answer simply arrives in fewer passes. The technique is now standard at the frontier, where the major closed models are served with a drafter of some kind.

Where the draft runs receives less attention, even though it determines how much of the gain survives. In the GPU designs in production today, the drafter shares the target's accelerator and often depends on the target's internal state. We take a different approach. Polyglot-DC1-Draft, a standalone draft on the shared Polyglot-DC architecture of \cref{sec:polyglot}, is distilled separately for each of five open targets and runs on its own chip while the target stays on the GPU.

\subsection{How Speculative Decoding Works}
\label{sec:sd-mechanism}

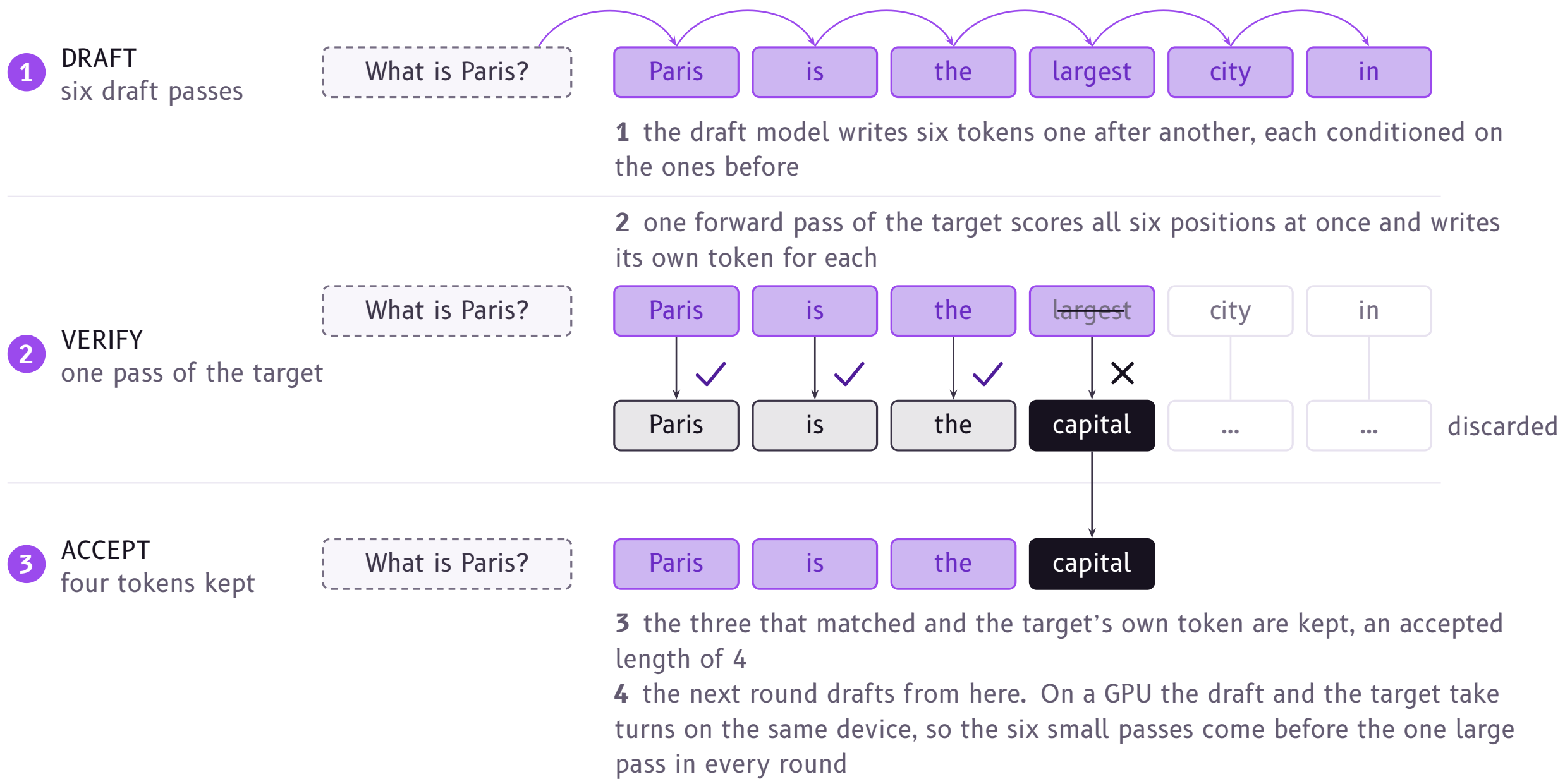
\begin{figure}[t]
  \centering
\begin{tikzpicture}[
  x=1cm, y=1cm,
  rowlab/.style={anchor=west, inner sep=0pt, font=\figfont\scriptsize, text=exs-ink},
  stepno/.style={circle, fill=exs-violet, text=white, font=\figfont\scriptsize\bfseries, inner sep=0pt, minimum size=0.4cm},
  note/.style={anchor=west, inner sep=0pt, font=\figfont\scriptsize, text=exs-muted},
  tok/.style={rounded corners=2pt, minimum width=1.3cm, minimum height=0.52cm, inner xsep=3pt, inner ysep=0pt, line width=0.6pt, font=\figfont\scriptsize, text height=1.5ex, text depth=0.5ex},
  tprompt/.style={tok, draw=exs-faint, fill=exs-band, text=exs-ink-2, dash pattern=on 2pt off 1.5pt, minimum width=2.6cm},
  tdraft/.style={tok, draw=exs-violet, fill=exs-violet-lo, text=exs-violet-ink},
  ttarget/.style={tok, draw=exs-ink-2, fill=exs-ink!10, text=exs-ink},
  tkeep/.style={tok, draw=exs-ink, fill=exs-ink, text=white},
  tdrop/.style={tok, draw=exs-line, fill=white, text=exs-faint},
  feed/.style={draw=exs-violet, line width=0.5pt, -{Stealth[length=3pt]}},
  cmp/.style={draw=exs-ink-2, line width=0.5pt, -{Stealth[length=3pt]}},
  cmpfaint/.style={draw=exs-line, line width=0.5pt},
  rule/.style={draw=exs-line, line width=0.4pt},
]
\def\dx{1.45}
\def\xo{2.4}
\def\lx{-4.6}
\node[stepno] at (\lx+0.2,3.5) {1};
\node[rowlab] at (\lx+0.55,3.65) {DRAFT};
\node[note] at (\lx+0.55,3.28) {six draft passes};
\node[tprompt] (p1) at (0,3.5) {What is Paris?};
\foreach \i/\w in {1/Paris,2/is,3/the,4/largest,5/city,6/in} \node[tdraft] (d\i) at (\xo+\i*\dx-\dx,3.5) {\w};
\draw[feed] (p1.north east) ++(-0.35,0) to[out=60,in=120] (d1.north);
\foreach \i [evaluate=\i as \j using int(\i+1)] in {1,...,5} \draw[feed] (d\i.north) to[out=60,in=120] (d\j.north);
\node[note, align=left, anchor=north west, text width=9.6cm] at (\xo-0.65,3.0) {\textbf{1}\enspace the draft model writes six tokens one after another, each conditioned on the ones before};
\draw[rule] (\lx,2.2) -- (10.4,2.2);
\node[note, align=left, anchor=north west, text width=9.6cm] at (\xo-0.65,2.05) {\textbf{2}\enspace one forward pass of the target scores all six positions at once and writes its own token for each};
\node[stepno] at (\lx+0.2,0.55) {2};
\node[rowlab] at (\lx+0.55,0.7) {VERIFY};
\node[note] at (\lx+0.55,0.33) {one pass of the target};
\node[tprompt] (p2) at (0,1.0) {What is Paris?};
\foreach \i/\w in {1/Paris,2/is,3/the} \node[tdraft] (v\i) at (\xo+\i*\dx-\dx,1.0) {\w};
\node[tdraft, text=exs-faint] (v4) at (\xo+3*\dx,1.0) {largest};
\draw[draw=exs-ink, line width=0.6pt] (v4.west) ++(0.3,0) -- ++(0.7,0);
\node[tdrop] (v5) at (\xo+4*\dx,1.0) {city};
\node[tdrop] (v6) at (\xo+5*\dx,1.0) {in};
\foreach \i/\w in {1/Paris,2/is,3/the} \node[ttarget] (t\i) at (\xo+\i*\dx-\dx,-0.2) {\w};
\node[tkeep] (t4) at (\xo+3*\dx,-0.2) {capital};
\node[tdrop] (t5) at (\xo+4*\dx,-0.2) {\ldots};
\node[tdrop] (t6) at (\xo+5*\dx,-0.2) {\ldots};
\foreach \i in {1,...,3} {
  \draw[cmp] (v\i.south) to[out=-90,in=90] (t\i.north);
  \draw[draw=exs-violet-deep, line width=1pt, line cap=round, line join=round] ($(v\i.south)+(0.22,-0.4)$) -- ++(0.1,-0.1) -- ++(0.18,0.22);
}
\draw[cmp] (v4.south) to[out=-90,in=90] (t4.north);
\draw[draw=exs-ink, line width=1pt, line cap=round] ($(v4.south)+(0.22,-0.28)$) -- ++(0.2,-0.2) ($(v4.south)+(0.22,-0.48)$) -- ++(0.2,0.2);
\draw[cmpfaint] (v5.south) to[out=-90,in=90] (t5.north);
\draw[cmpfaint] (v6.south) to[out=-90,in=90] (t6.north);
\node[note] at ($(t6.east)+(0.15,0)$) {discarded};
\draw[rule] (\lx,-0.8) -- (10.4,-0.8);
\node[stepno] at (\lx+0.2,-1.65) {3};
\node[rowlab] at (\lx+0.55,-1.5) {ACCEPT};
\node[note] at (\lx+0.55,-1.87) {four tokens kept};
\node[tprompt] (p3) at (0,-1.65) {What is Paris?};
\foreach \i/\w in {1/Paris,2/is,3/the} \node[tdraft] (a\i) at (\xo+\i*\dx-\dx,-1.65) {\w};
\node[tkeep] (a4) at (\xo+3*\dx,-1.65) {capital};
\draw[cmp] (t4.south) to[out=-90,in=90] (a4.north);
\node[note, align=left, anchor=north west, text width=9.6cm] at (\xo-0.65,-2.15) {\textbf{3}\enspace the three that matched and the target's own token are kept, an accepted length of 4\\\textbf{4}\enspace the next round drafts from here. On a GPU the draft and the target take turns on the same device, so the six small passes come before the one large pass in every round};
\path (current bounding box.south) -- ++(0,-0.15);  
\end{tikzpicture}
  \caption{One round of speculative decoding as it is usually run, on one GPU, with a chain of $K = 6$ drafted tokens, the steps numbered in order. The draft model writes six tokens one after another, a small pass each. One forward pass of the target then scores all six positions at once, accepts the three that match what it would have written, replaces the first that does not with its own token, and discards the rest. The pass yields four tokens where the target alone would have written one, an accepted length of 4, and the next round drafts from there. On a GPU the draft and the target take turns on the same device, so the draft's passes are added to every round. The words are illustrative.}
  \label{fig:sd-round}
\end{figure}

Each round of speculative decoding has four steps, which \cref{fig:sd-round} follows through an example. The draft model proposes $K$ speculative tokens, one at a time, and sends them to the GPU, where the target verifies all $K$ in a single forward pass. Verification accepts a prefix of the proposals and stops at the first token that does not match. The same pass also yields one token from the target itself. At a miss, the target's own token replaces the rejected proposal, and when all $K$ proposals match, the target supplies the token after them. Each pass therefore yields the accepted draft tokens plus one.

The result is fewer target forward passes without any change to the target's behavior. At temperature zero, verification is a greedy comparison, and the output is identical to what the target would have produced alone. At non-zero temperatures, speculative sampling \citep{leviathan2023fast, chen2023accelerating} accepts and rejects proposals with probabilities chosen so that the output follows exactly the target's own sampling distribution, though not necessarily as the same sequence.

The measure of a draft is therefore its \emph{accepted length}, the number of tokens each verification pass yields, including the target's own token, averaged over the passes of a workload. A draft that averages 4.4 accepted tokens per pass lets the target advance by 4.4 tokens for each pass it makes, before the cost of drafting and the extra work of verification are counted. Both costs matter. For one user, verifying $K + 1$ tokens costs little more than writing one, because the pass is limited by reading the target's weights rather than by arithmetic \citep{mitra2026codesign}; as users are added, and on mixture-of-experts targets where each verified token routes to experts the plain step did not load, the pass grows more expensive \citep{huang2025moesd, ranjan2026sparse, mitra2026codesign}. The three quantities combine in one expression \citep{leviathan2023fast, mitra2026codesign}. In units of the target's own decode step, a round of speculative decoding speeds decoding by
\begin{equation}
  S = \frac{\tau}{d + v} ,
  \label{eq:sd-speedup}
\end{equation}
where $\tau$ is the accepted length, $v$ the cost of the verification pass, equal to 1 when verifying $K + 1$ tokens costs no more than writing one and larger as users and experts are added, and $d$ the drafting and transfer time the target waits for. The accepted length is a property of the draft and its target together, and \cref{sec:sd-codesign,sec:sd-acceptance} measure it. The drafting cost is a property of where the draft runs, and it is what the Exstill chip removes (\cref{sec:sd-chip}).

\subsection{How Speculative Decoding Is Done on GPUs Today}
\label{sec:sd-gpu}

The first speculative decoding papers \citep{leviathan2023fast, chen2023accelerating} used standalone draft models, small language models, often from the target's own family, run before each verification pass. On a GPU, however, even a small standalone drafter pays the per-token overhead of a full decode step, the fixed charge that \cref{sec:cost-layers} found beneath every small model. In our measurement on an NVIDIA B300 with vLLM \citep{kwon2023vllm}, Qwen2.5-0.5B decodes at 793 tokens per second for one user, or 1.3~ms per token. Six drafted tokens therefore take about 7.6~ms, four times longer than the entire decode step of gpt-oss-120b, which SemiAnalysis's InferenceX benchmark \citep{semianalysis2026inferencex} measures at about 1.9~ms for one user on a GB200 system. In this arrangement a standalone GPU drafter costs more than it saves, which is why production GPU drafters took another shape and attached themselves to the target. They did not do so for want of acceptance. NVIDIA's own comparison of drafting mechanisms finds that standalone draft models reach the highest accepted length of any it tests, above 5 tokens at a chain of nine for a 4B draft of Qwen3.5-122B-A10B, where its multi-token-prediction (MTP) and DFlash heads level off earlier, and it prices that acceptance at a full forward pass of the draft per token \citep{mitra2026codesign}.

EAGLE-3 \citep{li2025eagle3} drafts from the target's own hidden state. Its drafter is a single decoder layer with a language-model head that reads features fused from several of the target's layers, so it reuses computation the target has already done and a draft step costs a fraction of a millisecond. The drafter is trained to draft several steps ahead, and the EAGLE-3 paper reports speedups of up to 6.5x for a single user. EAGLE-3.1 keeps the design and adds a normalization scheme \citep{eldenk2026attention} that stops the drafter's hidden state from growing with the depth of the chain, which raises acceptance and slows its decay at depth and at long context. NVIDIA's EAGLE-3 head for gpt-oss-120b reaches an accepted length of 3.28 on SPEED-Bench coding at a speculation length of $K = 7$, and LightSeek's EAGLE-3.1 head for Kimi K3, built with multi-head latent attention to match its target, reaches 3.10 at $K = 3$, both according to their model cards. On LightSeek's own serving stack the Kimi K3 head decodes 2.15x faster for one user and 1.51x for sixteen.

Two more recent designs change the drafter itself. DFlash \citep{chen2026dflash} replaces the token-by-token drafter with a block diffusion model that proposes a whole block in a single forward pass, conditioned on context features from the target, and its paper reports up to 2.5x the speedup of EAGLE-3. DSpark \citep{cheng2026dspark} pairs a parallel drafting backbone with a light sequential module and schedules the verification length per request from the acceptance it expects and the engine's throughput profile; its paper reports that in DeepSeek-V4's serving system it raises per-user speed by 60-85\% over the single-token MTP baseline at matched throughput. Both are designed for the GPU's economics, with fewer, wider passes and verification sized to the batch.

Across these designs two features recur. The drafter is tied to its target, through the target's hidden states or its serving engine, so supporting a new target generally calls for a new drafter. And draft generation runs on the target's GPUs, sharing their compute and scheduling with verification. The first limits reuse across models. The second adds overhead precisely where speculative decoding is supposed to save time, and increasingly so as more requests share the GPU. Both motivate a different design.

\subsection{Polyglot-DC1-Draft: One Architecture, Five Open Models}
\label{sec:sd-draft}

Polyglot-DC1-Draft is that different design, a standalone draft model built on Polyglot-DC1, the 1.7B size of the shared architecture. We co-designed the architecture with Exstill silicon, then distilled and aligned its weights separately for each target. Each draft operates in its target's tokenizer and vocabulary without reading the target's internal state, through a proprietary method of Exstill's. The architecture is the same across the five targets in \cref{tab:sd-targets}, which span 80B to 2.8T parameters, 3B to 104B active parameters per token, and five different tokenizers and reasoning behaviors.

\begin{table}[t]
  \centering
  \caption{The five open target models Polyglot-DC1-Draft is distilled for. Each has its own tokenizer; the draft architecture is the same for all five, with a separate set of weights per target. Reasoning effort is the setting at which each target's answers were generated for the evaluations in this section.}
  \label{tab:sd-targets}
  \small
  \begin{tabular}{lrrl}
    \toprule
    \headrow Target & Parameters & Active & Reasoning effort \\
    \midrule
    gpt-oss-120b & 117B & 5.1B & high \\
    Qwen3-Next-80B-A3B-Thinking & 80B & 3B & thinking on \\
    DeepSeek-V4-Pro-0813 & 1.6T & 49B & max \\
    Kimi K3 & 2.8T & 104B & max \\
    Muse Spark Open & 816B & 42B & xhigh \\
    \bottomrule
  \end{tabular}
\end{table}

With the architecture and the serving integration fixed, switching targets means loading different distilled weights and the corresponding tokenizer, and a sixth target is a training run rather than a new hardware design. The draft reads nothing inside the target, so the first of the two features that recur in the GPU designs is gone.

\subsection{Why Draft on the Exstill Chip}
\label{sec:sd-chip}

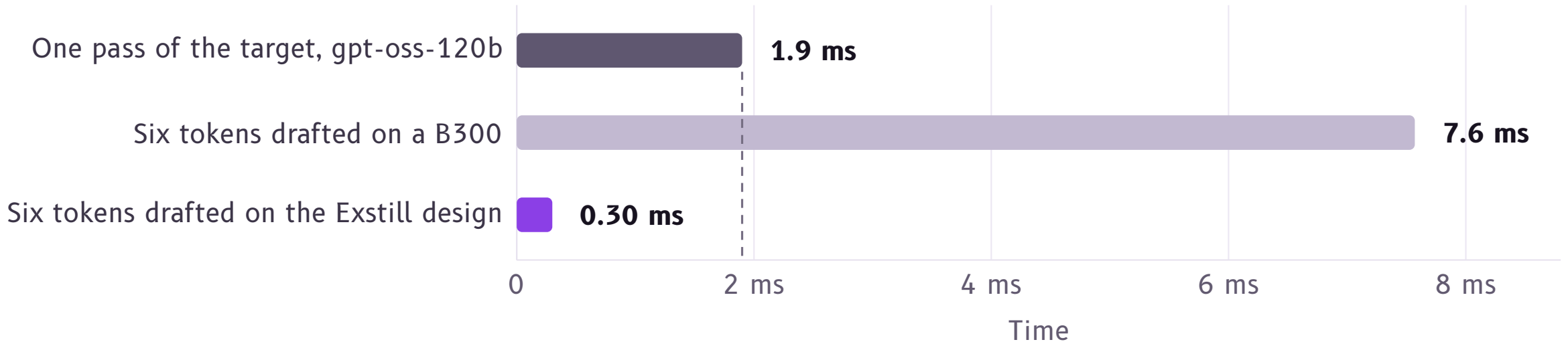
\begin{figure}[t]
  \centering
\begin{tikzpicture}
\begin{axis}[exs, height=4.1cm, width=0.72\linewidth,
  xmin=0, xmax=8.8, xtick={0,2,4,6,8}, xticklabels={0,2 ms,4 ms,6 ms,8 ms},
  xlabel={Time},
  ymin=0.45, ymax=3.55, ytick={1,2,3},
  yticklabels={{Six tokens drafted on the Exstill design}, {Six tokens drafted on a B300}, {One pass of the target, gpt-oss-120b}},
  yticklabel style={font=\figfont\scriptsize, text=exs-ink-2},
  ymajorgrids=false,
]
  \fill[exs-mark-gpu-strong, rounded corners=1.5pt] (axis cs:0,2.79) rectangle (axis cs:1.9,3.21);
  \fill[exs-mark-gpu-soft, rounded corners=1.5pt] (axis cs:0,1.79) rectangle (axis cs:7.57,2.21);
  \fill[exs-mark-us, rounded corners=1.5pt] (axis cs:0,0.79) rectangle (axis cs:0.30,1.21);
  \draw[exs refline] (axis cs:1.9,0.5) -- (axis cs:1.9,2.74);
  \node[exs strong, anchor=west, xshift=4pt] at (axis cs:1.9,3) {1.9 ms};
  \node[exs strong, anchor=west, xshift=4pt] at (axis cs:7.57,2) {7.6 ms};
  \node[exs strong, anchor=west, xshift=4pt] at (axis cs:0.30,1) {0.30 ms};
\end{axis}
\end{tikzpicture}
  \caption{What drafting costs against the pass it serves, on one linear time scale. The target's pass is one decode step of gpt-oss-120b for one user, as measured by InferenceX \citep{semianalysis2026inferencex}, which verifying a short chain barely lengthens (\cref{sec:sd-mechanism}); the dashed line carries it down for comparison. Six tokens drafted on a B300 is our measurement of Qwen2.5-0.5B under vLLM (\cref{sec:sd-gpu}), a model less than a third the size of Polyglot-DC1-Draft. Six tokens drafted on the Exstill data-center design is pre-silicon, at 50~\textmu s per token.}
  \label{fig:sd-draft-time}
\end{figure}

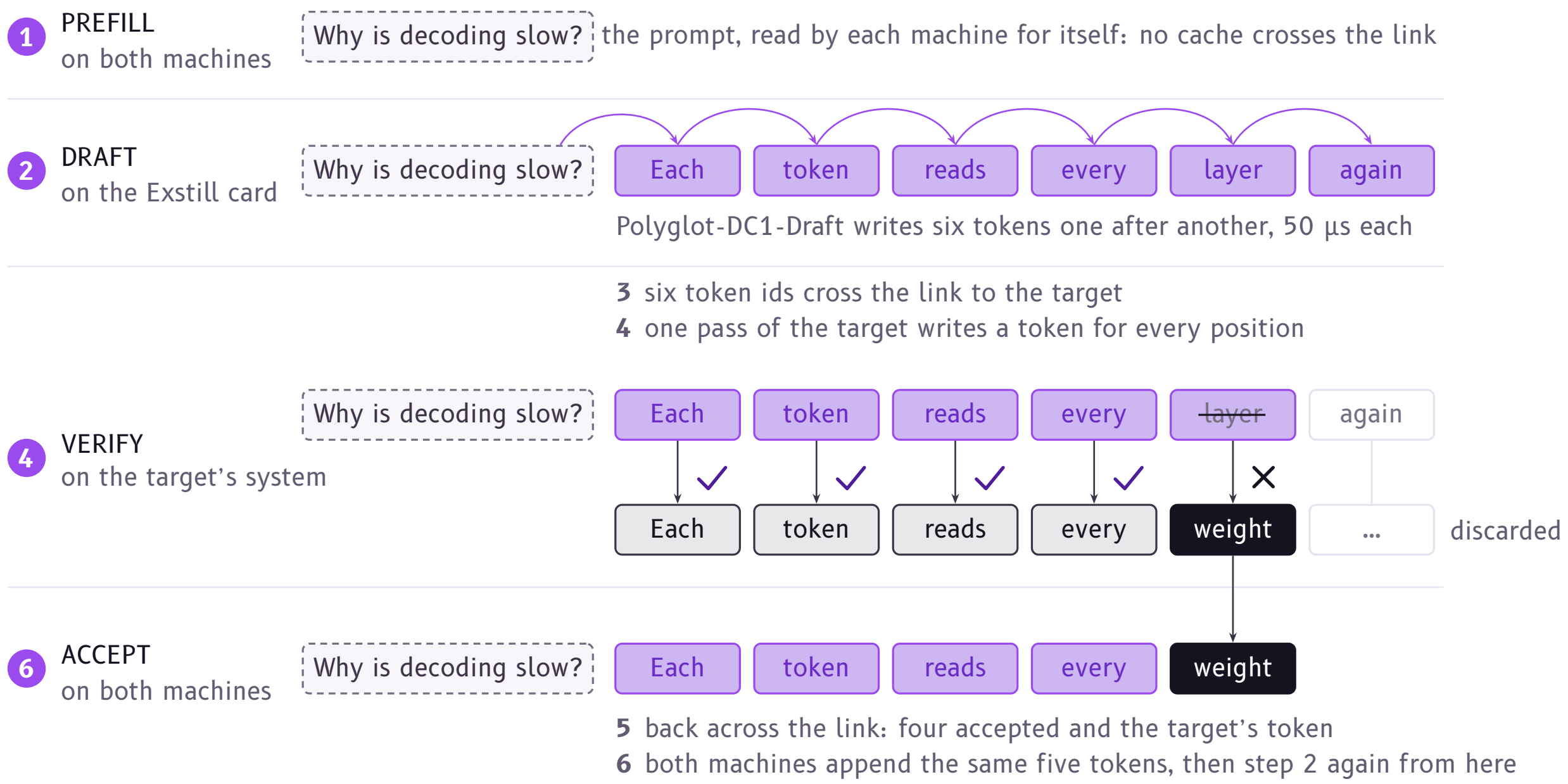
\begin{figure}[t]
  \centering
\begin{tikzpicture}[
  x=1cm, y=1cm,
  rowlab/.style={anchor=west, inner sep=0pt, font=\figfont\scriptsize, text=exs-ink},
  stepno/.style={circle, fill=exs-violet, text=white, font=\figfont\scriptsize\bfseries, inner sep=0pt, minimum size=0.4cm},
  note/.style={anchor=west, inner sep=0pt, font=\figfont\scriptsize, text=exs-muted},
  tok/.style={rounded corners=2pt, minimum width=1.3cm, minimum height=0.52cm, inner xsep=3pt, inner ysep=0pt, line width=0.6pt, font=\figfont\scriptsize,
    text height=1.5ex, text depth=0.5ex},  
  tprompt/.style={tok, draw=exs-faint, fill=exs-band, text=exs-ink-2, dash pattern=on 2pt off 1.5pt, minimum width=2.6cm},
  tdraft/.style={tok, draw=exs-violet, fill=exs-violet-lo, text=exs-violet-ink},
  ttarget/.style={tok, draw=exs-ink-2, fill=exs-ink!10, text=exs-ink},
  tkeep/.style={tok, draw=exs-ink, fill=exs-ink, text=white},
  tdrop/.style={tok, draw=exs-line, fill=white, text=exs-faint},
  feed/.style={draw=exs-violet, line width=0.5pt, -{Stealth[length=3pt]}},
  cmp/.style={draw=exs-ink-2, line width=0.5pt, -{Stealth[length=3pt]}},
  cmpfaint/.style={draw=exs-line, line width=0.5pt},
  rule/.style={draw=exs-line, line width=0.4pt},
]
\def\dx{1.45}
\def\xo{2.4}
\def\lx{-4.6}
\node[stepno] at (\lx+0.2,4.6) {1};
\node[rowlab] at (\lx+0.55,4.75) {PREFILL};
\node[note] at (\lx+0.55,4.38) {on both machines};
\node[tprompt] (p0) at (0,4.6) {Why is decoding slow?};
\node[note, align=left] at (1.6,4.6) {the prompt, read by each machine for itself: no cache crosses the link};
\draw[rule] (\lx,3.95) -- (10.4,3.95);
\node[stepno] at (\lx+0.2,3.2) {2};
\node[rowlab] at (\lx+0.55,3.35) {DRAFT};
\node[note] at (\lx+0.55,2.98) {on the Exstill card};
\node[tprompt] (p1) at (0,3.2) {Why is decoding slow?};
\foreach \i/\w in {1/Each,2/token,3/reads,4/every,5/layer,6/again} \node[tdraft] (d\i) at (\xo+\i*\dx-\dx,3.2) {\w};
\draw[feed] (p1.north east) ++(-0.35,0) to[out=60,in=120] (d1.north);
\foreach \i [evaluate=\i as \j using int(\i+1)] in {1,...,5} \draw[feed] (d\i.north) to[out=60,in=120] (d\j.north);
\node[note] at (\xo-0.65,2.6) {Polyglot-DC1-Draft writes six tokens one after another, 50 \textmu s each};
\draw[rule] (\lx,2.2) -- (10.4,2.2);
\node[note, align=left, anchor=north west] at (\xo-0.65,2.05) {\textbf{3}\enspace six token ids cross the link to the target\\\textbf{4}\enspace one pass of the target writes a token for every position};
\begin{scope}[yshift=-0.35cm]
\node[stepno] at (\lx+0.2,0.55) {4};
\node[rowlab] at (\lx+0.55,0.7) {VERIFY};
\node[note] at (\lx+0.55,0.33) {on the target's system};
\node[tprompt] (p2) at (0,1.0) {Why is decoding slow?};
\foreach \i/\w in {1/Each,2/token,3/reads,4/every} \node[tdraft] (v\i) at (\xo+\i*\dx-\dx,1.0) {\w};
\node[tdraft, text=exs-faint] (v5) at (\xo+4*\dx,1.0) {layer};
\draw[draw=exs-ink, line width=0.6pt] (v5.west) ++(0.3,0) -- ++(0.7,0);
\node[tdrop] (v6) at (\xo+5*\dx,1.0) {again};
\foreach \i/\w in {1/Each,2/token,3/reads,4/every} \node[ttarget] (t\i) at (\xo+\i*\dx-\dx,-0.2) {\w};
\node[tkeep] (t5) at (\xo+4*\dx,-0.2) {weight};
\node[tdrop] (t6) at (\xo+5*\dx,-0.2) {\ldots};
\foreach \i in {1,...,4} {
  \draw[cmp] (v\i.south) to[out=-90,in=90] (t\i.north);
  \draw[draw=exs-violet-deep, line width=1pt, line cap=round, line join=round] ($(v\i.south)+(0.22,-0.4)$) -- ++(0.1,-0.1) -- ++(0.18,0.22);
}
\draw[cmp] (v5.south) to[out=-90,in=90] (t5.north);
\draw[draw=exs-ink, line width=1pt, line cap=round] ($(v5.south)+(0.22,-0.28)$) -- ++(0.2,-0.2) ($(v5.south)+(0.22,-0.48)$) -- ++(0.2,0.2);
\draw[cmpfaint] (v6.south) to[out=-90,in=90] (t6.north);
\node[note] at ($(t6.east)+(0.15,0)$) {discarded};
\draw[rule] (\lx,-0.8) -- (10.4,-0.8);
\node[stepno] at (\lx+0.2,-1.65) {6};
\node[rowlab] at (\lx+0.55,-1.5) {ACCEPT};
\node[note] at (\lx+0.55,-1.87) {on both machines};
\node[tprompt] (p3) at (0,-1.65) {Why is decoding slow?};
\foreach \i/\w in {1/Each,2/token,3/reads,4/every} \node[tdraft] (a\i) at (\xo+\i*\dx-\dx,-1.65) {\w};
\node[tkeep] (a5) at (\xo+4*\dx,-1.65) {weight};
\draw[cmp] (t5.south) to[out=-90,in=90] (a5.north);
\node[note, align=left, anchor=north west] at (\xo-0.65,-2.15) {\textbf{5}\enspace back across the link: four accepted and the target's token\\\textbf{6}\enspace both machines append the same five tokens, then step 2 again from here};
\end{scope}
\path (current bounding box.south) -- ++(0,-0.15);  
\end{tikzpicture}
  \caption{Speculative decoding across two machines, in the draft-then-verify flow of \citet{xia2024survey}, the steps numbered in the order of one round. Each machine reads the prompt for itself and keeps its own cache, so no cache ever crosses the link between them. The Exstill card drafts six tokens one after another at about 50~\textmu s each, every token conditioned on the ones before it, and the six token identifiers cross to the target, gpt-oss-120b, Qwen3-Next, Kimi K3, DeepSeek-V4-Pro or Muse Spark Open, on whatever system serves it. One forward pass of the target writes its own token for every position. The drafts are kept while they match: four do, the fifth differs, so the target's own token takes its place and the sixth is discarded. What returns to the card is the count of drafts accepted and that one token; both machines append the same five tokens to their caches, and the next round drafts from there. Had all six matched, the target's seventh token would have come back as a bonus and the round would have yielded seven.}
  \label{fig:sd-silicon}
\end{figure}

The second recurring feature, drafting on the target's GPUs, is what the chip removes. In our simulation of the pre-silicon data-center design, each stream decodes at about 50~\textmu s per token, so a chain of a dozen tokens is drafted in well under a millisecond, a small fraction of the time the target takes to verify it. \Cref{fig:sd-draft-time} sets the two ways of drafting against the pass they serve, where six tokens take four times as long as the pass on a GPU and about a sixth of it on the chip. The chip can also prepare the next draft while the GPU verifies the current one, overlapping much of the drafting with verification. In the terms of \cref{eq:sd-speedup}, $d$ is about 4 for the GPU drafter of \cref{sec:sd-gpu} and about 0.16 for the chip, and the overlap takes it toward zero. \Cref{fig:sd-silicon} draws the arrangement, with the draft on its own card, the target on whatever system serves it, and only token identifiers crossing between them each round, the proposals in one direction and the verdict in the other, as the count accepted and the target's own token. \Cref{tab:sd-mechanisms} sets this arrangement beside the GPU drafters of \cref{sec:sd-gpu}, in the terms of NVIDIA's own comparison \citep{mitra2026codesign}. An independent analysis of draft disaggregation arrives at the same arrangement, the draft decoding on its own accelerator while the target stays on the GPU \citep{kuznetsov2026disaggregating}.

The draft needs none of the target's hidden states, so the traffic between the two machines is tens of bytes per round and drafting stays off the GPU entirely. To measure what sharing the target's GPU costs instead, we ran a 1.7B draft as the draft-model speculator of vLLM 0.31 alongside gpt-oss-120b on one B300. Decoding was 4.7-8.0x slower than the target alone for one user and 2.8-4.4x slower for eight, and slower still with the two models split across two B300s. The time went to the draft's serial passes inside the serving loop, each costing several times what the same pass costs when the draft is served alone, not to the target's own step.

In the same simulation, one card holds twelve streams at once, so a single card drafts for twelve people, for any of the five targets, by loading the weights for that target. Each stream's slot is sized for a 16K-token context, but the draft is not bound by it. Exstill's proprietary long-context method lets it keep drafting at contexts many times longer, for agents and long conversations, with its memory per stream fixed, so twelve long sessions fit the card as well as twelve short ones.

\begin{table}[tp]
  \centering
  \caption{Draft mechanisms compared, after NVIDIA's table \citep{mitra2026codesign}, with the Exstill arrangement added as the first column. The GPU mechanisms are as NVIDIA characterizes them. The first column is Polyglot-DC1-Draft on its own card, which needs only the target's token identifiers and drafts off the GPU. In NVIDIA's own measurements the standalone draft also reaches the highest accepted length of these mechanisms (\cref{sec:sd-gpu}).}
  \label{tab:sd-mechanisms}
  \footnotesize
  \setlength{\tabcolsep}{4pt}
  \renewcommand{\arraystretch}{1.25}
  \begin{tabularx}{\linewidth}{>{\raggedright\arraybackslash}p{2.05cm}>{\columncolor{exs-violet-lo!35}\raggedright\arraybackslash}X*{4}{>{\raggedright\arraybackslash}X}}
    \toprule
    \headrow & \cellcolor{exs-violet-lo!35}Polyglot-DC1-Draft on the Exstill card & EAGLE-3 & MTP & DFlash & DSpark \\
    \midrule
    Draft architecture & A standalone small model, one architecture, weights per target & A decoder layer with a linear projection & A decoder layer with a linear projection & Decoder layers with linear KV fusion & Decoder layers with KV fusion and a light sequential head \\
    Inputs & Token identifiers & The previous token's embedding and the target's hidden states & The target's final hidden state & The target's fused hidden states, as the draft's KV cache & The target's fused hidden states, as the draft's KV cache \\
    Generation & $K$ autoregressive passes, as a chain or a tree, pre-drafted while the target verifies & The module run $K$ times, serially & The module run $K$ times, serially & $K$ tokens in one parallel pass & $K$ tokens in one parallel pass, then a light serial correction \\
    Runs on & The Exstill card, beside the target's system & The target's GPUs & The target's GPUs & The target's GPUs & The target's GPUs \\
    Release method & A new set of weights, the same draft architecture and silicon for any target & A separate head, trained after the target & Part of the target's checkpoint & A separate head, trained after the target & A separate head, trained after the target \\
    Activation memory cost & Draft weights and a sparse KV cache & Weights and a small KV cache & Weights and a small KV cache & Weights and a small KV cache built from the target's & Weights and a small KV cache built from the target's \\
    Speculation cost per $K$ tokens & A full draft forward per token, on the card & One layer, serial & One layer, serial & About five layers, once & About five layers once, then the light head serially \\
    Best for & Exstill beside a GPU target. Not recommended on GPUs alone & Not recommended, lower acceptance than MTP & Larger models on GPUs & Smaller models at batch size 1 & Smaller models at batch size 1 \\
    \bottomrule
  \end{tabularx}
\end{table}

\subsection{Off the Shelf, or Co-Designed and Distilled for the Target}
\label{sec:sd-codesign}

Moving drafting off the GPU addresses where the work runs. The next question is how well the draft predicts its target, and the natural alternative to a purpose-built drafter is a small open model from the target's own family. For Qwen3-Next-80B-A3B-Thinking, Qwen3-1.7B is a particularly strong candidate. It shares the tokenizer, and it was distilled by Qwen from Qwen3-32B and Qwen3-235B-A22B, first off-policy and then on-policy, on top of 36 trillion tokens of pretraining \citep{yang2025qwen3}. No off-the-shelf model is better placed to draft for this target.

Polyglot-DC1-Draft for Qwen3-Next has the same parameter count as Qwen3-1.7B and shares nothing else with it. Its architecture is Polyglot-DC's, found by exploring the model and the chip as one design space, and its weights are distilled from Qwen3-Next's own outputs and aligned to its choices. \Cref{tab:sd-codesign} sets the two side by side, scored by us on the same cached answers in the same way with our evaluation harness, on SPEED-Bench \citep{abramovich2026speedbench}, a benchmark for speculative decoding that pools eighteen public datasets into eleven task categories, and on the nine datasets that DSpark reports on \citep{cheng2026dspark}: GSM8K \citep{cobbe2021gsm8k}, the 500-problem subset of MATH \citep{hendrycks2021math, lightman2023verify}, the 2025 American Invitational Mathematics Examination (AIME), HumanEval \citep{chen2021humaneval}, MBPP \citep{austin2021mbpp}, LiveCodeBench \citep{jain2024livecodebench}, MT-Bench \citep{zheng2023mtbench}, AlpacaEval \citep{dubois2024alpacaeval} and Arena-Hard v2 \citep{li2024arenahard}.

\begin{table}[t]
  \centering
  \caption{The same size, off the shelf or co-designed. Accepted length per verification pass for Qwen3-Next-80B-A3B-Thinking with a chain of $K = 6$ drafted tokens, the target's own token included, under verification of the target's cached answers. Both drafts have 1.7B parameters and the target's tokenizer. The last column is the relative gain of Polyglot-DC1-Draft over Qwen3-1.7B.}
  \label{tab:sd-codesign}
  \small
  \begin{tabular}{lrrr}
    \toprule
    \headrow Benchmark & Polyglot-DC1-Draft & Qwen3-1.7B & Gain \\
    \midrule
    SPEED-Bench, all & 3.68 & 3.16 & +16\% \\
    SPEED-Bench, coding & 4.86 & 3.81 & +28\% \\
    \midrule
    \multicolumn{4}{l}{\itshape Mathematical reasoning} \\
    GSM8K & 5.24 & 4.45 & +18\% \\
    MATH-500 & 5.12 & 4.18 & +22\% \\
    AIME 2025 & 4.57 & 3.43 & +33\% \\
    \addlinespace
    \multicolumn{4}{l}{\itshape Coding} \\
    HumanEval & 4.81 & 3.58 & +34\% \\
    MBPP & 4.58 & 3.09 & +48\% \\
    LiveCodeBench & 4.32 & 3.07 & +41\% \\
    \addlinespace
    \multicolumn{4}{l}{\itshape General chat} \\
    MT-Bench & 3.65 & 3.02 & +21\% \\
    AlpacaEval & 3.33 & 2.76 & +21\% \\
    Arena-Hard v2 & 3.62 & 3.17 & +14\% \\
    \midrule
    All nine datasets & 4.18 & 3.32 & +26\% \\
    \bottomrule
  \end{tabular}
\end{table}

Polyglot-DC1-Draft yields more tokens per pass on every row. Pooled over the nine datasets it averages 4.18 tokens per pass against 3.32, a gain of 26\%, and the gap is widest on code and competition mathematics, at 48\% on MBPP, 41\% on LiveCodeBench, 34\% on HumanEval and 33\% on AIME 2025. Both drafts are the same size and run at the same speed on the same chip, so every one of these differences carries straight through to the speedup.

\begin{figure}[t]
  \centering
\begin{tikzpicture}
\begin{axis}[exs, scale only axis, width=5.3cm, height=7.0cm, at={(0cm,0)},
  xmin=1, xmax=7.25, xtick={1,2,3,4,5,6,7}, xticklabels={1,2,3,4,5,6,7},
  xlabel={Accepted length, chain of six (tokens per pass)},
  ymin=-0.9, ymax=15.1, ytick={14.2,13.2,11.4,10.4,9.4,7.6,6.6,5.6,3.8,2.8,1.8,0},
  yticklabels={{SPEED-Bench, all},{SPEED-Bench, coding},{GSM8K},{MATH-500},{AIME 2025},{HumanEval},{MBPP},{LiveCodeBench},{MT-Bench},{AlpacaEval},{Arena-Hard v2},{\bfseries\color{exs-ink}All nine datasets}}, yticklabel style={font=\figfont\scriptsize, text=exs-ink-2},
  ymajorgrids=false,
]
  \draw[exs-mark-gpu-soft, line width=1.4pt] (axis cs:3.16,14.2) -- (axis cs:3.68,14.2);
  \draw[exs-mark-gpu-soft, line width=1.4pt] (axis cs:3.81,13.2) -- (axis cs:4.86,13.2);
  \draw[exs-mark-gpu-soft, line width=1.4pt] (axis cs:4.45,11.4) -- (axis cs:5.24,11.4);
  \draw[exs-mark-gpu-soft, line width=1.4pt] (axis cs:4.18,10.4) -- (axis cs:5.12,10.4);
  \draw[exs-mark-gpu-soft, line width=1.4pt] (axis cs:3.43,9.4) -- (axis cs:4.57,9.4);
  \draw[exs-mark-gpu-soft, line width=1.4pt] (axis cs:3.58,7.6) -- (axis cs:4.81,7.6);
  \draw[exs-mark-gpu-soft, line width=1.4pt] (axis cs:3.09,6.6) -- (axis cs:4.58,6.6);
  \draw[exs-mark-gpu-soft, line width=1.4pt] (axis cs:3.07,5.6) -- (axis cs:4.32,5.6);
  \draw[exs-mark-gpu-soft, line width=1.4pt] (axis cs:3.02,3.8) -- (axis cs:3.65,3.8);
  \draw[exs-mark-gpu-soft, line width=1.4pt] (axis cs:2.76,2.8) -- (axis cs:3.33,2.8);
  \draw[exs-mark-gpu-soft, line width=1.4pt] (axis cs:3.17,1.8) -- (axis cs:3.62,1.8);
  \draw[exs-mark-gpu-soft, line width=1.4pt] (axis cs:3.32,0) -- (axis cs:4.18,0);
  \addplot[only marks, mark=*, mark size=2.5pt, mark options={fill=exs-mark-gpu, draw=white, line width=0.8pt}] coordinates {(3.16,14.2) (3.81,13.2) (4.45,11.4) (4.18,10.4) (3.43,9.4) (3.58,7.6) (3.09,6.6) (3.07,5.6) (3.02,3.8) (2.76,2.8) (3.17,1.8) (3.32,0)};
  \addplot[only marks, mark=*, mark size=2.7pt, mark options={fill=exs-mark-us, draw=white, line width=0.8pt}] coordinates {(3.68,14.2) (4.86,13.2) (5.24,11.4) (5.12,10.4) (4.57,9.4) (4.81,7.6) (4.58,6.6) (4.32,5.6) (3.65,3.8) (3.33,2.8) (3.62,1.8) (4.18,0)};
  \node[exs strong, anchor=west, xshift=5pt] at (axis cs:4.18,0) {+26\%};
  \node[exs strong, anchor=west, xshift=5pt] at (axis cs:4.58,6.6) {+48\%};
\end{axis}
\begin{axis}[exs, scale only axis, width=5.3cm, height=7.0cm, at={(6.3cm,0)},
  xmin=1, xmax=7.25, xtick={1,2,3,4,5,6,7}, xticklabels={1x,2x,3x,4x,5x,6x,7x},
  xlabel={Speedup over the target alone, best chain},
  ymin=-0.9, ymax=15.1, ytick={14.2,13.2,11.4,10.4,9.4,7.6,6.6,5.6,3.8,2.8,1.8,0},
  yticklabels={},
  ymajorgrids=false,
]
  \draw[exs-mark-gpu-soft, line width=1.4pt] (axis cs:2.94,14.2) -- (axis cs:3.57,14.2);
  \draw[exs-mark-gpu-soft, line width=1.4pt] (axis cs:3.81,13.2) -- (axis cs:5.68,13.2);
  \draw[exs-mark-gpu-soft, line width=1.4pt] (axis cs:4.83,11.4) -- (axis cs:6.57,11.4);
  \draw[exs-mark-gpu-soft, line width=1.4pt] (axis cs:4.42,10.4) -- (axis cs:6.27,10.4);
  \draw[exs-mark-gpu-soft, line width=1.4pt] (axis cs:3.29,9.4) -- (axis cs:5.06,9.4);
  \draw[exs-mark-gpu-soft, line width=1.4pt] (axis cs:3.48,7.6) -- (axis cs:5.55,7.6);
  \draw[exs-mark-gpu-soft, line width=1.4pt] (axis cs:2.86,6.6) -- (axis cs:5.06,6.6);
  \draw[exs-mark-gpu-soft, line width=1.4pt] (axis cs:2.84,5.6) -- (axis cs:4.59,5.6);
  \draw[exs-mark-gpu-soft, line width=1.4pt] (axis cs:2.77,3.8) -- (axis cs:3.53,3.8);
  \draw[exs-mark-gpu-soft, line width=1.4pt] (axis cs:2.48,2.8) -- (axis cs:3.1,2.8);
  \draw[exs-mark-gpu-soft, line width=1.4pt] (axis cs:2.94,1.8) -- (axis cs:3.5,1.8);
  \draw[exs-mark-gpu-soft, line width=1.4pt] (axis cs:3.13,0) -- (axis cs:4.37,0);
  \addplot[only marks, mark=*, mark size=2.5pt, mark options={fill=exs-mark-gpu, draw=white, line width=0.8pt}] coordinates {(2.94,14.2) (3.81,13.2) (4.83,11.4) (4.42,10.4) (3.29,9.4) (3.48,7.6) (2.86,6.6) (2.84,5.6) (2.77,3.8) (2.48,2.8) (2.94,1.8) (3.13,0)};
  \addplot[only marks, mark=*, mark size=2.7pt, mark options={fill=exs-mark-us, draw=white, line width=0.8pt}] coordinates {(3.57,14.2) (5.68,13.2) (6.57,11.4) (6.27,10.4) (5.06,9.4) (5.55,7.6) (5.06,6.6) (4.59,5.6) (3.53,3.8) (3.1,2.8) (3.5,1.8) (4.37,0)};
  \node[exs strong, anchor=west, xshift=5pt] at (axis cs:4.37,0) {+40\%};
  \node[exs strong, anchor=west, xshift=5pt] at (axis cs:5.06,6.6) {+77\%};
\end{axis}
\fill[exs-mark-gpu] (0.12,7.42) circle (2.5pt);
\node[exs label, anchor=west] at (0.26,7.42) {Qwen3-1.7B, off the shelf};
\fill[exs-mark-us] (4.75,7.42) circle (2.7pt);
\node[exs label, anchor=west] at (4.89,7.42) {Polyglot-DC1-Draft, co-designed and distilled for the target};
\end{tikzpicture}
  \caption{Off the shelf or co-designed, benchmark by benchmark, for Qwen3-Next-80B-A3B-Thinking. Left, accepted length per verification pass at a chain of $K = 6$ (\cref{tab:sd-codesign}, measured), on a scale that runs from 1, the target's own token alone, to 7, all six proposals accepted. Right, end-to-end speedup over the target alone at each draft's best chain length (\cref{tab:sd-codesign-speedup}, simulated). The benchmarks are grouped as in the tables, and the printed gains are those of the pooled row and of MBPP, the largest.}
  \label{fig:sd-codesign}
\end{figure}
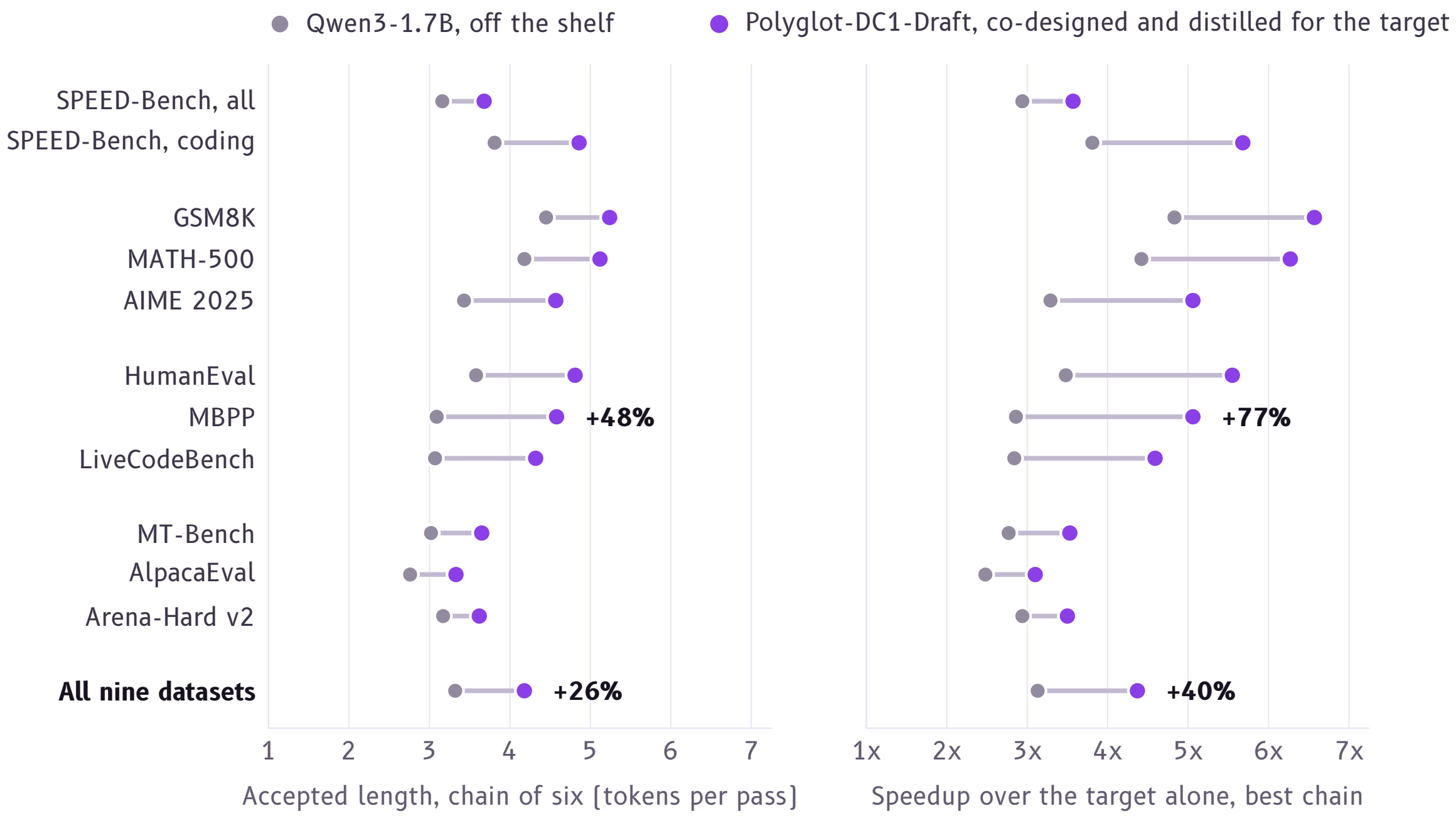

\Cref{tab:sd-codesign-speedup} shows how far. It prices each draft on the Exstill data-center design with our simulator at the chain length that gives it the highest end-to-end speedup, with the chip pre-drafting the next round while the GPU verifies the current one. Qwen3-Next alone decodes at 274 tokens per second for one user, measured by us on one B300 with vLLM and taken for a DGX B300 at tensor parallel 8. The gains in speed exceed the gains in accepted length because a draft whose proposals survive longer can profitably run a longer chain, and because the fixed cost of a round is spread over more accepted tokens. Pooled over the nine datasets, Polyglot-DC1-Draft decodes 4.37x faster than the target alone against 3.13x for Qwen3-1.7B, a 40\% advantage, and on code the advantage reaches 77\% on MBPP and 62\% on LiveCodeBench. \Cref{fig:sd-codesign} draws the two tables on a common scale, where the widening from accepted length to speed is visible benchmark by benchmark.

\begin{table}[t]
  \centering
  \caption{End-to-end speedup over Qwen3-Next-80B-A3B-Thinking alone for one user, simulated on the pre-silicon Exstill data-center design, with each draft at the chain length $K$ that gives it the highest speedup, the next round pre-drafted during verification, and the GPU's verification cost taken from our B300 measurement. A request is a 2,000-token prompt and a 512-token answer. Accepted lengths are measured as in \cref{tab:sd-codesign}; the speedups are simulated. The last column is the relative gain of Polyglot-DC1-Draft over Qwen3-1.7B.}
  \label{tab:sd-codesign-speedup}
  \small
  \begin{tabular}{lrrrrr}
    \toprule
    \headrow & \multicolumn{2}{c}{Polyglot-DC1-Draft} & \multicolumn{2}{c}{Qwen3-1.7B} & \\
    \cmidrule(lr){2-3}\cmidrule(lr){4-5}
    \headrow Benchmark & Speedup & $K$ & Speedup & $K$ & Gain \\
    \midrule
    SPEED-Bench, all & 3.57x & 13 & 2.94x & 11 & +21\% \\
    SPEED-Bench, coding & 5.68x & 16 & 3.81x & 14 & +49\% \\
    \midrule
    \multicolumn{6}{l}{\itshape Mathematical reasoning} \\
    GSM8K & 6.57x & 16 & 4.83x & 15 & +36\% \\
    MATH-500 & 6.27x & 16 & 4.42x & 15 & +42\% \\
    AIME 2025 & 5.06x & 15 & 3.29x & 14 & +54\% \\
    \addlinespace
    \multicolumn{6}{l}{\itshape Coding} \\
    HumanEval & 5.55x & 16 & 3.48x & 14 & +59\% \\
    MBPP & 5.06x & 15 & 2.86x & 11 & +77\% \\
    LiveCodeBench & 4.59x & 15 & 2.84x & 11 & +62\% \\
    \addlinespace
    \multicolumn{6}{l}{\itshape General chat} \\
    MT-Bench & 3.53x & 14 & 2.77x & 11 & +27\% \\
    AlpacaEval & 3.10x & 11 & 2.48x & 9 & +25\% \\
    Arena-Hard v2 & 3.50x & 14 & 2.94x & 11 & +19\% \\
    \midrule
    All nine datasets & 4.37x & 15 & 3.13x & 12 & +40\% \\
    \bottomrule
  \end{tabular}
\end{table}

With the same size, the same tokenizer and the same target, what separates the two is the model, its architecture and what it was trained to do. A draft is not judged by its own quality but by how often its proposals survive the target's verification, which is a property of the pair. Qwen3-1.7B was distilled to be a good model in its own right. A drafter is distilled to agree with one target, token by token, which the literature frames as aligning the draft's distribution to the target's under the verification rule \citep{zhou2024distillspec}. Exstill's own distillation and alignment methods are built for that objective and remain proprietary.

Co-design changes not only what the draft proposes but what silicon can be built around it. In our pre-silicon estimates, a chip built around Qwen3-1.7B as released would spend 3,041~\textmu J per drafted token, against 523~\textmu J for the Exstill data-center design running Polyglot-DC1-Draft, or 5.8x as much energy. These are the two Exstill points of \cref{fig:chip-floor}, costed within the same envelope, and \cref{sec:chip-capacity} sets out what else separates them.

\begin{insight}{A draft is judged by agreement, not by quality}
  Two drafts of the same size and tokenizer can differ by a quarter to a half in the tokens their target accepts. What matters is not how good the draft is as a language model but how closely its next-token choices track one specific target, and that is what target-specific distillation and alignment buy.
\end{insight}

\subsection{Accepted Length Across Benchmarks}
\label{sec:sd-acceptance}

\Cref{tab:sd-acceptance} reports the same measurement, made by us with the same harness, for all five targets. Every figure is an accepted length at $K = 6$. Each target's answers are generated once and cached, then the draft's proposals are verified against them token for token, and the tokens each pass yields are averaged over every pass, the target's own token included. SPEED-Bench is its first hundred qualitative conversations, with responses of up to 4,096 tokens; the nine datasets are sampled as DSpark's evaluation samples them, 3,030 prompts in all, with responses of up to 2,048 tokens. Each target answers at the reasoning effort in \cref{tab:sd-targets}.

\begin{table}[t]
  \centering
  \caption{Accepted length per verification pass with a chain of $K = 6$ drafted tokens, the target's own token included, for Polyglot-DC1-Draft against each of its five targets, under verification of the target's cached answers. The targets are those of \cref{tab:sd-targets}, with Qwen3-Next-80B-A3B-Thinking abbreviated. SPEED-Bench's rows are its qualitative split; the nine datasets below are those of \cref{tab:sd-codesign}.}
  \label{tab:sd-acceptance}
  \small
  \setlength{\tabcolsep}{4pt}
  \begin{tabular}{lrrrrr}
    \toprule
    \headrow Benchmark & gpt-oss-120b & Qwen3-Next & \begin{tabular}[t]{@{}r@{}}DeepSeek-\\V4-Pro-0813\end{tabular} & Kimi K3 & \begin{tabular}[t]{@{}r@{}}Muse-\\Spark-Open\end{tabular} \\
    \midrule
    SPEED-Bench, all & 4.04 & 3.68 & 3.22 & 2.92 & 2.97 \\
    SPEED-Bench, coding & 4.36 & 4.86 & 4.04 & 4.18 & 3.92 \\
    \midrule
    \multicolumn{6}{l}{\itshape Mathematical reasoning} \\
    GSM8K & 4.90 & 5.24 & 4.41 & 4.96 & 4.32 \\
    MATH-500 & 4.78 & 5.12 & 4.23 & 3.55 & 4.12 \\
    AIME 2025 & 4.31 & 4.57 & 3.76 & 2.92 & 3.75 \\
    \addlinespace
    \multicolumn{6}{l}{\itshape Coding} \\
    HumanEval & 4.15 & 4.81 & 3.84 & 4.91 & 4.07 \\
    MBPP & 4.49 & 4.58 & 4.01 & 4.49 & 4.16 \\
    LiveCodeBench & 3.84 & 4.32 & 3.25 & 3.26 & 3.26 \\
    \addlinespace
    \multicolumn{6}{l}{\itshape General chat} \\
    MT-Bench & 3.98 & 3.65 & 3.19 & 3.08 & 3.28 \\
    AlpacaEval & 3.88 & 3.33 & 3.06 & 3.07 & 3.03 \\
    Arena-Hard v2 & 3.68 & 3.62 & 3.08 & 2.90 & 2.89 \\
    \bottomrule
  \end{tabular}
\end{table}

Three patterns hold across the table. First, every draft is strongest on mathematics and code and weakest on open-ended conversation. On GSM8K the five drafts yield 4.3 to 5.2 tokens per pass, and on HumanEval and MBPP 3.8 to 4.9, whereas on AlpacaEval and Arena-Hard v2 they yield 2.9 to 3.9. Structured text, with its formulas, identifiers and boilerplate, is more predictable than free prose, and a chain of six proposals is correspondingly more often accepted in full. Second, the spread between benchmarks differs by target. Kimi K3 shows the widest range, from 4.9 tokens on HumanEval and GSM8K to 2.9 on AIME 2025 and Arena-Hard v2, while gpt-oss-120b is the most even, within 3.7 to 4.9 everywhere. Third, no single target leads throughout. Qwen3-Next is drafted best on most of the datasets, Kimi K3 edges it on GSM8K and HumanEval, and gpt-oss-120b leads on each of the three conversational benchmarks. These differences reflect the targets' own regularity as much as the drafts, since a target whose next token is more often determined by the preceding ones is easier to draft for, whatever the draft.

Two conventions bear on how the table is read. The SPEED-Bench and nine-dataset answers for Muse Spark Open and for Qwen3-Next were generated on different serving stacks, whose outputs differ after a few tokens, so each target's two blocks were scored against answers that are not identical in every token. And accepted length depends on the sampling settings at which a target answers, so figures from other settings are not directly comparable with these.

\Cref{tab:sd-speedup} turns those accepted lengths into decoding speed. As in \cref{tab:sd-codesign-speedup}, each draft runs at the chain length that gives it the highest end-to-end speedup for one user, with the next round pre-drafted while the GPU verifies the current one. That is the rule NVIDIA's guidance gives for the latency-bound regime, to lengthen the chain only while the gain in accepted length justifies the added draft cost \citep{mitra2026codesign}. Each target alone decodes at the speed its own measurements give for one user: gpt-oss-120b at 526 tokens per second, measured without speculation on a GB200 system by SemiAnalysis's InferenceX \citep{semianalysis2026inferencex}; Qwen3-Next at 274, measured by us on one B300 with vLLM; DeepSeek-V4-Pro-0813 at 100, measured on a DGX B300 with vLLM by InferenceX, and Kimi K3 at 101, measured the same way by Open Zeka \citep{openzeka2026kimik3}; and Muse Spark Open at 150, derived from the two DGX B300 measurements.

\begin{table}[t]
  \centering
  \caption{End-to-end speedup over each target alone for one user, simulated on the pre-silicon Exstill data-center design, with Polyglot-DC1-Draft at the chain length that gives the highest speedup and the next round pre-drafted during verification. For gpt-oss-120b and Qwen3-Next the GPU's verification cost is taken from our B300 measurement; for the three targets no single B300 can hold it is modeled from their published expert counts and sizes. A request is a 2,000-token prompt and a 512-token answer. Accepted lengths are measured as in \cref{tab:sd-acceptance}; the speedups are simulated. Target names are abbreviated as in \cref{tab:sd-acceptance}.}
  \label{tab:sd-speedup}
  \small
  \setlength{\tabcolsep}{4pt}
  \begin{tabular}{lrrrrr}
    \toprule
    \headrow Benchmark & gpt-oss-120b & Qwen3-Next & \begin{tabular}[t]{@{}r@{}}DeepSeek-\\V4-Pro-0813\end{tabular} & Kimi K3 & \begin{tabular}[t]{@{}r@{}}Muse-\\Spark-Open\end{tabular} \\
    \midrule
    SPEED-Bench, all & 3.77x & 3.57x & 2.92x & 2.33x & 2.50x \\
    SPEED-Bench, coding & 4.19x & 5.68x & 3.90x & 3.53x & 3.48x \\
    \midrule
    \multicolumn{6}{l}{\itshape Mathematical reasoning} \\
    GSM8K & 5.06x & 6.57x & 4.50x & 4.57x & 4.02x \\
    MATH-500 & 4.86x & 6.27x & 4.23x & 2.88x & 3.76x \\
    AIME 2025 & 4.11x & 5.06x & 3.57x & 2.33x & 3.29x \\
    \addlinespace
    \multicolumn{6}{l}{\itshape Coding} \\
    HumanEval & 3.88x & 5.55x & 3.63x & 4.51x & 3.68x \\
    MBPP & 4.40x & 5.06x & 3.88x & 3.89x & 3.78x \\
    LiveCodeBench & 3.49x & 4.59x & 2.95x & 2.61x & 2.76x \\
    \addlinespace
    \multicolumn{6}{l}{\itshape General chat} \\
    MT-Bench & 3.68x & 3.53x & 2.88x & 2.45x & 2.79x \\
    AlpacaEval & 3.57x & 3.10x & 2.74x & 2.45x & 2.55x \\
    Arena-Hard v2 & 3.33x & 3.50x & 2.77x & 2.31x & 2.43x \\
    \midrule
    All nine datasets & 3.82x & 4.37x & 3.13x & 2.65x & 2.92x \\
    \bottomrule
  \end{tabular}
\end{table}

The speedups follow the accepted lengths, with two further patterns. The first is that the best chain length tracks acceptance: where a draft is accepted deep into the chain, on mathematics and code, the optimum lies at 12 to 16 tokens, and where acceptance falls off early, on conversation, it lies at 6 to 11. The second is that the optimum is shorter for the trillion-parameter targets at any acceptance, because each of their verified tokens routes to more expert weights than a plain step loads, so beyond a point an extra proposal costs the GPU more than it is likely to earn. Kimi K3 is the clearest case, at a chain of 6 on every conversational benchmark and 15 on GSM8K and HumanEval, where it accepts most. For one user the gains run from 2.3x to 6.6x, with every target above 3.4x on coding and above 2.3x across SPEED-Bench as a whole.

These gains bring together the two quantities the section set out to separate. The accepted lengths belong to each draft and its target, the result of distilling one architecture for each of the five, and they carry through to decoding speed because the draft runs on its own silicon, where drafting adds little to the round it serves.


\section{Guardrails in the Critical Path}
\label{sec:guard}

A language model in production rarely runs alone. Beside it runs a second model whose only job is to decide what the first may read and what it may write. The reason is that safety training is a tendency, not a guarantee. Simple adaptive attacks, a random search over an appended suffix or a reply begun on the model's behalf, reached 100\% attack success against a dozen leading safety-aligned models, GPT-4o and every Claude model among them \citep{andriushchenko2025jailbreaking}, and published defenses have not closed the gap. An adaptive attacker bypassed twelve recent ones, most of them at attack success above 90\%, though most had originally reported attack success near zero \citep{nasr2025attacker}. Agents raise the stakes. An agent that reads web pages, email or documents on a user's behalf can be steered by instructions planted in them, and agents are open to such prompt injection whenever they handle untrusted data \citep{debenedetti2026defeating}. The danger is greatest when the agent can call tools or touch sensitive information \citep{beurerkellner2025patterns}, because a manipulated answer becomes a manipulated action. OWASP's 2026 list ranks prompt injection first among the risks of language-model applications, noting that the model itself draws no built-in line between instructions and data \citep{owasp2026top10}. Even the newest frontier models are exposed. In Gray Swan's indirect prompt-injection arena, OpenAI's GPT-6 Astra, with its safeguards enabled, still fell to an estimated 8.5\% of attacks across fifteen attempts per scenario, against 27.0\% for its predecessor, GPT-5.6 Sol, according to its system card \citep{openai2026astra}.

What the main model's training cannot guarantee is left to the guard, a separate model that reads each prompt before the main model does and each reply before the user does. Its job has three parts. It must catch harm, including the attacks the main model's own training misses. It must do so in every language the product's users write, whereas multilingual moderation has long concentrated on a small set of languages, English and Chinese chief among them \citep{kumar2025polyguard}. And it must leave harmless requests alone. Over-refusal is now measured as carefully as harm, with OR-Bench's 80,000 prompts that only sound harmful, tested across 32 models \citep{cui2025orbench}, and FalseReject's 16,000, on which 29 current models still over-refuse \citep{zhang2025falsereject}. A guard that flags too many harmless requests quickly becomes a liability.

The guard also faces a constraint unrelated to what it detects. It sits in the inference critical path, where every check adds compute and delays the response it screens, and much of how production screening is built follows from that cost.

\subsection{How the Frontier Screens Traffic Today}
\label{sec:gd-production}

Anthropic, Google DeepMind and OpenAI have each described how they screen production traffic, and the open guards follow the same pattern. Anthropic guards Claude with Constitutional Classifiers, classifiers trained on synthetic data generated from a written constitution of what is allowed and what is not, one screening the input and one the output \citep{sharma2025constitutional}. The output classifier scores the reply token by token as it streams and halts generation the moment the score crosses a threshold; preserving time to first token and streaming was a design requirement. In Anthropic's evaluation, jailbreak success on 10,000 synthetic jailbreaks fell from 86\% to 4.4\%, at the price of a 0.38\% absolute rise in refusals on production Claude.ai traffic and a 23.7\% inference overhead. Anthropic deployed the classifiers with Claude Opus 4 under its AI Safety Level 3 protections in May 2025 \citep{anthropic2025asl3}. The next generation is built around cost. A linear probe on Claude's own activations screens all traffic and escalates what it flags to an ensemble of a probe and a classifier that judges each reply together with its prompt. Anthropic estimates the overhead at roughly 1\% of compute on Claude Opus 4.0 traffic, and over a month on Claude Sonnet 4.5 traffic the system refused 0.05\% of harmless queries \citep{cunningham2026constitutional, anthropic2026classifiers}.

Google DeepMind has deployed probes on Gemini's activations to catch cyber-offensive misuse in user-facing Gemini. In its cascade, the probe defers to Gemini 2.5 Flash on about 8\% of traffic and beats Flash alone at roughly a fiftieth of the cost; the hardest part was the move from short inputs to long ones \citep{kramar2026probes}. OpenAI describes two tiers of real-time oversight against biological misuse on GPT-5's thinking models, a fast topical classifier followed by a reasoning model that decides whether a response can be shown, running on all of that traffic and scanning user messages, tool calls and final output \citep{openai2025gpt5card}. OpenAI's open-weight gpt-oss-safeguard reasons from a policy given in its prompt, and its report is direct about the cost. Such reasoning is time- and compute-intensive, so OpenAI's own Safety Reasoner relies on smaller, faster classifiers to decide what to assess and is sometimes run asynchronously to keep latency low \citep{openai2025safeguard}. With GPT-6 Astra the screened surface grew again. Alongside classifiers in production, OpenAI added misalignment monitoring to all tool-using inference in Astra's external deployment and notes that it comes with significant compute cost \citep{openai2026astra}.

The open guards are built the same way. Meta's Llama Guard 4, a 12B model, classifies prompts and responses \citep{meta2025llamaguard4}, and its Prompt Guard 2 catches injections and jailbreaks with 86M and 22M models that take 92.4~ms and 19.3~ms per check on an A100 at 512 tokens, according to its model card \citep{meta2025promptguard}. Qwen3Guard, from 0.6B to 8B parameters, labels content Safe, Controversial or Unsafe, and its stream variant classifies every token of a reply as it is generated \citep{zhao2025qwen3guard}. NVIDIA's NeMo Guardrails checks a streamed reply in chunks, 200 tokens by default, and its documentation states the trade-off plainly, that larger chunks mean fewer checks but a longer wait for the first output \citep{nvidia2026nemostreaming}.

Despite their differences, these systems converge on one shape. Something cheap screens everything; something stronger judges what it flags; and both the prompt and the streaming reply are screened, increasingly along with every tool call an agent makes. Nearly every design choice serves the latency and compute budget, and each buys its saving with a loss elsewhere. A cascade's strong classifier sees only the fraction of traffic the cheap screen flags. A larger chunk either delays the reply or lets more of it reach the user before it is checked. An asynchronous check rules after the user may already have read the text. Probes, the cheapest screen of all, are cheap because they reuse activations the model has already computed, and that brings limits of its own. A probe needs access to the served model's internals and must be trained for each model it guards, so an application built on another provider's API has nothing to read, and Google DeepMind reports that its probe techniques do not significantly reduce the success of adaptive attacks. We asked instead what screening would look like if it were fast enough that its cost no longer dictated when and how often it could run.

\subsection{Why the Guard Belongs on Its Own Silicon}
\label{sec:gd-chip}

A guard sits in each user's critical path twice, before the model sees the prompt and on every chunk of the reply before the user sees it. Its verdict is short, typically 10 to 15 tokens, so each check is a small burst of decoding for a single conversation. That is the regime in which a GPU is weakest. To write each token it pays the fixed charge of the layers and reads the whole model in from memory, as \cref{sec:cost} showed, and a short verdict leaves no time to spread either across a batch. \Cref{tab:gd-speed} puts numbers on the gap. The B300 figures are measured, on one GPU serving an open model of each guard's size with vLLM in FP8 on a screening-shaped load, and the Exstill figures are pre-silicon, for the design built for a model of that size: the data-center design of \cref{sec:chip-dc} for the 1.7B and the DC2 design of \cref{sec:chip-dc2} for the 4B.

\begin{table}[t]
  \centering
  \caption{Decoding a model of each guard's size on the pre-silicon Exstill design built for it and on one NVIDIA B300. The B300 serves Qwen3-1.7B and Qwen3-4B-Thinking-2507 in FP8 with vLLM, measured on a screening-shaped load of 4K tokens in and 256 out; the data-center design serves twelve users for the 1.7B's size and the DC2 design four users for the 4B's, each with a 16K-token context. A verdict is 14 tokens. Decoding only.}
  \label{tab:gd-speed}
  \small
  \begin{tabular}{lrrr}
    \toprule
    \headrow & Exstill design & NVIDIA B300 & Ratio \\
    \midrule
    \multicolumn{4}{l}{\itshape A model of Polyglot-DC1-Guard's size (1.7B)} \\
    One user, tokens per second & 19,921 & 522.9 & 38x \\
    Twelve users at once, tokens per second each & 19,921 & 342.5 & 58x \\
    Twelve users at once, tokens per second in total & 239,056 & 3,654.9 & 65x \\
    Tokens per joule, twelve users & 1,912 & 6.2 & \\
    Power, twelve users & 125~W & 585~W & \\
    One 14-token verdict, one user & 0.70~ms & 26.8~ms & 38x \\
    \addlinespace
    \multicolumn{4}{l}{\itshape A model of Polyglot-DC2-Guard's size (4B)} \\
    One user, tokens per second & 13,998 & 359.1 & 39x \\
    Four users at once, tokens per second each & 13,998 & 320.7 & 44x \\
    Four users at once, tokens per second in total & 55,992 & 1,162.8 & 48x \\
    Tokens per joule, four users & 531 & 2.3 & \\
    Power, four users & 106~W & 506~W & \\
    One 14-token verdict, one user & 1.00~ms & 39.0~ms & 39x \\
    \bottomrule
  \end{tabular}
\end{table}

For one user, the B300 decodes a model of the 1.7B guard's size at 522.9 tokens per second, so a 14-token verdict takes 26.8~ms and four checks on one long answer take 107~ms. The data-center design decodes a model of the same size at 19,921 tokens per second for each of twelve conversations at once, 0.70~ms per verdict and 2.8~ms for the four. A model of the 4B guard's size is slower on both, 39.0~ms per verdict on the B300 against 1.00~ms on the DC2 design, a ratio of 39x. Screening everything is cheap in energy as well as in time. At twelve conversations at once the data-center design draws 125~W and delivers 1,912 tokens per joule, against 6.2 for the B300 at 585~W, roughly 300x less energy per verdict. The DC2 design, at four conversations at once, draws 106~W and delivers 531 tokens per joule against 2.3, roughly 230x less.

At about a millisecond per check, a guard can screen every prompt and every chunk of every reply rather than the fraction a cheaper screen escalates, the chunks can be small, and no check has to rule after the user has read the text. A deployment may still send hard cases to a larger model, as Anthropic and Google DeepMind do, but its first line no longer has to be a compromise.

\begin{insight}{Screening is rationed by its cost, and the cost is where it runs}
  Every production safeguard cascades, chunks or defers its checks because each check is a short single-user decode, the workload a GPU serves worst. On silicon built for that workload a verdict costs about a millisecond, and the rationing is no longer necessary.
\end{insight}

\subsection{Polyglot-DC1-Guard and Polyglot-DC2-Guard}
\label{sec:gd-model}

The guard is the safety role of the Polyglot-DC architecture of \cref{sec:polyglot}, in two sizes, Polyglot-DC1-Guard with 1.7B parameters and Polyglot-DC2-Guard with 4B. Each reads a conversation, classifies its last turn, the user's message or the assistant's reply, and answers in three short lines:

\nopagebreak
\begin{samepage}  
\begin{verbatim}
Safety:     Safe | Controversial | Unsafe
Categories: None | one or more categories
Refusal:    Yes | No            (replies only)
\end{verbatim}
\end{samepage}

The three-level safety label follows Qwen3Guard's \citep{zhao2025qwen3guard}. Controversial marks harm that depends on context, audience or jurisdiction, for a deployment to send to review or to resolve under its own policy, and it keeps borderline content from being forced into a binary verdict in either direction. The thirteen categories run from violence, weapons and illegal acts through personal data, self-harm, hate, political sensitivity, copyright and specialized advice to jailbreak attempts, the last for prompts only. The refusal line records whether the assistant declined, which separates a safe refusal from a safe answer and gives a deployment the signal for auditing an assistant that refuses too much or too little. The guard reads English, French, Spanish, Portuguese, German and Chinese with no translation step.

In deployment a single message, a prompt or a complete reply, is checked once, and a streamed reply is checked as it is written. A long answer from a reasoning model, its thinking included, streams to the user and is re-checked at intervals of a few hundred tokens; an answer allowed at every check reaches the user whole, one sent for review at its first check can be held, and one judged unsafe at a later check stops there, with nothing after it shown.

\subsection{Detection Quality and False Alarms}
\label{sec:gd-quality}

\begin{table}[t]
  \centering
  \caption{F1 on the Unsafe class, in best mode, on the English benchmarks, measured by us on the same held-out items for every model. Prompt classification judges the user's message; response classification judges the assistant's reply given its prompt. Bold marks the best value in each row.}
  \label{tab:gd-english}
  \footnotesize
  \setlength{\tabcolsep}{4pt}
  \begin{tabular}{lrrrrrr}
    \toprule
    \headrow Benchmark & {\scriptsize\begin{tabular}[t]{@{}r@{}}Polyglot-\\DC1-Guard\\(1.7B)\end{tabular}} & {\scriptsize\begin{tabular}[t]{@{}r@{}}Polyglot-\\DC2-Guard\\(4B)\end{tabular}} & {\scriptsize\begin{tabular}[t]{@{}r@{}}Qwen3Guard-\\Gen-0.6B\end{tabular}} & {\scriptsize\begin{tabular}[t]{@{}r@{}}Qwen3Guard-\\Gen-4B\end{tabular}} & {\scriptsize\begin{tabular}[t]{@{}r@{}}Llama\\Guard 4\\(12B)\end{tabular}} & {\scriptsize\begin{tabular}[t]{@{}r@{}}gpt-oss-\\safeguard-\\20b\end{tabular}} \\
    \midrule
    \multicolumn{7}{l}{\itshape Prompt classification} \\
    ToxicChat & 79.5 & 82.0 & 77.9 & \textbf{83.1} & 50.9 & 74.0 \\
    OpenAI Moderation & 79.9 & 80.5 & 77.5 & \textbf{80.8} & 73.9 & 78.5 \\
    Aegis 1.0 & 85.3 & 87.6 & 90.9 & \textbf{91.1} & 67.8 & 84.4 \\
    Aegis 2.0 & 86.3 & \textbf{86.5} & 85.0 & 85.8 & 70.8 & 82.4 \\
    SimpleSafetyTests & 99.0 & 99.0 & 99.0 & \textbf{99.5} & 98.5 & \textbf{99.5} \\
    HarmBench & 98.1 & 98.3 & 98.3 & \textbf{100.0} & 97.6 & \textbf{100.0} \\
    WildGuardTest & 87.9 & 88.2 & 88.5 & \textbf{89.2} & 71.1 & 85.4 \\
    \addlinespace
    \multicolumn{7}{l}{\itshape Response classification} \\
    HarmBench & 86.5 & \textbf{87.7} & 84.7 & 86.7 & 82.4 & 86.4 \\
    SafeRLHF & 65.1 & 67.7 & 66.7 & \textbf{69.6} & 43.7 & 64.9 \\
    BeaverTails & 85.2 & 85.9 & 86.3 & \textbf{86.5} & 69.8 & 83.2 \\
    XSTest & 94.7 & \textbf{94.9} & 90.6 & 93.0 & 89.0 & 90.6 \\
    Aegis 2.0 & 82.3 & 82.7 & 84.4 & \textbf{86.5} & 64.7 & 79.3 \\
    WildGuardTest & 79.0 & \textbf{79.9} & 76.9 & \textbf{79.9} & 66.4 & 75.4 \\
    Qwen3GuardTest, thinking & 85.4 & \textbf{86.1} & \textbf{86.1} & 84.0 & 60.6 & 85.7 \\
    \bottomrule
  \end{tabular}
\end{table}

Speed alone does not make a production guard, so we evaluated both sizes against four public guard models on 72,229 held-out items from 21 public benchmarks in six languages, 32,519 of them English and the rest French, Spanish, Portuguese, German and Chinese. The English sets are the fourteen on which Qwen3Guard reports, named with \cref{tab:gd-english}. The other five languages are covered by the test split of NVIDIA's Nemotron Safety Guard dataset, the Aegis 2.0 test adapted into twelve languages with the CultureGuard pipeline \citep{joshi2025cultureguard, nvidia2026nemotronsafetyguard}, by PolyGuardPrompts, which is WildGuardTest translated into seventeen languages \citep{kumar2025polyguard}, by RTP-LX, a set of toxic prompts translated and adapted by native speakers in 28 languages \citep{dewynter2025rtplx}, and, for Chinese, by MultiJail \citep{deng2024multijail}. Portuguese, which the Nemotron set omits, rests on PolyGuardPrompts and RTP-LX alone. Every model runs on the same items, each in its own prompt format. Scores are F1, the harmonic mean of precision and recall, on the Unsafe class in best mode, meaning that for each benchmark the better of two readings is taken, one counting Controversial as unsafe and one counting it as safe, as Qwen3Guard reports its own results. The baselines are Qwen3Guard-Gen-0.6B and Qwen3Guard-Gen-4B \citep{zhao2025qwen3guard}, Meta's Llama Guard 4 \citep{meta2025llamaguard4}, a 12B model which answers safe or unsafe only and has no refusal output, and OpenAI's gpt-oss-safeguard-20b \citep{openai2025safeguard, openai2025gptoss}, a 21B-parameter mixture-of-experts model with 3.6B active, given the same policy in its prompt and run at medium reasoning effort.

\begin{table}[t]
  \centering
  \caption{Detection quality and false alarms of the Polyglot-DC guards against four public guards, measured by us on the same held-out items. F1 is on the Unsafe class in best mode, averaged over the benchmarks of each group; the English averages are over the prompt and response benchmarks that Qwen3Guard reports. The public guards are Qwen3Guard-Gen \citep{zhao2025qwen3guard}, Llama Guard 4 \citep{meta2025llamaguard4} and gpt-oss-safeguard-20b \citep{openai2025safeguard}. The last two rows are the share of harmless prompts flagged as unsafe, lower being better.}
  \label{tab:gd-headline}
  \footnotesize
  \setlength{\tabcolsep}{4pt}
  \begin{tabular}{lrrrrrr}
    \toprule
    \headrow & {\scriptsize\begin{tabular}[t]{@{}r@{}}Polyglot-\\DC1-Guard\\(1.7B)\end{tabular}} & {\scriptsize\begin{tabular}[t]{@{}r@{}}Polyglot-\\DC2-Guard\\(4B)\end{tabular}} & {\scriptsize\begin{tabular}[t]{@{}r@{}}Qwen3Guard-\\Gen-0.6B\end{tabular}} & {\scriptsize\begin{tabular}[t]{@{}r@{}}Qwen3Guard-\\Gen-4B\end{tabular}} & {\scriptsize\begin{tabular}[t]{@{}r@{}}Llama\\Guard 4\\(12B)\end{tabular}} & {\scriptsize\begin{tabular}[t]{@{}r@{}}gpt-oss-\\safeguard-\\20b\end{tabular}} \\
    \midrule
    \multicolumn{7}{l}{\itshape F1 on unsafe content, higher is better} \\
    English prompts & 88.0 & 88.9 & 88.1 & 89.9 & 75.8 & 86.3 \\
    English responses & 82.6 & 83.5 & 82.2 & 83.7 & 68.1 & 80.8 \\
    Other five languages, prompts & 86.5 & 87.7 & 85.5 & 88.3 & 56.3 & 83.4 \\
    Other five languages, responses & 80.4 & 81.5 & 79.6 & 82.5 & 60.1 & 79.0 \\
    \addlinespace
    \multicolumn{7}{l}{\itshape Harmless prompts flagged, lower is better} \\
    PHTest & 20.0\% & 9.8\% & 29.2\% & 23.8\% & 6.8\% & 17.8\% \\
    OR-Bench hard-1k & 14.9\% & 7.2\% & 72.7\% & 82.2\% & 10.6\% & 90.5\% \\
    \bottomrule
  \end{tabular}
\end{table}

\Cref{tab:gd-headline} gives the four aggregates and the two false-alarm measures. Polyglot-DC2-Guard comes within one F1 point of Qwen3Guard-Gen-4B, the strongest model in the comparison, on all four aggregates, trailing it by 1.0 on English prompts, 0.2 on English responses, 0.6 on prompts in the other five languages and 1.0 on their responses, and it is ahead of Qwen3Guard-Gen-0.6B, Llama Guard 4 and gpt-oss-safeguard-20b on every one of them. Polyglot-DC1-Guard, with under half of Qwen3Guard-Gen-4B's parameters, is level with Qwen3Guard-Gen-0.6B in English, 88.0 and 82.6 against 88.1 and 82.2, and ahead of it in the other five languages. Llama Guard 4, at 12B parameters, trails the 4B guard by 13 to 15 F1 points in English and by more outside it.

The more striking difference is how often the models flag harmless content. PHTest is a set of harmless prompts that sound harmful \citep{an2024phtest}. Polyglot-DC2-Guard flags 9.8\% of them, against 17.8\% to 29.2\% for gpt-oss-safeguard and the two Qwen3Guard models, under half the rate of Qwen3Guard-Gen-4B, the guard it matches within a point on harm. Only Llama Guard 4 flags fewer, at 6.8\%, and its English F1 is 13 to 15 points lower. On OR-Bench hard-1k \citep{cui2025orbench}, the thousand hardest of OR-Bench's seemingly toxic prompts, the gap widens further, with the Polyglot-DC guards flagging 7.2\% and 14.9\% where Qwen3Guard flags 72.7\% to 82.2\% and gpt-oss-safeguard 90.5\%.

\Cref{tab:gd-english} breaks the English aggregates into their fourteen benchmarks: for prompts, ToxicChat \citep{lin2023toxicchat}, the OpenAI Moderation set \citep{markov2023moderation}, Aegis 1.0 and 2.0 \citep{ghosh2024aegis, ghosh2025aegis2}, SimpleSafetyTests \citep{vidgen2023simplesafetytests}, HarmBench \citep{mazeika2024harmbench} and WildGuardTest \citep{han2024wildguard}; for responses, HarmBench, SafeRLHF \citep{ji2025pkusaferlhf}, BeaverTails \citep{ji2023beavertails}, XSTest \citep{rottger2024xstest}, Aegis 2.0, WildGuardTest and the thinking split of Qwen3GuardTest \citep{zhao2025qwen3guard}. The pattern is consistent rather than dominated by any one set. Qwen3Guard-Gen-4B leads on six of the seven prompt sets, mostly by about a point. Polyglot-DC2-Guard leads on Aegis 2.0 prompts and on the HarmBench and XSTest responses, XSTest being the set on which exaggerated safety is the failure under test. It ties the leader on Qwen3GuardTest responses, and the two 4B guards are tied on WildGuardTest responses. The places the Polyglot-DC guards trail clearly are Aegis 1.0 prompts and Aegis 2.0 responses, by about four points. Llama Guard 4 trails everywhere, most of all on the conversational sets, and gpt-oss-safeguard-20b, despite reasoning over an explicit policy at every check, sits between it and the smaller Qwen3Guard.

\begin{table}[t]
  \centering
  \caption{F1 on the Unsafe class in the five languages other than English, in best mode, averaged over the benchmarks in each language, measured by us on the same held-out items for every model. The benchmarks are the Nemotron Safety Guard test split \citep{joshi2025cultureguard}, PolyGuardPrompts \citep{kumar2025polyguard} and RTP-LX \citep{dewynter2025rtplx} in every language, and MultiJail \citep{deng2024multijail} in Chinese; Portuguese has PolyGuardPrompts and RTP-LX only. Bold marks the best value in each row.}
  \label{tab:gd-languages}
  \footnotesize
  \setlength{\tabcolsep}{4pt}
  \begin{tabular}{llrrrrrr}
    \toprule
    \headrow Language & Task & {\scriptsize\begin{tabular}[t]{@{}r@{}}Polyglot-\\DC1-Guard\\(1.7B)\end{tabular}} & {\scriptsize\begin{tabular}[t]{@{}r@{}}Polyglot-\\DC2-Guard\\(4B)\end{tabular}} & {\scriptsize\begin{tabular}[t]{@{}r@{}}Qwen3Guard-\\Gen-0.6B\end{tabular}} & {\scriptsize\begin{tabular}[t]{@{}r@{}}Qwen3Guard-\\Gen-4B\end{tabular}} & {\scriptsize\begin{tabular}[t]{@{}r@{}}Llama\\Guard 4\\(12B)\end{tabular}} & {\scriptsize\begin{tabular}[t]{@{}r@{}}gpt-oss-\\safeguard-\\20b\end{tabular}} \\
    \midrule
    French & prompts & 86.4 & \textbf{87.9} & 85.2 & 87.7 & 56.6 & 82.4 \\
           & responses & 81.3 & 82.4 & 80.5 & \textbf{83.6} & 60.1 & 80.3 \\
    \addlinespace
    Spanish & prompts & 85.8 & \textbf{87.1} & 85.1 & 86.8 & 56.2 & 83.2 \\
            & responses & 81.6 & 82.4 & 81.8 & \textbf{83.8} & 60.7 & 79.6 \\
    \addlinespace
    Portuguese & prompts & 86.7 & 87.5 & 84.7 & \textbf{89.3} & 50.2 & 82.0 \\
               & responses & 77.0 & 78.2 & 75.3 & \textbf{78.6} & 59.9 & 77.7 \\
    \addlinespace
    German & prompts & 85.2 & 86.0 & 84.3 & \textbf{87.6} & 57.5 & 82.5 \\
           & responses & 80.4 & 82.2 & 79.9 & \textbf{83.3} & 61.4 & 79.0 \\
    \addlinespace
    Chinese & prompts & 88.6 & \textbf{90.0} & 87.9 & 89.9 & 61.0 & 86.7 \\
            & responses & 81.6 & 82.5 & 80.5 & \textbf{83.1} & 58.7 & 78.6 \\
    \bottomrule
  \end{tabular}
\end{table}

\Cref{tab:gd-languages} shows the same comparison in the other five languages. The Polyglot-DC guards hold their English standing in each of them. Polyglot-DC2-Guard leads on French, Spanish and Chinese prompts and is within a point or two of Qwen3Guard-Gen-4B everywhere else, and the ordering of the public guards is unchanged, with Llama Guard 4 falling furthest, by roughly 30 points on prompts, once the language is not English. Portuguese responses, which come from PolyGuardPrompts alone, are the weakest set for five of the six models, and the only language in which Polyglot-DC1-Guard drops below 80 F1. The refusal flag, which only the Polyglot-DC guards, Qwen3Guard and gpt-oss-safeguard produce, is detected at 89 to 95 F1 on WildGuardTest, XSTest and PolyGuardPrompts \citep{kumar2025polyguard} by both sizes, level with Qwen3Guard, and within one to three points of gpt-oss-safeguard, which leads on two of the three sets and trails on XSTest.

\subsection{Choosing the Size}
\label{sec:gd-size}

Of the two sizes, Polyglot-DC2-Guard is the more accurate. It gains about a point of F1 on every aggregate over Polyglot-DC1-Guard, and it flags under half as many of PHTest's harmless prompts, 9.8\% against 20.0\%, which for a deployment is the difference between a guard that is occasionally over-cautious and one that gets in the way. On a GPU, choosing the more accurate size also raises the latency of every check. One B300 takes 39.0~ms for a 14-token verdict from a model of the 4B's size, against 26.8~ms from one of the 1.7B's, an extra 12.2~ms on each of the several checks a long answer receives. On the Exstill designs the same verdicts decode in 1.00~ms and 0.70~ms, so the more accurate guard costs 0.30~ms more per check, and a deployment can choose its guard for accuracy rather than for speed.

\begin{insight}{Choose the guard for accuracy, not for latency}
  On a GPU, the more accurate of two guards adds over ten milliseconds to every check, so the smaller one is chosen. On silicon where either verdict takes about a millisecond or less, the trade disappears and the more accurate guard is the obvious choice.
\end{insight}

The frontier labs have shown what a production guard must do, which is to screen the prompt and the streaming reply, catch the attacks the model's own training misses, and leave harmless requests alone. They have also shown that every safeguard is shaped by its cost. On Exstill silicon a deployment can screen every prompt, every chunk of every reply and every step of an agent's work without the screening becoming the inference bottleneck.


\section{Agents and the Cost of a Turn}
\label{sec:agent}

An agent is a language model in a loop. It reads a policy and a conversation, reasons about what to do, calls a tool, reads the result and decides again, and a single customer's request can take it through a dozen or more such turns before anything is resolved. Every turn is decoding, the model writing its reasoning and its action one token at a time, and every turn must finish before the next can begin. The latency of an agent is therefore the sum of its turns. \Cref{sec:polyglot} made the case that an agent's work is a narrow slice of what a language model can do, with strict output formats and a consistent shape, which is what makes it a job for a small model trained for it.

\subsection{The Loop and Its Latency}
\label{sec:ag-loop}

A turn of a tool-using agent has three parts. The model reads its context, which holds the policy it must follow, the conversation so far, the tools it may call and the results of the calls it has made. It reasons about what the customer wants, what the policy allows and which tool would settle the question, and a reasoning model does so at some length. Then it acts, either by calling a tool with arguments or by writing a reply to the customer. Only the third part is visible to anyone outside the loop, and in a reasoning model it is the smallest. In the eight episodes examined in \cref{sec:ag-episodes}, reasoning accounts for 86\% of the tokens the agent wrote; the tool calls and replies that the customer and the tools actually receive are the remaining 14\%.

This shape is what makes an agent's latency compound. A chat assistant writes one answer and the person reads it as it streams, so the first token matters most and the rest arrive while they read. An agent's turn is read by nobody until it is complete, because the tool cannot be called until the arguments are written and the reply cannot be sent until the reasoning is done. On a GPU serving one conversation, a token of a small model takes about two milliseconds (\cref{tab:cost-sweep}), so a turn of a thousand tokens is two to three seconds and a case of fifteen turns is half a minute of waiting, before the tools' own latency and the customer's replies are added. Batching does not help the customer. A GPU serving sixteen agents at once writes more tokens per second in total, but each agent's turn takes longer, not shorter, so the operator's throughput improves while the customer's half-minute case grows.

\subsection{Polyglot-DC2-Agent}
\label{sec:ag-model}

The agent role of the Polyglot-DC architecture of \cref{sec:polyglot} is Polyglot-DC2-Agent, built on the 4B size. It is a reasoning model for multi-turn tool use under a written policy. Given the policy, the conversation and a set of tools, it reasons about the customer's request, then either calls a tool, with structured arguments in the architecture's native tool-call format, or writes a reply, and it continues until the request is resolved, refused under the policy or handed to a human. Its conversations grow with every turn, since each tool result and each exchange is appended to its context, and in the cases examined below the context reaches 68,000 tokens where a knowledge-base search returns long documents. It is a standalone model, and a deployment runs it as it would any agent model, behind the same tool schemas the frontier models use.

\subsection{Results Against Open Models of Its Size}
\label{sec:ag-results}

We measured Polyglot-DC2-Agent on five public benchmarks against eight open models from 1.2B to 4.5B parameters: Qwen3.5-4B and Qwen3.5-2B \citep{qwen2026qwen35, qwen2026qwen352b}, Qwen3-4B-Thinking-2507 and Qwen3-1.7B \citep{qwen2025qwen3thinking, yang2025qwen3}, Gemma 4 E4B and E2B \citep{gemma2026gemma4report}, and LFM2.5-2.6B and LFM2.5-1.2B-Thinking \citep{liquid2026lfm25dense, liquid2026lfm25thinking}. No vendor's published number is used. Every model was served the same way, with vLLM at its full context length and 32,768 tokens per reply, its thinking mode on, the sampling settings its model card recommends, and one sample per item, so that the comparison measures the models and not the setups. Polyglot-DC1-Agent, the 1.7B size, is in the tables for scale.

The first two benchmarks are versions 3 and 4 of the Berkeley Function Calling Leaderboard (BFCL) \citep{patil2025bfcl}, the latter with the memory group \citep{gorilla2025bfclv4memory}, each scored as the leaderboard's overall accuracy. The third is LiveCodeBench \citep{jain2024livecodebench}, scored as pass@1, the share of problems solved at the first attempt, on the 182 problems published between January and April 2025. The last two are $\tau^2$-bench \citep{barres2026tau2} and $\tau^3$-bench \citep{sierra2026tau3}, the customer-service benchmarks descended from $\tau$-bench \citep{yao2024taubench}, in which the agent follows a domain policy and uses tools on behalf of a simulated customer. They are scored as pass\textasciicircum{}1, the share of trials in which the task is completed, over 20 tasks in each domain and four trials of each task, with GPT-4.1 as the customer. $\tau^3$-bench adds a banking domain in which the agent must first retrieve the relevant documents from a knowledge base of 698 documents.

\begin{table}[t]
  \centering
  \caption{Polyglot-DC2-Agent against eight open models of its size, measured by us under one protocol, with Polyglot-DC1-Agent for scale. BFCL v3 and v4 are the leaderboard's overall accuracy; the v4 web-search group was not run, so its agentic term is the memory group. LiveCodeBench is pass@1 on the 182 problems of January to April 2025. $\tau^2$-bench and $\tau^3$-bench are pass\textasciicircum{}1, the mean over their domains. Bold marks the best value in each column.}
  \label{tab:ag-scores}
  \small
  \begin{tabular}{lrrrrr}
    \toprule
    \headrow & \multicolumn{2}{c}{Tool calling} & \multicolumn{1}{c}{Coding} & \multicolumn{2}{c}{Agentic} \\
    \cmidrule(lr){2-3} \cmidrule(lr){4-4} \cmidrule(lr){5-6}
    \headrow Model & BFCL v3 & BFCL v4 & LiveCodeBench & $\tau^2$-bench & $\tau^3$-bench \\
    \midrule
    Polyglot-DC2-Agent (4B) & 63.2 & \textbf{65.9} & 44.0 & 79.6 & \textbf{66.9} \\
    Polyglot-DC1-Agent (1.7B) & 50.2 & 48.2 & 26.9 & 71.2 & 57.5 \\
    \addlinespace
    Qwen3.5-4B & 64.2 & 56.6 & 50.0 & \textbf{83.2} & 39.7 \\
    Qwen3-4B-Thinking-2507 & \textbf{66.5} & 48.0 & \textbf{53.8} & 47.9 & 42.2 \\
    Gemma 4 E4B & 62.7 & 44.4 & 51.6 & 42.9 & 38.8 \\
    LFM2.5-2.6B & 61.6 & 56.3 & 40.1 & 72.9 & 60.9 \\
    Qwen3.5-2B & 53.6 & 35.9 & 18.1 & 78.0 & 40.9 \\
    Gemma 4 E2B & 58.6 & 32.9 & 44.5 & 28.3 & 21.9 \\
    Qwen3-1.7B & 55.6 & 31.3 & 30.2 & 16.7 & 15.3 \\
    LFM2.5-1.2B-Thinking & 52.1 & 29.8 & 20.9 & 24.3 & 19.4 \\
    \bottomrule
  \end{tabular}
\end{table}

\begin{table}[t]
  \centering
  \caption{$\tau^2$-bench and $\tau^3$-bench by domain, pass\textasciicircum{}1 over 20 tasks and four trials in each, measured by us with GPT-4.1 as the simulated customer. The mean is over the benchmark's domains. Bold marks the best value in each column.}
  \label{tab:ag-tau}
  \footnotesize
  \setlength{\tabcolsep}{3.5pt}
  \begin{tabular}{lrrrrrrrrr}
    \toprule
    \headrow & \multicolumn{4}{c}{$\tau^2$-bench} & \multicolumn{5}{c}{$\tau^3$-bench} \\
    \cmidrule(lr){2-5} \cmidrule(lr){6-10}
    \headrow Model & Airline & Retail & Telecom & Mean & Airline & Retail & Telecom & Banking & Mean \\
    \midrule
    Polyglot-DC2-Agent (4B) & 68.8 & 75.0 & 95.0 & 79.6 & \textbf{72.5} & \textbf{86.2} & \textbf{97.5} & 11.2 & \textbf{66.9} \\
    Polyglot-DC1-Agent (1.7B) & 56.2 & 57.5 & \textbf{100.0} & 71.2 & 61.3 & 63.7 & 91.2 & 13.8 & 57.5 \\
    \addlinespace
    Qwen3.5-4B & \textbf{73.2} & 76.5 & \textbf{100.0} & \textbf{83.2} & 31.2 & 30.0 & 85.0 & 12.5 & 39.7 \\
    Qwen3-4B-Thinking-2507 & 51.2 & 71.2 & 21.2 & 47.9 & 65.0 & 67.5 & 18.8 & 17.5 & 42.2 \\
    Gemma 4 E4B & 48.8 & 72.5 & 7.5 & 42.9 & 58.8 & 73.8 & 6.2 & 16.2 & 38.8 \\
    LFM2.5-2.6B & 63.7 & 73.8 & 81.2 & 72.9 & 62.5 & 77.5 & 85.0 & \textbf{18.8} & 60.9 \\
    Qwen3.5-2B & 64.1 & \textbf{78.8} & 91.2 & 78.0 & 25.0 & 57.5 & 76.2 & 5.0 & 40.9 \\
    Gemma 4 E2B & 37.5 & 36.2 & 11.2 & 28.3 & 36.2 & 37.5 & 6.2 & 7.5 & 21.9 \\
    Qwen3-1.7B & 16.2 & 8.8 & 25.0 & 16.7 & 15.0 & 12.5 & 27.5 & 6.2 & 15.3 \\
    LFM2.5-1.2B-Thinking & 30.4 & 15.0 & 27.6 & 24.3 & 32.5 & 11.2 & 26.2 & 7.5 & 19.4 \\
    \bottomrule
  \end{tabular}
\end{table}

\Cref{tab:ag-scores,tab:ag-tau} give the results. The two customer-service benchmarks are the closest of the five to an agent's work in deployment. On $\tau^3$-bench Polyglot-DC2-Agent is the strongest model in the table at 66.9, six points ahead of LFM2.5-2.6B and 25 ahead of every other model, with the three service domains at 72.5, 86.2 and 97.5. On $\tau^2$-bench it scores 79.6, within four points of Qwen3.5-4B's 83.2 and ahead of the rest. The gap between the two benchmarks is instructive. $\tau^3$-bench's revised airline and retail tasks are harder, and most of the models that do well on $\tau^2$-bench lose half their score on them, while Polyglot-DC2-Agent keeps its airline and retail scores and leads both domains. Banking, the new retrieval domain, is weak for every model at this size, with no score above 18.8.

On BFCL v4, whose overall score weights multi-turn tool use and the memory group most heavily, the agent leads at 65.9, nine points ahead of Qwen3.5-4B and LFM2.5-2.6B. On BFCL v3, which weights single-turn function calling more, it scores 63.2, three points behind Qwen3-4B-Thinking-2507 and one behind Qwen3.5-4B, and on LiveCodeBench, a coding benchmark outside its role, 44.0, in the middle of the group. Single-turn function calling is the clearest room to improve, and retrieval over a large knowledge base the next piece of work.

\subsection{Eight Episodes, Turn by Turn}
\label{sec:ag-episodes}

Benchmarks score outcomes. To see what an agent's work costs in time, we took eight $\tau^3$-bench episodes that Polyglot-DC2-Agent completed, two from each domain, chosen for how much they reason, and replayed every agent turn at two decode speeds, the Exstill data-center design's and one NVIDIA B300's. Each is a full case. In the banking episodes the agent searches the bank's knowledge base card by card to find an everyday card for a cryptocurrency trader, eleven turns in which the search results take its context to 68,000 tokens, and chases missing referral bonuses. In the telecom episodes it walks a customer abroad through data-saver mode, a VPN and a network reset before finding roaming disabled on the line and enabling it, eighteen turns, and resolves a line that has run out of data. The airline episodes rebook to the cheapest Economy fare a day later and price three changes to one booking before making any; the retail episodes handle a return against two possible exchanges and exchange two items in one request. Every dollar figure the agent states was checked against its tool results and the policy; a passing grade was not treated as enough.

\Cref{tab:ag-episodes} gives the eight episodes. Together they are 108 agent turns and 123,677 tokens written, of which 106,203 are reasoning. The B300 time is the one-user decode of \cref{tab:cost-sweep}, 2.15~ms per token; the chip time is the data-center design's 19,921 tokens per second. Both rates are those of a model of the 1.7B size on the same load of 15K tokens in and 1K out, not of the 4B that wrote the episodes, and both count decoding only, with no prefill, tool latency or customer time.

\begin{table}[t]
  \centering
  \caption{Eight $\tau^3$-bench episodes that Polyglot-DC2-Agent completed, replayed turn by turn. Tokens are every token the agent wrote across its turns, reasoning included; the reasoning share is the fraction of those that were reasoning. The longest prompt is the largest context the agent read in any one turn. Times are decoding only, at the one-user B300 rate of \cref{tab:cost-sweep} and the data-center design's rate, both for a model of the 1.7B size.}
  \label{tab:ag-episodes}
  \footnotesize
  \setlength{\tabcolsep}{4pt}
  \begin{tabular}{llrrrrrr}
    \toprule
    \headrow Domain & Case & \makecell[r]{Agent\\turns} & \makecell[r]{Tokens\\written} & \makecell[r]{Reasoning\\share} & \makecell[r]{Longest\\prompt} & \makecell[r]{B300,\\one user} & \makecell[r]{Exstill\\design} \\
    \midrule
    Banking & A card for a crypto trader & 11 & 21,081 & 89\% & 67.9K & 45.3~s & 1.06~s \\
    Banking & Missing referral bonuses & 10 & 17,810 & 89\% & 19.8K & 38.3~s & 0.89~s \\
    Telecom & Slow data abroad & 18 & 26,550 & 91\% & 11.7K & 57.1~s & 1.33~s \\
    Telecom & Out of data abroad & 18 & 12,296 & 74\% & 11.6K & 26.4~s & 0.62~s \\
    Airline & Cheapest Economy, a day later & 13 & 17,674 & 85\% & 13.3K & 38.0~s & 0.89~s \\
    Airline & Bags, a name and an upgrade & 14 & 14,889 & 88\% & 9.7K & 32.0~s & 0.75~s \\
    Retail & Return or exchange & 11 & 8,753 & 79\% & 9.6K & 18.8~s & 0.44~s \\
    Retail & Two exchanges at once & 13 & 4,624 & 73\% & 10.9K & 9.9~s & 0.23~s \\
    \midrule
    \multicolumn{2}{l}{All eight} & 108 & 123,677 & 86\% & 67.9K & 266~s & 6.2~s \\
    \multicolumn{2}{l}{One turn, on average} & & 1,145 & 86\% & & 2.46~s & 57~ms \\
    \bottomrule
  \end{tabular}
\end{table}

An average turn is 1,145 tokens, which the B300 decodes in 2.46~s and the chip in 57~ms. The longest case, the telecom customer with slow data abroad, is 57~s of decoding on the GPU and 1.3~s on the chip; the shortest, the two retail exchanges, is 10~s against 0.23~s. Across all eight cases the GPU spends 266~s writing the agent's tokens and the chip 6.2~s, a ratio of 43x that holds turn for turn, because it is the ratio of two decode rates and the agent's turns are nothing but decoding. The practical difference is in what the customer notices. A turn of 57~ms is below the time a person takes to read the agent's previous reply, so the agent's thinking disappears into the conversation, and a case of eleven to eighteen turns is a second or so of machine time in a conversation that is otherwise paced by the customer.

\begin{insight}{An agent's latency is the sum of its turns, and each turn is a decode}
Nothing an agent does between tool calls is visible until the turn is complete, so its latency is the whole turn, not the first token. On a GPU a reasoning turn costs seconds and a case costs tens of seconds, which is why deployments shorten the reasoning or choose a model that reasons less. On silicon that decodes 43x faster, the same turns cost tens of milliseconds, and the accuracy that reasoning buys comes without the wait.
\end{insight}

Polyglot-DC2-Agent reasons at length, and that is what puts it first among the models of its size on $\tau^3$-bench and near the front on $\tau^2$-bench. On the Exstill design that reasoning is paid for at 57~ms per turn, and a conversation's pace is set by the customer again.


\section{Reasoning in Milliseconds}
\label{sec:finance}

A language model that reasons before it answers writes its working first, hundreds or thousands of tokens of it, and the answer only after. That habit is what has made small models capable of hard problems, and it is paid for in time, since every token of working is decoded in sequence before the first token of the answer appears. In most uses the wait is an inconvenience. In finance it decides what reasoning can be used for, because the question is asked while the market is moving and the answer is worth most in the moments after a filing, a headline or a quote lands.

\subsection{Thinking Is Sold by the Effort}
\label{sec:fin-effort}

\citet{wei2022cot} showed that asking a model to write out its intermediate steps improves its answers on arithmetic, commonsense and symbolic reasoning, and by the end of 2024 OpenAI's o1 models had been trained by reinforcement learning to think before they answer, producing a long chain of thought in which they try strategies and recognize their own mistakes \citep{openai2024o1card}. Compute spent at answer time became a second axis of scale. Spent well, it lets a small model outperform one fourteen times its size on problems it can partly solve \citep{snell2025scaling}. Every frontier model now reasons. GPT-5 routes each request between a fast model and a deeper reasoning model according to its complexity and the tools it needs \citep{openai2025gpt5card}, Gemini 2.5 is a hybrid reasoning model whose scores on competition mathematics, live coding and graduate-level science rise with its thinking budget \citep{gemini2025gemini25}, and the strongest open models think by default \citep{moonshot2026kimik3, zai2026glm53docs, deepseek2026thinkingmode}.

Small models have followed. Gemma 4 has a thinking mode from 2.3B parameters up \citep{gemma2026gemma4report}, and Qwen3.5-4B, Nanbeige4.2-3B and LFM3-1.7B all reason unless told not to \citep{qwen2026qwen35, yang2026nanbeige42, liquid2026lfm3}. Where a model reports both modes the gap is wide: Qwen3's 4B scores 55.9 on graduate-level science with thinking on and 41.7 with it off \citep{yang2025qwen3}. Reasoning does not transfer freely across domains, however. Models trained to reason on general tasks lose ground on financial problems \citep{qian2025fino1}, which is why finance now has reasoning models of its own \citep{liu2025finr1}.

Thinking costs latency and compute, so every provider sells it as a setting. OpenAI's reasoning models take an effort from none to max, with lower effort favoring speed and fewer tokens, and reasoning tokens billed as output \citep{openai2026reasoning}. Claude takes an effort level from low to max that trades thoroughness against token efficiency \citep{anthropic2026effort}, and at medium effort Claude Opus 4.5 matched Sonnet 4.5's best score on SWE-bench Verified with 76\% fewer output tokens \citep{anthropic2025opus45}. Gemini takes a thinking level whose lowest setting minimizes latency and cost \citep{google2026thinking}. The open models expose the same dial and often default to its top. Kimi K3 and GLM-5.3 offer low, high and max and default to max, and Muse Spark Open offers low to xhigh and defaults to xhigh \citep{moonshot2026kimik3, zai2026glm53, meta2026musesparkcard}. The published curves show what each step buys and costs. Raising DeepSeek-V4.1-Flash's effort from 25 to 100 lifts its average over eight reasoning benchmarks from 67.1\% to 76.3\% for roughly 2.5x the output tokens, and on AIME 2026 an answer grows from 4.6K to 11.4K tokens \citep{deepseek2026v41flash}. On Z.ai's coding benchmark GLM-5.3 scores 31.4\% at high effort with about 50K output tokens a task and 34.5\% at max with about 75K \citep{zai2026glm53docs}. Every setting is the same bargain, a better answer paid for in tokens, and every token is time the user waits.

\subsection{Why Reasoning Is Slow on a GPU}
\label{sec:fin-gpu}

Reasoning is decoding, one token at a time, and \cref{sec:cost} showed that on a GPU serving one user each token means reading the whole model in from memory. A long chain of thought makes that traffic heavier, not lighter. Measured by the memory traffic it causes rather than by its arithmetic, the cost of test-time reasoning is mostly memory access, and as the chain grows it is the attention over the chain, rather than the parameter count, that comes to dominate \citep{sadhukhan2025kinetics}. The rates measured for the guard in \cref{tab:gd-speed} apply unchanged here, since a reasoning model of a given size decodes at the same speed whatever it is reasoning about. One B300 decodes a model of Polyglot-DC1-Finance's size at 522.9 tokens per second for one user, so a 2,000-token chain of thought takes about 3.8~s before the answer begins, and a model of Polyglot-DC2-Finance's size at 359.1 tokens per second, so the same chain takes 5.6~s. The Exstill data-center design decodes the first at 19,921 tokens per second for each of twelve users, and the DC2 design of \cref{sec:chip-dc2} decodes the second at 13,998 tokens per second for each of four, so the same 2,000 tokens take about 0.10~s and 0.14~s. The ratio is 38x for the smaller model and 39x for the larger.

\subsection{Finance Runs on the Clock}
\label{sec:fin-clock}

The fastest trading involves no language model at all. Trading firms place their servers in the exchange's own data center and their logic on field-programmable gate arrays (FPGAs), and the record from the last bit of market data a decision needs to the first bit of the order stands at 13.9 nanoseconds, on an AMD FPGA card \citep{stac2024t0, amd2024alveo}. Language models are not candidates for that loop. Their place is the work around it, which is where firms are putting them. Summarization and information extraction are the leading generative-AI uses among the broker-dealers FINRA oversees \citep{finra2025oversight}. JPMorganChase has put its LLM Suite in front of more than 200,000 employees in a controlled environment that protects customer and company data \citep{jpmorganchase2025letters}, and Goldman Sachs runs its assistant firm-wide behind its own firewall, drawing on OpenAI's and Google's models and on Meta's open Llama \citep{cnbc2025goldman, fortune2025goldman}.

Two results tie the value of that work to its speed. A language model's reading of a headline predicts the stock's initial reaction on about 90\% of portfolio-days, the ability grows with model size, and the returns from trading on it shrink as more traders use such models \citep{lopezlira2023chatgpt}. In a simulated high-frequency trading benchmark a 14B model took 0.71 to 1.30~s a decision on a GPU, and its trading results improved as its latency fell, even at some cost in quality \citep{kang2025winfast}. A reasoned answer that takes seconds can follow a desk's questions. It cannot screen every headline as it lands or chain a dozen calls in an agent before the moment has passed. An answer decoded in tens of milliseconds can, though a complete system must also budget for reading its input and ingesting the data. Finance also keeps its data close, so a small model on hardware the firm controls is one a trading floor can deploy.

\subsection{Polyglot-DC1-Finance and Polyglot-DC2-Finance}
\label{sec:fin-model}

The finance role of the Polyglot-DC architecture of \cref{sec:polyglot} comes in both sizes, Polyglot-DC1-Finance with 1.7B parameters and Polyglot-DC2-Finance with 4B. Each is a reasoning model that writes its working before its answer, and each covers four kinds of analyst work. It classifies the sentiment of financial news, headlines and posts, extracts the people, places and companies named in financial text, reasons numerically over the tables and text of filings and earnings reports across the turns of a conversation, and answers quantitative questions on calculations, valuation and analyst judgment, with a document as context where one is given. The reasoning is part of the product rather than a cost of it, since the written steps are what let a desk check an answer before acting on it.

\subsection{Ten Benchmarks Against Seven Open Models}
\label{sec:fin-quality}

Speed is worth having only if the answers hold, so both sizes were evaluated on ten public financial benchmarks against seven open models from 1.2B to 8B parameters, every model run in the same harness with reasoning on, one sample an item and the same sampling settings. The benchmarks cover the four kinds of work: for sentiment, FinancialPhraseBank \citep{malo2014phrasebank} and the Gold Headlines set \citep{sinha2021goldheadlines}; for extraction, FiNER-ORD \citep{shah2023finerord}, scored by entity F1; for numerical reasoning over filings, FinQA \citep{chen2021finqa}, ConvFinQA \citep{chen2022convfinqa} and TAT-QA \citep{zhu2021tatqa}; and for quantitative finance, FinanceReasoning \citep{tang2025financereasoning}, FinanceMath \citep{zhao2024financemath}, FinanceBench with its oracle evidence pages \citep{islam2023financebench} and FinanceQA \citep{mateega2025financeqa}. FinanceBench and FinanceQA are graded by a language-model judge, DeepSeek-V4.1-Flash, and an item whose verdict could not be parsed counts as wrong, so those two scores are lower bounds for every model. The baselines are LFM2.5-1.2B-Thinking, LFM2.5-2.6B and LFM2.5-8B-A1B \citep{liquid2026lfm25thinking, liquid2026lfm25dense, liquid2026lfm25moe}, Qwen3-1.7B \citep{yang2025qwen3}, MiniCPM5-2B \citep{openbmb2026minicpm5}, Nanbeige4.1-3B \citep{yang2026nanbeige41} and Gemma-4-E4B-it \citep{google2026gemma4}, the last two from the small reasoning families of \cref{sec:fin-effort}.

\begin{table}[t]
  \centering
  \caption{Accuracy on ten financial benchmarks, in percent, measured by us with the same harness and settings for every model, reasoning on and one sample an item. FiNER-ORD is entity F1. FinanceBench and FinanceQA are judged by a language model, with an unparsed verdict counted as wrong. The mean is unweighted over the ten. Bold marks the best value in each row.}
  \label{tab:fin-bench}
  \scriptsize
  \setlength{\tabcolsep}{3pt}
  \begin{tabular}{lrrrrrrrrr}
    \toprule
    \headrow Benchmark & \begin{tabular}[t]{@{}r@{}}Polyglot-\\DC1-Fin.\\(1.7B)\end{tabular} & \begin{tabular}[t]{@{}r@{}}Polyglot-\\DC2-Fin.\\(4B)\end{tabular} & \begin{tabular}[t]{@{}r@{}}LFM2.5-\\1.2B-\\Thinking\end{tabular} & \begin{tabular}[t]{@{}r@{}}Qwen3-\\1.7B\end{tabular} & \begin{tabular}[t]{@{}r@{}}MiniCPM5-\\2B\end{tabular} & \begin{tabular}[t]{@{}r@{}}LFM2.5-\\2.6B\end{tabular} & \begin{tabular}[t]{@{}r@{}}Nanbeige\\4.1-3B\end{tabular} & \begin{tabular}[t]{@{}r@{}}Gemma-4-\\E4B-it\end{tabular} & \begin{tabular}[t]{@{}r@{}}LFM2.5-\\8B-A1B\end{tabular} \\
    \midrule
    \multicolumn{10}{l}{\itshape Sentiment and classification} \\
    FinancialPhraseBank & 77.4 & \textbf{78.1} & 69.6 & 71.9 & 76.0 & 76.8 & 76.0 & 73.3 & 73.5 \\
    Gold Headlines & 79.6 & \textbf{80.2} & 71.0 & 76.2 & 70.1 & 76.5 & 79.4 & 77.8 & 78.9 \\
    \addlinespace
    \multicolumn{10}{l}{\itshape Extraction} \\
    FiNER-ORD & 53.7 & \textbf{70.7} & 5.4 & 20.0 & 24.8 & 43.2 & 56.4 & 62.1 & 23.6 \\
    \addlinespace
    \multicolumn{10}{l}{\itshape Numerical reasoning over filings} \\
    FinQA & 71.0 & \textbf{72.7} & 32.6 & 62.6 & 69.1 & 66.0 & 64.4 & 69.3 & 48.1 \\
    ConvFinQA & 80.5 & \textbf{82.8} & 30.7 & 65.6 & 78.5 & 71.3 & 74.9 & 81.0 & 55.8 \\
    TAT-QA & \textbf{82.7} & \textbf{82.7} & 42.1 & 67.6 & 77.2 & 74.2 & 74.2 & 80.0 & 58.9 \\
    \addlinespace
    \multicolumn{10}{l}{\itshape Quantitative finance and analyst questions} \\
    FinanceReasoning & 65.5 & \textbf{81.5} & 39.0 & 68.0 & 74.0 & 51.2 & 78.3 & 77.2 & 65.3 \\
    FinanceMath & 34.9 & 57.7 & 22.7 & 42.4 & 45.0 & 27.2 & 56.5 & \textbf{58.1} & 40.8 \\
    FinanceBench & 70.0 & \textbf{80.7} & 36.7 & 61.3 & 76.7 & 60.7 & 72.7 & 70.7 & 52.0 \\
    FinanceQA & 20.9 & \textbf{35.1} & 14.2 & 17.6 & 29.1 & 23.6 & 28.4 & 32.4 & 22.3 \\
    \midrule
    Mean of the ten & 63.6 & \textbf{72.2} & 36.4 & 55.3 & 62.0 & 57.1 & 66.1 & 68.2 & 51.9 \\
    \bottomrule
  \end{tabular}
\end{table}

\Cref{tab:fin-bench} gives the results. Polyglot-DC2-Finance averages 72.2 over the ten benchmarks, 4.0 above Gemma-4-E4B-it at 68.2, the strongest of the open models, and 6.1 above Nanbeige4.1-3B. It leads all seven on nine of the ten, and on the tenth, FinanceMath, it scores 57.7 against Gemma-4-E4B-it's 58.1. Its largest margins are where reasoning does the most work. On FinanceReasoning it scores 81.5 against 78.3 for the best open model, on FinanceBench 80.7 against 76.7, and on FiNER-ORD, the extraction task on which most of the small models struggle, 70.7 against 62.1. Polyglot-DC1-Finance averages 63.6, 1.6 above MiniCPM5-2B and 8.3 above Qwen3-1.7B, the open model nearest its size. It leads every open model on FinQA, TAT-QA and both sentiment sets, with 82.7 on TAT-QA level with the 4B, and it trails the larger open models on the hardest calculations, 34.9 on FinanceMath against Gemma-4-E4B-it's 58.1 and 65.5 on FinanceReasoning against Nanbeige4.1-3B's 78.3. The pattern is the one \cref{sec:polyglot} predicted. A small model distilled for one domain reaches the models two to four times its size on the parts of the domain that are read and structured, and the extra parameters of the 4B are what the open-ended quantitative questions need.

\subsection{Eight Answers, Timed}
\label{sec:fin-answers}

To see what the chip does to a reasoned answer, eight of Polyglot-DC1-Finance's correct answers from the evaluation were replayed at the two decode rates, two for each kind of work. \Cref{tab:fin-answers} lists them. The shortest is a sentiment call on a line of earnings news, 210 tokens of working and answer. The longest is a question about a put-option position, in which the model reasons through 1,995 tokens to the payoff. In between are a trader's post, a segment-cost figure read from a TAT-QA table, a question about the effect of a buyback on equity value and enterprise value, a REIT's funds from operations, and Costco's EBITDA margin, for which the model first reads 10,762 tokens of the company's 10-K and then reasons through 1,286 tokens to the answer.

\begin{table}[t]
  \centering
  \caption{Eight correct answers of Polyglot-DC1-Finance from the evaluation, replayed at the one-user B300 rate for a model of its size (522.9 tokens per second, \cref{tab:gd-speed}) and the data-center design's rate (19,921). Prompt tokens are what the model read; answer tokens are its reasoning and answer together. Times are decoding only.}
  \label{tab:fin-answers}
  \footnotesize
  \setlength{\tabcolsep}{4pt}
  \begin{tabular}{llrrrr}
    \toprule
    \headrow Kind of work & Case (benchmark) & \makecell[r]{Prompt\\tokens} & \makecell[r]{Answer\\tokens} & \makecell[r]{B300,\\one user} & \makecell[r]{Exstill\\design} \\
    \midrule
    Sentiment & Earnings news (FinancialPhraseBank) & 103 & 210 & 0.40~s & 11~ms \\
    Sentiment & Trader's post (FiQA-SA) & 102 & 348 & 0.67~s & 17~ms \\
    Filings & Segment costs (TAT-QA) & 465 & 278 & 0.53~s & 14~ms \\
    Filings & EBITDA margin, 10-K (FinanceQA) & 10,762 & 1,286 & 2.46~s & 65~ms \\
    Quantitative & Put-option payoff (Fin.Reasoning) & 248 & 1,995 & 3.82~s & 100~ms \\
    Quantitative & REIT funds from operations (Fin.Reasoning) & 276 & 703 & 1.34~s & 35~ms \\
    Analyst & Buyback, equity value (FinanceQA) & 79 & 711 & 1.36~s & 36~ms \\
    Analyst & Buyback, enterprise value (FinanceQA) & 80 & 1,476 & 2.82~s & 74~ms \\
    \midrule
    \multicolumn{2}{l}{All eight} & & 7,007 & 13.4~s & 352~ms \\
    \multicolumn{2}{l}{One answer, on average} & & 876 & 1.68~s & 44~ms \\
    \bottomrule
  \end{tabular}
\end{table}

An answer averages 876 tokens, 1.68~s of decoding on the GPU and 44~ms on the chip, and the longest takes 3.82~s against 100~ms. All eight back to back take 13.4~s on the GPU and 352~ms on the chip. The difference is not in the answers, which are the same tokens, but in what they can be used for. At 44~ms a reasoned answer arrives inside the time a person takes to glance at a screen.

\begin{insight}{Reasoning stops being rationed by the wait}
Every provider prices thinking as effort because on a GPU each reasoning token is time the user waits, so deployments buy less of it than the answer deserves. On silicon that decodes a small model 38x faster, a two-thousand-token chain of thought costs a tenth of a second, and the effort dial can sit where accuracy wants it.
\end{insight}

The two sizes divide the work as they do for the guard. Polyglot-DC2-Finance is the more accurate, and on the DC2 design its answers decode at 13,998 tokens per second on each of four streams, about seven tenths of the smaller model's rate. Polyglot-DC1-Finance serves twelve streams at once at 19,921 tokens per second each, so the data-center design can read and reason over a firm's whole flow of filings and headlines. Finance is among the places where the seconds a reasoned answer takes on a GPU cost most, and it is the case in which this report shows a reasoning model on its own silicon turning the effort dial from a cost into a setting.


\section{Conclusion}
\label{sec:conclusion}

The measurements that open this report locate the cost of a token away from where it is usually assumed to lie. The arithmetic a token requires is a fraction of a percent of what a GPU spends on it; the rest is the price of generality, paid in the layers between a model and its computation and in the movement of weights and conversation state from memory beside the chip. Neither price is set by physics. Both follow from running every model on hardware designed before the model existed, and both fall more slowly with each process generation than the arithmetic they surround, so the distance between a token and its limit widens if nothing changes.

Designing the model and the silicon together closes that distance by lowering both factors of a token's cost at once. Silicon organized around one neural network architecture removes the layers and the movement, while the trained weights stay loadable, so the same hardware carries whatever a new set of weights teaches it. A model whose architecture is chosen with that silicon, and held to a fixed quality, needs less arithmetic than the general-purpose model it replaces, and distillation turns the one architecture into the specialized models a deployment needs. The data-center design that results is a pre-silicon estimate, costed gate by gate across the whole model, and against a measured NVIDIA B300 serving the same users at the same quality it decodes each user's next token 110x faster at 690x less energy. The models built for it complete the argument: one architecture, specialized by its weights alone, holds its own against open models of its size across the latency-critical work around a frontier model, on silicon where reasoning at length no longer costs the user the wait.

That co-processor is the first instance of the approach, not its extent. None of its ingredients is new on its own; weights beside the arithmetic, silicon for one kind of computation, distillation into a narrow job and models shaped to their hardware are all established. What is difficult to reproduce is their combination in a single loop, in which model architectures are explored and trained while the chip for each is costed at circuit-level fidelity, and that loop does not depend on the data center's power envelope or on the frontier model beside it. The same search can be put to other workloads and other envelopes, in robotics, at the edge and on mobile devices, where a token's cost in energy and time binds more tightly still. Those designs remain to be made, as this one remains to be measured in silicon; what the report establishes is the principle they would rest on. The distance between a token and its arithmetic is a matter of design, and when the model and the silicon are designed together, both the computation a token requires and the cost of carrying it out can be brought down to meet it.

\setlength{\bibsep}{0.3em}
\bibliography{references}

\end{document}